\documentclass[aps,prb, floatfix,notitlepage,nofootinbib,twocolumn,superscriptaddress,reprint,preprintnumbers]{revtex4-1}
\usepackage{amsmath}
\usepackage{amssymb}
\usepackage{graphicx}
\usepackage[tight]{subfigure}
\usepackage{enumitem}
\usepackage{tikz}
\usetikzlibrary{calc,fadings,decorations.pathreplacing,shapes,shapes.multipart,arrows,shapes.misc,intersections,positioning}
\usepackage{wrapfig}
\usepackage{bm}
\usepackage[autostyle, english = american]{csquotes}
\usepackage{xifthen}
\usepackage{xargs}
\usepackage[tight]{subfigure}
\MakeOuterQuote{"}
\usepackage{pifont}
\usepackage{array}
\usepackage{multirow}

\usepackage[english]{babel}
\makeatletter
\def\bbl@set@language#1{%
  \edef\languagename{%
    \ifnum\escapechar=\expandafter`\string#1\@empty
    \else\string#1\@empty\fi}%
  \@ifundefined{babel@language@alias@\languagename}{}{%
    \edef\languagename{\@nameuse{babel@language@alias@\languagename}}%
  }%
  \select@language{\languagename}%
  \expandafter\ifx\csname date\languagename\endcsname\relax\else
    \if@filesw
      \protected@write\@auxout{}{\string\select@language{\languagename}}%
      \bbl@for\bbl@tempa\BabelContentsFiles{%
        \addtocontents{\bbl@tempa}{\xstring\select@language{\languagename}}}%
      \bbl@usehooks{write}{}%
    \fi
  \fi}
\newcommand{\DeclareLanguageAlias}[2]{%
  \global\@namedef{babel@language@alias@#1}{#2}%
}
\makeatother
\DeclareLanguageAlias{en}{english}

\newcolumntype{C}{>{$\displaystyle} c <{$}}

\newcommand{\etc}{\textit{etc.} }

\newcommand{\up}{\mathord{\uparrow}}

\newcommand{\I}{\mathbb{I}}

\usepackage{bm}
\renewcommand{\vec}[1]{\boldsymbol{\mathbf{#1}}}

\newcommand{\tr}[0]{\text{tr}}

\usepackage[colorlinks=true,citecolor=blue,linkcolor=magenta]{hyperref}

\graphicspath{{./}{./images/}}

\usepackage{xifthen}
\usepackage{xargs}

\newcommandx{\drawbox}[6][1=0,2=0,3=1,4=1,5=,6=]{
  \ifthenelse{\equal{#5}{}}{
    \draw[line width = 0.5pt] (#1,#2) rectangle (#3,#4);
  }{
    \draw[line width = 0.5pt, fill = #5] (#1,#2) rectangle (#3,#4);
  }
  \node () at (#1*0.5+#3*0.5,#2*0.5+#4*0.5) {#6};
}

\newcommandx{\regbox}[4][1=0,2=0,3=,4=]{
  \begin{scope}[shift={(#1,#2)}]
    \drawbox[0][0][1.5][1][#3][#4];
  \end{scope}
}

\newcommandx{\smallarc}[2][1=,2=]{
  \ifthenelse{\equal{#2}{r}}{
    \pgfmathsetmacro{\flag}{-1};
  }{
    \pgfmathsetmacro{\flag}{1};
  }
  \ifthenelse{\equal{#1}{}}{
    \draw[line width = 0.5pt] (0,0) to[out=\flag*90,in=180] (0.5,\flag*0.5) to[out=0,in=\flag*90] (1,0);
  }{
    \draw[line width = 0.5pt] (0,0) to[out=\flag*90,in=190] (0.5-0.2,\flag*0.45);
    \draw[line width = 0.5pt] (0.5+0.2,\flag*0.45) to[out=-10,in=\flag*90] (1,0);
    \node () at (0.5,\flag*0.45) {#1};
  }
}

\newcommandx{\lowarc}[2][1=,2=]{
  \ifthenelse{\equal{#2}{r}}{
    \pgfmathsetmacro{\flag}{-1};
  }{
    \pgfmathsetmacro{\flag}{1};
  }
  \ifthenelse{\equal{#1}{}}{
    \draw[line width = 0.5pt] (0,0) to[out=\flag*90,in=180] (0.3,\flag*0.3) to[out=0,in=\flag*90] (0.6,0);
  }{
    \draw[line width = 0.5pt] (0,0) to[out=\flag*90,in=190] (0.3-0.2,\flag*0.28);
    \draw[line width = 0.5pt] (0.3+0.2,\flag*0.28) to[out=-10,in=\flag*90] (0.6,0);
    \node () at (0.3,\flag*0.28) {#1};
  }
}

\newcommandx{\higharc}[2][1=,2=]{
  \ifthenelse{\equal{#2}{r}}{
    \pgfmathsetmacro{\flag}{-1};
  }{
    \pgfmathsetmacro{\flag}{1};
  }
  \ifthenelse{\equal{#1}{}}{
    \draw[line width = 0.5pt] (0,0) to[out=\flag*90,in=180] (1,\flag*0.5) to[out=0,in=\flag*90] (2,0);
  }{
    \draw[line width = 0.5pt] (0,0) to[out=\flag*90,in=180] (1-0.2,\flag*0.55);
    \draw[line width = 0.5pt] (1+0.2,\flag*0.55) to[out=0,in=\flag*90] (2,0);
    \node () at (1,\flag*0.55) {#1};
  }
}

\newcommandx{\idst}[5][1=0,2=0,3=,4=,5=]{
  \begin{scope}[shift={(#1,#2)}]
    \smallarc[#3][#5]
    \begin{scope}[shift={(1.5,0)}]
      \smallarc[#4][#5]
    \end{scope}
  \end{scope}
}

\newcommandx{\swapst}[5][1=0,2=0,3=,4=,5=]{
  \begin{scope}[shift={(#1,#2)}]
    \higharc[#3][#5]
    \begin{scope}[shift={(0.75,0)}]
      \lowarc[#4][#5]
    \end{scope}
  \end{scope}
}

\newcommandx{\fineq}[4][1=-.8ex,2=1,3=1]{
  \begin{tikzpicture}[baseline={([yshift=#1]current  bounding  box.center)}, scale = #2, every node/.style={scale = #3}]
    #4
  \end{tikzpicture}
}

\newcommandx{\ugate}[5][1=0,2=0,3=,4=,5=]{
  \begin{scope}[shift={(#1,#2)}]
      \draw[line width = 0.5pt] (0,0)--++(0,0.25);
      \draw[line width = 0.5pt] (0,1.25)--++(0,0.25);
      \draw[line width = 0.5pt] (1,0)--++(0,0.25);
      \draw[line width = 0.5pt] (1,1.25)--++(0,0.25);
      \regbox[-0.25][0.25][#3][#4];
      \ifthenelse{\equal{#5}{p}}{
        \draw[line width = 2pt] (-0.1,0)--(1.1,0);
        \draw (0,0)--++(0,-0.1);
        \draw (1,0)--++(0,-0.1);
      }{}
  \end{scope}
}

\begin{document}

\title{Operator Growth from Global Out-of-time-order Correlators}

\author{Tianci Zhou}
\email{tzhou13@mit.edu}
\affiliation{Kavli Institute for Theoretical Physics, University of California, Santa Barbara, CA 93106, USA}
\affiliation{Center for Theoretical Physics, Massachusetts Institute of Technology, Cambridge, Massachusetts 02139, USA}
\author{Brian Swingle}
\affiliation{Brandeis University, Waltham, MA 02453, USA}
\date{\today}

\begin{abstract}
In the context of chaotic quantum many-body systems, we show that operator growth, as diagnosed by out-of-time-order correlators of local operators, also leaves a sharp imprint in out-of-time-order correlators of global operators. In particular, the characteristic spacetime shape of growing local operators can be accessed using global measurements without any local control or readout. Building on an earlier conjectured phase diagram for operator growth in chaotic systems with power-law interactions, we show that existing nuclear spin data for out-of-time-order correlators of global operators are well fit by our theory. We also predict super-polynomial operator growth in dipolar systems in 3d and discuss the potential observation of this physics in future experiments with nuclear spins and ultra-cold polar molecules.
\end{abstract}
\preprint{MIT-CTP/5368}
\maketitle

\section{Introduction}
\label{sec:intro}

Out-of-time order correlators (OTOCs) play an important role in the study of quantum chaos. Although these objects first appeared in the literature many years ago\cite{larkin_quasiclassical_1969}, interest in them was recently reignited by studies of large-$N$ holographic systems where an initial period of exponential growth was argued to provide a quantum generalization of a classical Lyapunov exponent\cite{shenker_black_2014,kitaev2015,maldacena_bound_2016,shenker_stringy_2014}. Since then, a large body of work has explored OTOCs in a wide variety of contexts\cite{nahum_operator_2018,von_keyserlingk_operator_2018,aleiner_microscopic_2016,xu_locality_2018,davison_thermoelectric_2017,gu_energy_2017,liao_nonlinear_2018,zhou_operator_2019}, with one major conclusion being that the Lyapunov behavior seen in large-$N$ models is non-generic in spatially local models. Motivated by various experimental systems\cite{Monroe13,yan_observation_2013,Britton12,Blatt12,Lukin17,Bollinger17}, particularly experiments with nuclear spins\cite{PhysRevA.80.052323,wei_exploring_2018,wei_emergent_2019,sanchez_clustering_2014,alvarez_localization-delocalization_2015,sanchez_perturbation_2020}, the community has also considered OTOCs in systems with power-law interactions\cite{zhou_operator_2020,chen_quantum_2019}.

\begin{figure}[h]
\centering
\includegraphics[width=\columnwidth]{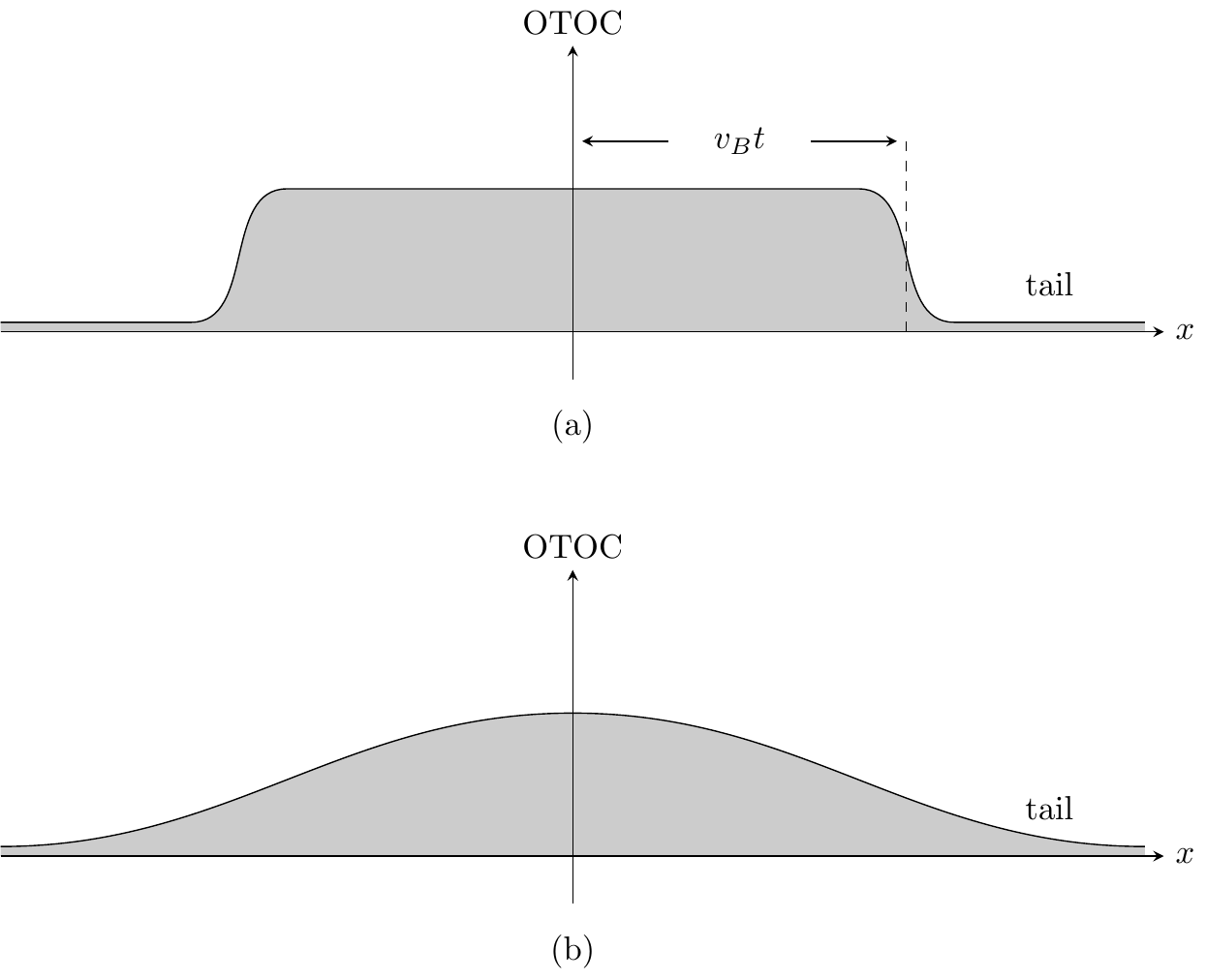}
\caption{The local OTOC probes the expansion of the time evolved operator. The global OTOC is approximately the area under the local OTOC curve. Here $x$ denotes the separation between the operators. (a) System with local interactions, where there is a characteristic velocity (the butterfly velocity) $v_B$ for the expansion. (b) System with long-range interactions, where the OTOC curve may have long tails and the light cone may not be sharp. }
\label{fig:h_area}
\end{figure}

The microphysics underlying the dynamics of OTOCs is the growth of Heisenberg operators\cite{mezei_entanglement_2016,nahum_operator_2018,von_keyserlingk_operator_2018,xu_locality_2018,chen_quantum_2019,zhou_operator_2019,qi_quantum_2018,roberts_operator_2018}. To illustrate the physics, consider a spin system defined on a spatial grid and let $Z_i$ be a Pauli-$z$ operator at site $i$ and $Z_i(t) = e^{i H t} Z_i e^{- i H t}$ be the corresponding Heisenberg operator. The infinite temperature OTOC between $Z_i$ and the Pauli-$x$ operator at site $j$ is $\tr( [ Z_i(t) , X_j] [ Z_i(t) , X_j]^\dagger)/ \tr( \I )$. At time zero, the OTOC is a delta function in space, since $Z_i$ and $X_j$ commute unless $i=j$. At later times, the operator $Z_i(t)$ grows in complexity and spreads in space leading the OTOC to become non-zero when $j$ is within a ball of time-dependent radius centered at $i$. This behavior is illustrated in the top panel of Fig.~\ref{fig:h_area} where ballistic operator growth characteristic of local interactions is shown.

There is considerable interest in exploring the physics of OTOCs in experiments, both to verify the overall physical picture in concrete systems and because experiments can shed additional light on the physics beyond the relatively small class of solvable models that are accessible analytically or numerically. For example, drawing on the tight connections between operator growth and holographic models of quantum gravity~\cite{shenker_black_2014,kitaev2015,maldacena_bound_2016,shenker_stringy_2014}, experiments probing OTOCs might point to the way to new models with holographic duals. However, experiments in this area are typically quite challenging, as they require either time evolution with both $H$ (forward) and $-H$ (backward)\cite{swingle_measuring_2016,yao_interferometric_2016-1} or a large number of randomized measurements\cite{vermersch_probing_2019} or precision measurements of a small purity-like signal~\cite{yao_interferometric_2016-1}. Local control and readout are also often required depending on the precise setup. 

Remarkably, there is a class of experimental systems in which backward time evolution is approximately possible: magnetic resonance experiments with nuclear spins\cite{baum_nmr_1986,yen_multiplequantum_1983,munowitz_multiple-quantum_1986}. The nuclear magnetic resonance (NMR) community has in effect been measuring relatives of many-body OTOCs since at least the 1970s with the development of the magic echo technique\cite{ernst_time-reversal_1998,rhim_time-reversal_1971,suter_multiple_1987,baum_nmr_1986,yen_multiplequantum_1983,munowitz_multiple-quantum_1986,schnell_high-resolution_2001}. However, there is a major complication: one typically has no local control or readout in these experiments, so what one gains access to are OTOCs of global operators, meaning operators which are sums of local operators. Again taking the spin system as an example, this means it is possible to measure OTOCs between global spin operators, such as $\tr( [ Z(t) , X] [ Z(t) , X]^\dagger)$, where $Z = \sum_i Z_i$ and $X = \sum_i X_i$ are the total $Z$ and $X$ spin operators. It is not clear a priori how these global measurements relate to the local OTOCs that are more commonly studied. 

Motivated by these observations, here we argue that the key physical property probed by local OTOCs, namely the size of growing operators, is also diagnosed by global OTOCs\cite{keselman_scrambling_2021,kukuljan_weak_2017}. Under a few conditions which we expect are generic to chaotic evolutions and which are verifiable in various toy models, we show that the global OTOC is proportional to the ``area under the local OTOC'', i.e. the grey region in Fig.~\ref{fig:h_area}. When the interaction is local, the OTOC has a sharp front, which defines a butterfly light cone. In that case, the global OTOC just measures the volume of the butterfly light cone. When the interaction is power-law, as with dipolar-coupled nuclear spins, the interpretation of the global OTOC as the area under the local OTOC still holds, but the scaling with time depends on the particular form of the local OTOC for that system. 

We apply the above results to the case of nuclear spins where data on global OTOCs are already available. We examine in detail a nuclear magnetic resonance experiment performed on the material adamantane in the presence of a strong Zeeman field~\cite{sanchez_perturbation_2020,sanchez_clustering_2014}. Because of its peculiar structure, this system can be well described as clusters of $16$ nuclear spins placed on the sites of a three-dimensional lattice with the clusters interacting via magnetic dipole interactions (see parameters in Ref.~\onlinecite{schnell_high-resolution_2001} for a magic angle spinning measurement). As we review below, global OTOCs of the total spin have already been measured in this compound in the context of what are called multiple quantum coherences\cite{munowitz_multiple-quantum_1986}. Previous works interpreted such global OTOCs in terms of an effective size of interacting spin clusters and used a stochastic model to understand the cluster growth\cite{munowitz_multiplequantum_1987}. However, in these stochastic models, the spatial structure of the interaction, namely the $\frac{1}{r^3}$ decay of the dipole interaction, was ignored.  

We propose our own stochastic model in which the spatial structure of the interaction is taken into account. An order of magnitude estimate confirms the experimental scale of the spin cluster size, with some experiments reaching a cluster size of roughly $10^4$~\cite{alvarez_localization-delocalization_2015}. Then, based on our previously conjectured phase diagram for operator growth in power-law systems, assuming the relevant dipolar Hamiltonian is quantum chaotic, we predict the asymptotic scaling of the global OTOC growth. Current experimental data is restricted to the short time regime, so we also compare the experimental data to a numerical simulation of our stochastic model and obtain excellent agreement (Fig.~\ref{fig:fcc_data}).

Finally, we consider the possibility of similar experiments using ultra-cold polar molecules\cite{Ye13,yan_observation_2013,hazzard_many-body_2014}. While many of the experimental ingredients have already been demonstrated, there are several interesting differences. We discuss these and propose an experimental protocol that should be able to probe the novel fast operator growth produced by dipolar interactions given modest gains in density and coherence time.

The remainder of this paper is organized as follows. In Sec.~\ref{sec:otoc_lc} we establish one of our main technical claims, that a global OTOC in a chaotic many-body system is equal to the area under the corresponding local OTOC. In Sec.~\ref{sec:nmr_otoc} we analyze experiments in adamantane using the result of Sec.~\ref{sec:otoc_lc} and a Brownian circuit model of the dipole dynamics. In Sec.~\ref{sec:polar_m} we discuss the possible extension of these experiments to systems of ultra-cold polar molecules, focusing on the novel features present in that setting. We conclude with a summary and outlook. Technical details are available in several appendices.


\section{Global OTOCs as the area under the local OTOCs}
\label{sec:otoc_lc}

As outlined above, in some quantum simulation platforms (Sec.~\ref{sec:nmr_otoc}, Sec.~\ref{sec:polar_m}), it is possible to measure a global version of the OTOC thanks to the ability to evolve with both $H$ and $-H$. These systems can be thought of as spin models, but where the spin degrees of freedom arise from nuclear spins or some internal orbital degrees of freedom. Let $X_a$, $Y_a$, $Z_a$ be the Pauli matrices for spin $a$. We consider infinite temperature global OTOCs built from commutators of the {\it total spin}, for example, we can take the commutator of the total $z$ spin $Z = \sum_a Z_a$ and its time evolved form, $Z(t) = e^{ i Ht} Z e^{-iHt}$,
\begin{equation}
  C_{g}(t) = -\frac{\tr( [Z(t), Z]^2 ) }{2^N },
\end{equation}
where $N$ is the total number of spins. In $d$ dimensions, a system with linear size $L$ has $N \sim L^d$. 

By contrast, the quantum chaos literature primarily studies {\it local OTOCs}, which only involve commutators of {\it local spins}. One example is
\begin{equation}
C_{ab}(t) = - \frac{\tr( [Z_a(t), Z_b]^2 )}{2^N},
\end{equation}
which depends on two spin labels $a$ and $b$. To relate the global and local OTOCs, we expand the global OTOC as a sum of local terms, 
\begin{equation}
\label{eq:abcd_expand}
\begin{aligned}
&C_g =- \sum_{abcd} \frac{\tr( [Z_a(t), Z_b][Z_c(t), Z_d] )}{2^N} \\
&= \sum_{ab} - \frac{\tr( [Z_a(t), Z_b]^2 )}{2^N} \\
&- \sum_{a\ne c \text{ or } b\ne d} \frac{\tr( [Z_a(t), Z_b][Z_c(t), Z_d] )}{2^N}.
\end{aligned}
\end{equation}
Suppose, as we will shortly argue, that we can neglect all but the first term in this expansion. How does the global OTOC behave in this approximation?

The first term (the diagonal term) restricts to the case of $a = c$ and $b = d$, which reduces to a sum of local OTOCs. Each local OTOC can be interpreted as follows. The operator $Z_a$ is initially localized at site $a$\footnote{$a$ can label the spatial coordinate as well as internal degrees of freedom}, and commutes with $Z_b$. The time evolution expands the support of $Z_a$ away from $a$, so that it no longer commute with $Z_b$ at site $b$. Thus the local OTOC probes the expansion of the time evolved operator $Z_a(t)$. Fixing $a$ and summing over $b$ gives us an integral of the local OTOC, which is the ``area'' (literally, in 1d) under the local OTOC curve. 

In a locally interacting system, there is a typical velocity $v_B$ that characterizes the spreading of $Z_a(t)$, such that it roughly takes time $|\vec{x}_b - \vec{x}_a|/v_B$ to reach site $b$. The local OTOC is almost $0$ when $|\vec{x}_b - \vec{x}_a| \gg v_B t$, and approaches an order unity value in a chaotic system when $|\vec{x}_b - \vec{x}_a| \ll  v_B t$, see Fig.~\ref{fig:h_area} (a). Hence, $v_B t$ is the characteristic size of the butterfly light cone. Assuming translation symmetry and ignoring edge effects, we have 
\begin{equation}
\begin{aligned}
-\sum_{a,b} \frac{\tr( [Z_a(t), Z_b]^2 )}{2^N}  &\sim - L^d \sum_{b}\frac{\tr( [Z_0(t), Z_b]^2 )}{2^N} \\
& \sim  L^d (v_B t )^d .
\end{aligned}
\end{equation}

In long-range interacting systems, the light cone cutoff may no longer be sharp, see Fig.~\ref{fig:h_area} (b); but the interpretation as the area under the local OTOC curve still applies. 

Now, what about the off-diagonal terms in Eq.~\eqref{eq:abcd_expand}? We argue that they are negligible compared to the sum of the local OTOCs. Let us consider the case when $a= c$, $b \ne d$. The OTOC can be rewritten as
\begin{equation}
\tr( [[Z_a(t)/2^{N/2}, Z_b],Z_d] Z_a(t)/2^{N/2}).
\end{equation}
The operator $Z_a(t)/2^{N/2}$ is normalized according to the operator inner product $(A,B)=\tr(A^\dagger B)$. When expanding it in terms of the Pauli string basis $B_\mu$,
\begin{equation}
Z_a(t)/2^{N/2} =  \sum_\mu a_{\mu}  B_\mu ,
\end{equation}
the amplitude squared $|a_{\mu}|^2$ can be regarded as the probability. At sufficiently long times, the operator $Z_a(t)$ is scrambled and can be regarded as a random operator supported on $N_{op}(t)$ spins, with $N_{op}(t)\sim (v_B t)^d$ in a system with local interactions. We model the effective randomness by treating the $\alpha_{\mu}$ as real random numbers (real because $Z_a(t)$ is Hermitian). There are $4^{N_{op}}$ of them, and $\sum_{\mu} |a_{\mu}|^2 = 1$. So the typical size of $a_{\mu}$ is $\sqrt{\frac{1}{4^{N_{op}}}} = 2^{-N_{op}}$. 

The double commutator interchanges Pauli strings depending on the operators in the string located at sites $b$ and $d$. Strings with $XX$ are exchanged with strings with $YY$, and similarly for $XY$ and $YX$. Other strings commute with at least one of $Z_b$ and $Z_d$. Writing the corresponding amplitudes as $a_{\mu}|_{XX}$, $a_\mu|_{YY}$, $a_\mu|_{XY}$, $a_\mu|_{YX}$, the off-diagonal OTOC is 
\begin{equation}
8\sum_{\mu} [\text{Re}(a_\mu|_{XX} a^*_\mu|_{YY}) +  \text{Re}(a_\mu|_{XY} a^*_\mu|_{YX} )].
\end{equation}
There are at most $2^{N_{op}}$ terms in the summation. Assuming they are uncorrelated, the amplitude of the sum is estimated from a random walk to be $\sqrt{2^{N_{op}}} \times 2^{-N_{op}} \sim 2^{- N_{op}/2}$, which is negligible compared to contributions from the diagonal local OTOCs. In App.~\ref{app:cross_term}, we refine this argument and do the computation for evolution with a circuit of local gates. The sum of the off-diagonal terms is indeed negligible. Then we argue that the same should hold for long-range interactions. Numerical computations in small systems confirm this observation (App.~\ref{app:off_diag}). 

Hence, we conclude that the global OTOC is approximately equal to the sum of local OTOCs, which measures the area under the corresponding local OTOC curve.


\section{Global OTOC in nuclear magnetic resonance experiments}
\label{sec:nmr_otoc}

In this section, we apply our theory to measurements of global OTOCs in nuclear magnetic resonance experiments. We first review the experimental situation and prior works seeking to explain the experimental observations. One of the goals of the review part of this section is to translate some NMR concepts into the language of many-body chaos. We then propose and analyze a stochastic model of operator spreading in adamantane and compare the results to existing experimental data.

\subsection{Review of NMR}

NMR experiments use nuclear spins to form interacting quantum magnets. External radio-frequency waves can excite the spin states and thus are tools to control the global spin variables. We give a broad sketch of the relevant concepts here, with many details in App.~\ref{app:nmr_review}. 

The largest energy scale is provided by a strong Zeeman field which defines the $z$ direction. The spins also experience long-range dipolar interactions which, in the rotating frame defined by the Zeeman field, are well approximated by a secular form,
\begin{equation}
\label{eq:YY_int}
\begin{aligned}
  H &= \sum_{a\ne b} D_{ab} ( Z_a Z_b - X_a X_b - Y_a Y_b  ), \\
\end{aligned}
\end{equation}
where $a$, $b$ label the spins and
\begin{equation}
\label{eq:D_ij}
D_{ab} \propto \frac{(3 \cos^2 \theta_{ab} -1 )}{2 r_{ab}^3}.
\end{equation}
Here $r_{ab}$ is the distance between the two spins, $r_{ab} = |\vec{r}_a - \vec{r}_b|$, and $\theta_{ab}$ is the angle between $\vec{r}_a - \vec{r}_b$ and the $z$ direction. The key features are the angular dependence, the power-law character of the interactions, and the fact that the interaction commutes with total $Z$ as a consequence of the secular approximation.

In addition to this dipole interaction, one can apply various radio-frequency pulses to the sample. Considerable effort is devoted to the design of radio-frequency pulse sequences that, when combined with time evolution under the basic dipolar interaction, can give rise to a variety of effective Hamiltonians. For example, in the adamantane experiments we discuss below, researchers use the double quantum Hamiltonian:
\begin{equation}
\label{eq:H_dq}
\begin{aligned}
  H_{\rm DQ} &= \sum_{a \ne b} D_{ab} ( X_a X_b - Y_a Y_b ) \\
  &= \sum_{a\ne b} \frac{D_{ab}}{2} (\sigma^+_a\sigma^+_b + \sigma^-_a \sigma^-_b   ),
\end{aligned}
\end{equation}
where each term changes the total $z$ spin by $\pm 2$. They also use the dipolar Hamiltonian, but in the ``Y'' convention, so that it does not commute with $Z$:
\begin{equation}
\label{eq:YY_int}
\begin{aligned}
  H_{\rm YY} &= \sum_{a\ne b} D_{ab} ( Y_a Y_b - X_a X_b - Z_a Z_b  ).\\
\end{aligned}
\end{equation}

Moreover, for each of these Hamiltonians, one can design pulse sequences that correspond to evolving with both $H$ and $-H$. As reviewed in App.~\ref{app:nmr_review}, this enables measurement of the global OTOC,
\begin{equation}
C_g = -  \tr( [ e^{ - iH t} Z e^{ i Ht} , Z]^2 )/ 2^N.
\end{equation}
Of course, the ability to evolve with $H$ and $-H$ is an approximate capability. It is interesting to consider the effects of additional subleading terms in the true Hamiltonian as well as environmental effects, but here we focus on the ideal situation.

In this work, we choose the material adamantane as an example, for both its long history in the NMR community and because recent global OTOC data is available. Adamantane is a solid polycrystal at room temperature. The crystal structure is face-centered cubic (fcc) with one adamantane molecule ($C_{10} H_{16}$) at each lattice site. The Hydrogen protons comprise the active nuclear spins, so there are $16$ spin-1/2s per lattice site. Adamantane also has the peculiar feature that the molecules tumble in place in the lattice at relevant temperatures due to their nearly spherical nature.

The measurement of global OTOCs in adamantane molecules dates back to the 1980s under the name of multiple quantum coherences\cite{yen_multiplequantum_1983,baum_multiplequantum_1985,baum_nmr_1986,cho_h_1996,schnell_high-resolution_2001,prigogine_principles_2007,sanchez_time_2007}, although at that time only a handful of coherent spins were involved~\cite{munowitz_multiplequantum_1987}. More recently, thanks to improved coherence times\cite{sanchez_time_2007} and the scaled Hamiltonian technique\cite{sanchez_evolution_2017}, the number of coherent spins can be as large as $10^4$~\cite{alvarez_localization-delocalization_2015}. Loschmidt echoes, a related class of observables that also probe time reversal effects, have also been studied~\cite{PhysRevLett.86.2490}.

Next, we review the popular $Kn$ space approach~\cite{munowitz_multiplequantum_1987} adopted by the NMR community to understand multiple quantum coherence. This approach does not take into account the spatial structure of the interaction, and not surprisingly, it predicts an exponential growth of the global OTOC in time. As an alternative to this approach, we apply our theory to a simple stochastic model to estimate the global OTOC. 

\subsection{The $Kn$ space approach}
\label{subsec:kn-space}
The experiments of interest do not measure the global OTOC directly, but rather extract it from multiple quantum coherences (MQC) defined as follows~\cite{prigogine_principles_2007}. Let $\rho$ be the time evolved operator $\rho = Z(t)$. In NMR, this represents the density matrix neglecting the identity part that does not participate in the dynamics (see App.~\ref{app:nmr_review} and the discussion in Sec.~\ref{subsec:weakly_polarized}). The operator $Z$ appears because at high temperature in thermal equilibrium the sample is weakly polarized due to the Zeeman field. 

The density operator $\rho$ can be decomposed as
\begin{equation}
\rho = \sum_n \rho_n ,
\end{equation}
where $\rho_n$ increases the total spin $z$ quantum number by $n$. Operators like $X$ and $Y$ change the total spin $z$ quantum number by $\pm 1$. They are an example of a single quantum coherence. When $|n| > 1$, $\rho_n \neq 0$ are called multiple quantum coherences, as they indicate the structure in density matrix further away from the diagonal. Formally, the operator $\rho_n$ satisfies
\begin{equation}
e^{i \phi Z } \rho_n e^{ - i \phi Z} = \rho_n e^{i n \phi },
\end{equation}
and the ``intensity''
\begin{equation}
g_n = \frac{1}{\tr( Z Z)} \tr( \rho_n \rho_{-n} ) 
\end{equation}
defines the multiple quantum coherence. The second moment of the MQC is proportional to the global OTOC\cite{khitrin_growth_1997}(see App.~\ref{app:nmr_review}),
\begin{equation}
\sum_n n^2 g_n = - \frac{1}{\tr( ZZ )} \tr( [Z, Z(t)]^2 ) .
\end{equation}

So far what we have said in this subsection is general. Next, we review some standard intuition for the MQC based on a simple counting argument. A $n$-quantum coherence maps a state with $x$ up spins to states with $n + x$ up spins. For a system of $K$ spins, there are 
\begin{equation}
\sum_{x=0}^{K} { K \choose x } { K \choose n + x}  = { 2K \choose K + n }  \sim 2^{2K} \exp\left( - \frac{n^2}{K}  \right) 
\end{equation}
operators that belong to the $n$-quantum coherence space. If all these operators are equally likely, then $g_n$ will be roughly a Gaussian function, $g_n \sim \exp( - \frac{n^2}{K} )$. Therefore, the second moment of $g_n$, the global OTOC, will scale as $K$, the number of spins in the system. 

In the dynamical setting, this idea is generalized by allowing $K$ to be a function of time. $K(t)$ represents the effective size of the spin cluster supporting the $n$-quantum coherence operators. In practice, $K(t)$ is obtained by a Gaussian fit from the MQC. Hence, the fluctuations of the MQC as a function $n$ indicate the dynamically growing cluster size and give the growth of the global OTOC. Our result above in essence shows that $K(t) \propto \int d^d r C(r,t)$, where $C(r,t)$ is the corresponding local OTOC.

One approach to model the dynamics of $K(t)$ is to replace the full quantum dynamics with a stochastic process in the $Kn$ plane~\cite{munowitz_multiplequantum_1987}. The transition rate between two points in the $Kn$ plane is taken to be proportional to the number of interaction terms that cause such a transition. We use the transition rates in Ref.~\onlinecite{munowitz_multiplequantum_1987} and reproduce the results for the double quantum Hamiltonian in Fig.~\ref{fig:N6} and Fig.~\ref{fig:N21} in the appendix.

\begin{figure}[h]
\centering
\includegraphics[width=\columnwidth]{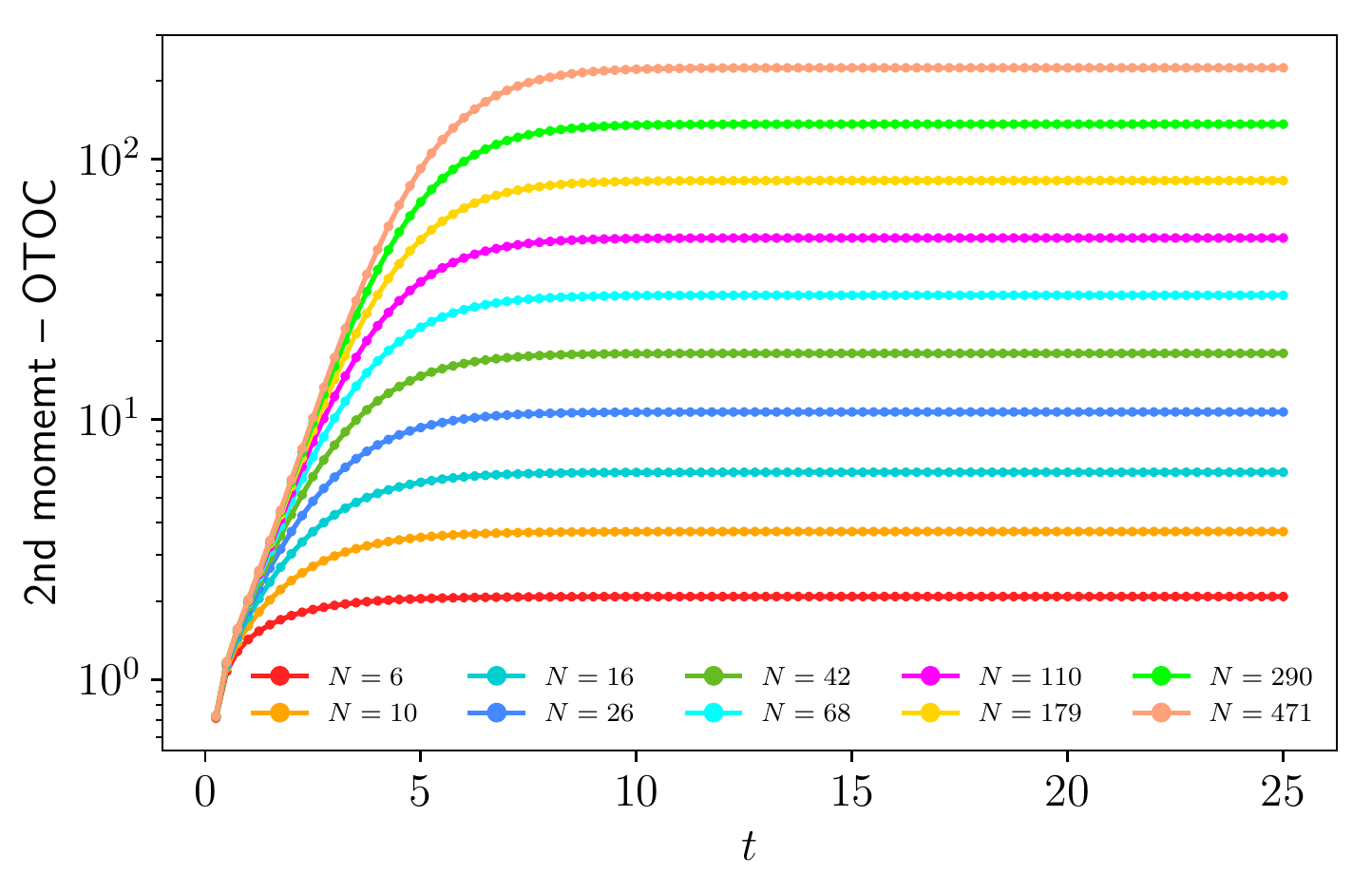}
\caption{The global OTOC computed from the stochastic motion in the $Kn$ space approach for $N \le 600$. Without the spatial struction of the interaction, the curve takes an exponential growth before saturation. }
\label{fig:Kn}
\end{figure}

However, this particular stochastic approach does not take the spatial dependence of the interactions into account. It implicitly treats all the sites on an equal footing. Hence, we expect this model to show an exponential growth of the OTOC, a phenomenology common to systems with all-to-all interactions. We verify this expectation in Fig.~\ref{fig:Kn}\footnote{Other variants that take the distribution to be a superposition of Gaussian functions with different cluster sizes also predict exponential growth~\cite{sanchez_quantum_2016,sanchez_clustering_2014-1}.}. Note that when this approach was proposed, the experimentally accessible system sizes were relatively small ($N \lesssim 21$) and the model compared favorably with data. 

In addition to the stochastic $Kn$ approach, there are also other non-stochastic effective models such as the Levy-Gleason model and recent variants\cite{levy_multiple_1992,dominguez_dynamics_2021}. In most of these models, the locality of the interaction is not incorporated. As experiments push to larger sizes and longer times, the spatial structure of the interactions becomes important. 

\subsection{Global OTOC estimated from local OTOCs}

\begin{figure}[htb]
\centering
\subfigure[]{
  \label{fig:alvarez_data}	
  \includegraphics[width=0.95\columnwidth]{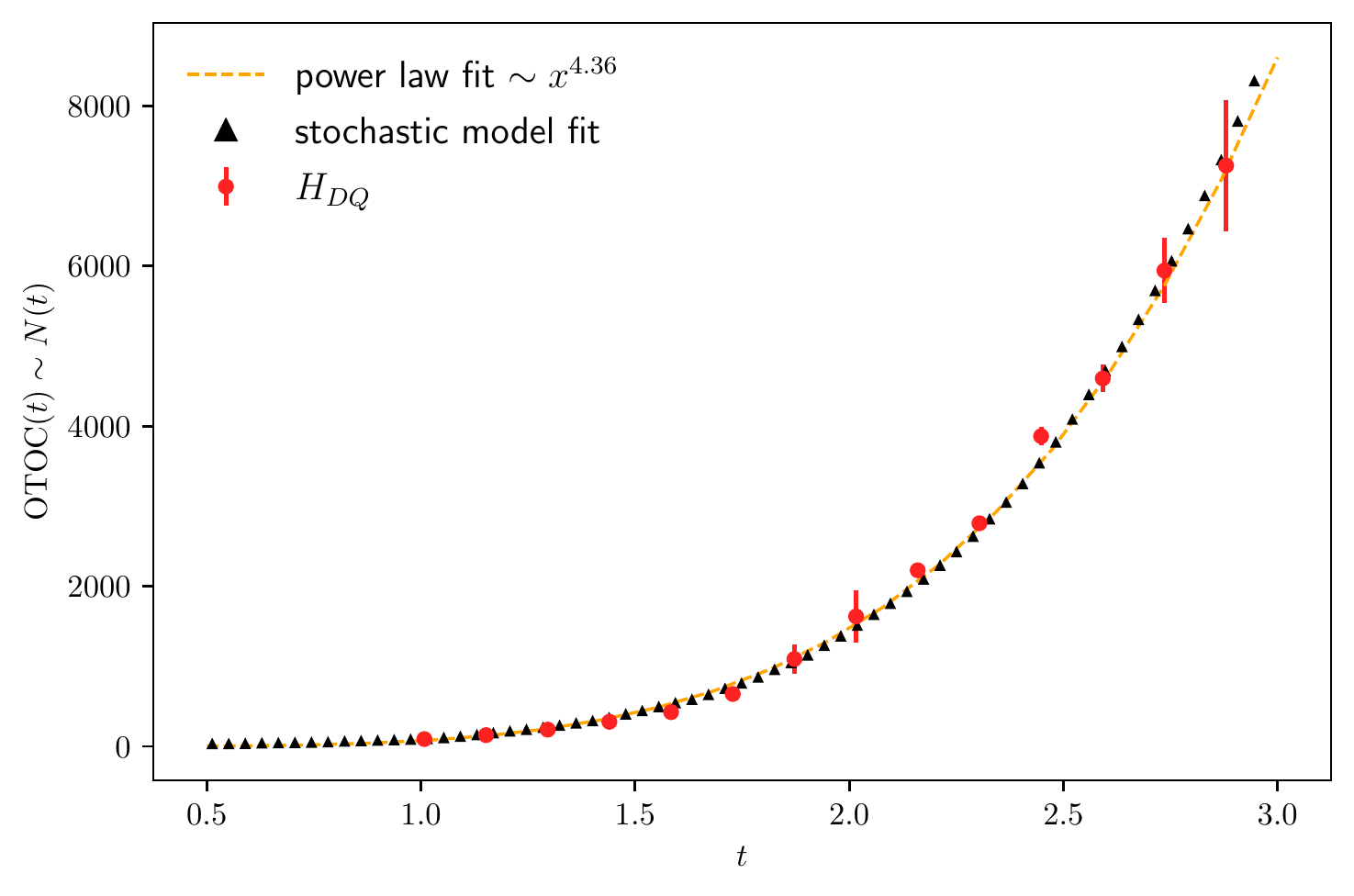}
}
\subfigure[]{
  \label{fig:sanchez_data}	
  \includegraphics[width=0.95\columnwidth]{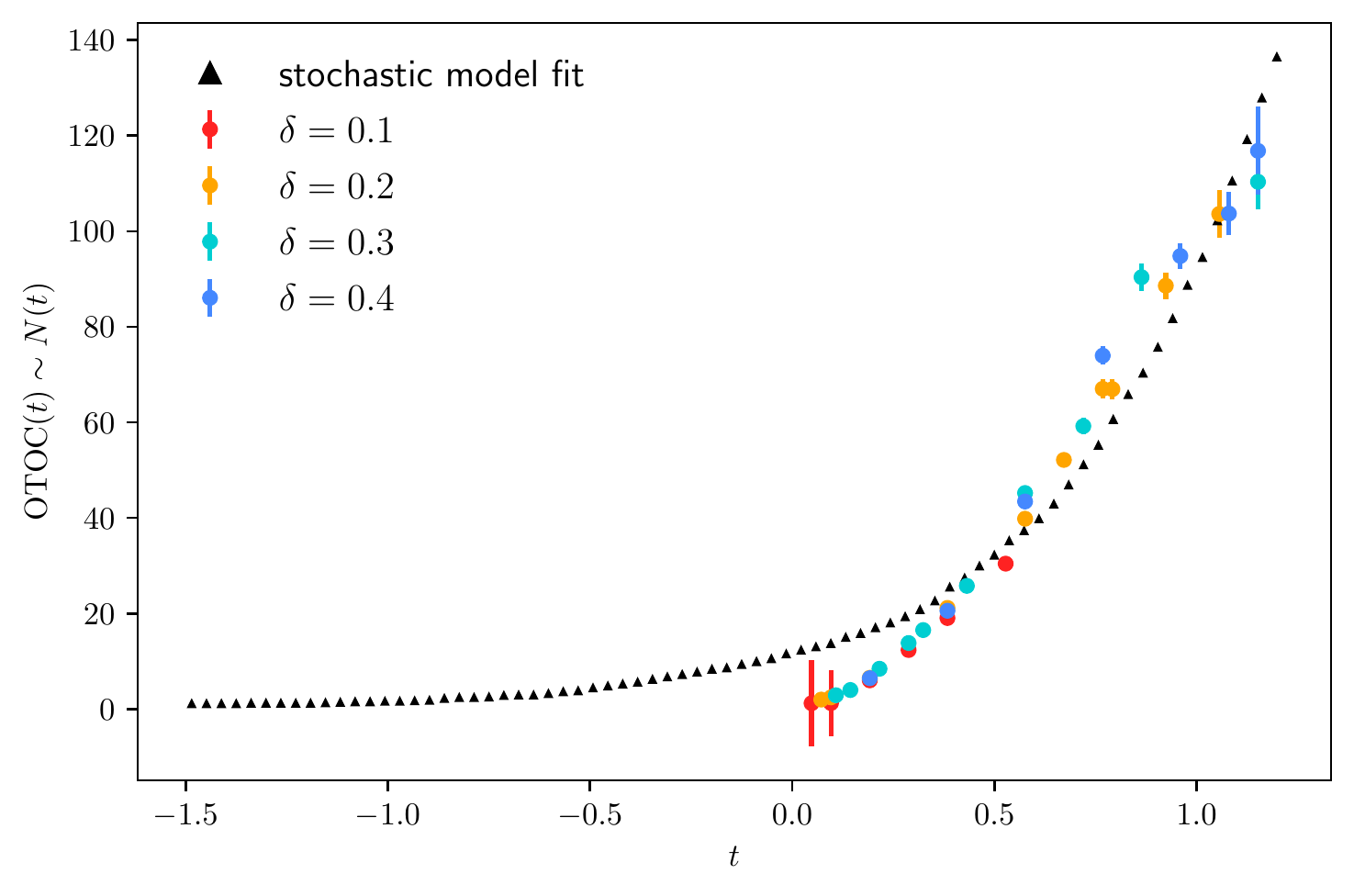}
}
\caption{A two-parameter fitting of our stochastic model results with the adamantane measurements of the global OTOCs. Experimental data is displayed with the permission of the authors. The unit of time here is 0.4 ms. 
(a) Data for double quantum Hamiltonian $H_{DQ}$ evolution taken from Fig. 10 of Ref.~\onlinecite{alvarez_quantum_2013}. The cluster size reaches almost $10^4$ with a power law fit of a mysterious exponent 4.36. The stochastic model has a time shift of $-0.87$ and $J \approx 1.76$. 
(b) Data for dipolar Hamiltonian $H_{YY}$ evolution taken from Fig. 3 of Ref.~\onlinecite{sanchez_perturbation_2020}. The stochastic model has a time shift of $-1.48$ and $J \approx 2.7$. }
\label{fig:fcc_data}
\end{figure}

Having reviewed the NMR developments, in this subsection, we propose a simple model that better accounts for the spatial structure of the dipolar interactions. We first state our approximations and assumptions and then postulate an effective Hamiltonian.

The long-range dipole interactions in the adamantane molecule exist between any pair of proton nuclear spins. There are both inter-molecular and intra-molecular interactions. An unusual feature of adamantane is that the molecules in the lattice constantly tumble in place. The tumbling time scale is much shorter than the time scale of the intra-molecular dipolar interaction. Now from Eq.~\eqref{eq:YY_int}, we see that the dipole interaction has an angle dependence $3 \cos^2 \theta_{ab} - 1$, where $\theta_{ab}$ the angle between the $z$ axis and vector from $a$ to $b$. Hence, if the molecule is rapidly tumbling, then for any two protons within a molecule, the vector from $a$ to $b$ will vary randomly over the sphere and the coupling will average to zero. In other words, $\theta_{ab}$ will change rapidly in time in a random fashion such that the time average of $ 3 \cos^2 \theta_{ab} - 1$ is zero. This means that all intra-molecular couplings average to zero. For a similar reason, all nearest-neighbor inter-molecular couplings time-average to the same value and decay as $\frac{1}{r^3}$. 

Thanks to this simplification, there is a simple model of the spin dynamics in adamantane. At each site $i$ of an fcc lattice, we place $M=16$ spin-1/2s. The total spin of each site interacts directly with the total spin of any other site and there are no interactions within a site. Denote the total spin $Z$ on site $i$ as $Z_i^{{}_{\sum}} = \sum_a Z_i^a$ and similarly for $X$ and $Y$. With this notation the effective double quantum Hamiltonian is
\begin{equation}
H_{DQ,{\rm eff}} = \sum_{i\ne j} D_{ij} ( X_i^{{}_{\sum}} X_j^{{}_{\sum}} - Y_i^{{}_{\sum}} Y_j^{{}_{\sum}} )
\end{equation}
and the dipolar Hamiltonian is
\begin{equation}
\label{eq:H_YY_eff}
H_{\rm YY, eff} = \sum_{i\ne j} D_{ij} ( Y_i^{{}_{\sum}} Y_j^{{}_{\sum}} - X_i^{{}_{\sum}} X_j^{{}_{\sum}} - Z_i^{{}_{\sum}} Z_j^{{}_{\sum}} ).
\end{equation}

Symbols like $Z_i^{{}_{\sum}}$ denote the total spin on a physical site. Hence, the interactions should be understood as interactions between different spin representations on each site. The spin representation on each site is a conserved quantity, and an average with respect to the infinite temperature state includes an average over the different spin representations. 

Thanks to the large number of conserved quantities, the resulting model is still very complicated, so we propose a further simplification that is expected to capture the leading time-dependence of operator growth. We emphasize that there is a physical assumption here, that the dipole interaction in three dimensions is non-integrable. In support of this assumption, we note that chaotic features have been observed in exact diagonalization studies in similar models in two dimensions\cite{keles_scrambling_2019} for the spectral statistics and the OTOC. We also note that one-dimensional systems are typically special, and various dipolar Hamiltonians could well be close to integrable in one dimension~\cite{PhysRevA.80.052323}.

In any event, our chaos assumption motivates a modification of the Hamiltonian that enhances chaos and makes the model more tractable theoretically. To this end, we consider a different model where the interactions between $i$ and $j$ include all possible terms,
\begin{equation}
\label{eq:H_D_ab}
H = \sum_{ij} D'_{ij} \sum_{a,b=1}^{16} \sum_{\mu,\nu =0}^3 (B_{\mu\nu})_{ij}^{ab} (\sigma^\mu)_i^a(\sigma^\nu)_j^b.
\end{equation}
This modification destroys most of the symmetries, including any spin rotation symmetries and the symmetry arising from permuting spins within a site, but the model retains (1) the essential structure of the long-range interaction and (2) a large number of spins (in this case, $M=16$) on each site. The interaction coefficient is chosen to be $D_{ij}' = D_{ij} \sqrt{\frac{3}{16}}$. This is because there are 3 types of spin-spin interactions terms (i.e. $YY, XX, ZZ$) in Eq.~\eqref{eq:H_YY_eff}, while there are 16 in Eq.~\eqref{eq:H_D_ab}. The adjustment by the factor $\sqrt{\frac{3}{16}}$ equates their operator norms.

In our previous works, we analyzed the asymptotic light cone structure of local OTOCs in generic models like Eq.~\eqref{eq:H_D_ab}\cite{zhou_operator_2020,xu_locality_2018,chen_quantum_2019}. The interactions were taken to decay as $\frac{1}{r^\alpha}$, with the system defined on a lattice in $d$-dimensional space. To give a solvable model, the coefficients $ (B_{\mu\nu})_{ij}^{ab} $ were taken to be independent white-noise-correlated random variables. This enabled us to map the operator spreading problem to a stochastic process somewhat similar to those encountered in the $Kn$ space approach, but retaining the spatial structure of the underlying quantum problem. The state space of the stochastic model is labeled by a choice of empty (identity operator) or occupied (non-trivial Pauli operator) for each spin in the system. The initial condition is given by a single occupied spin, corresponding to an initial operator which is a single Pauli operator on that spin. At each time step, there is a probability of $\frac{1}{r^{2\alpha}}$ for each occupied spin to fill an empty spin a distance $r$ away. This dynamical rule then leads to various scalings for the local OTOC, as collected in Table~\ref{tab:model1p}. In Ref.~\onlinecite{zhou_operator_2020}, we argued that general quantum chaotic models with power-law interactions and at high energy density would reside in the same universality class as the stochastic model at long times thanks to an effective dephasing of the quantum dynamics. Hence, in Ref.~\onlinecite{zhou_operator_2020} we conjectured that Table~\ref{tab:model1p} is generic across a broad class of systems, and we provided numerical evidence for this conjecture in the context of 1d spin chains with power-law interactions.

\begin{table}
\centering
\begin{tabular}{ |C|C|C|C| } 
 \hline
  \alpha & $light cone$  & $scaling function$ & $tail$ \\ \hline
  [\frac{d}{2},d) & \exp( B t^{\eta}) & C( \frac{r}{\exp( B t^{\eta})} ) & \multirow{4}{*}{$\frac{1}{r^{2\alpha}}*$} \\ \cline{1-3}
  d & \exp( \frac{(\ln t)^2}{4d \ln 2} ) & C( \frac{r}{t^{ \frac{1}{4d} \log_2 t }} ) &  \\ \cline{1-3} 
  (d,d+\frac{1}{2}) & t^{\frac{1}{2\alpha - 2d}} & C( \frac{r}{t^{\frac{1}{2\alpha - 2d}} } ) &  \\ \hline
  d + \frac{1}{2} & t \ln t & C( \frac{r}{t \ln t} ) &  \\ \hline
  (d + \frac{1}{2}, d+1) & v_B t & C( \frac{r-v_B t}{t^{\frac{1}{2\alpha - 2d}}} ) & \frac{1}{r^{2\alpha-2d}}*  \\ \hline
  d+1 & v_B t & C( \frac{r-v_B t}{(t \ln t)^{\frac{1}{2}}} ) & $erf$ \\ \hline
  [d+1,\infty) & v_B t & C( \frac{r-v_B t}{t^{\frac{1}{2}}}  ) & $erf$ \\ \hline
\end{tabular}
\caption{The scalings of the local OTOC predicted by the long range Brownian circuit model, see Ref.~\onlinecite{zhou_operator_2020,hallatschek_acceleration_2014,chatterjee_multiple_2013}. Parameters: $B = \frac{d\ln 2}{2( \alpha - d)^2} $, $\eta  = \log_2 \frac{d}{\alpha}$. The tail scalings with $*$ only has numerical support for $d = 1$ along with a few general scaling conjectures. }
\label{tab:model1p}
\end{table}

Returning to the experimental situation with adamantane, we start by estimating the basic timescales. The nearest neighbor distance between adamantane molecules is $0.67$ nm, which yields a value of $J \sim 2\pi \times 410 $ Hz $\sim 2500 $ Hz frequency for the nearest neighbor dipole coupling (also see Ref.~\onlinecite{schnell_high-resolution_2001}). This translates to a timescale of $0.4$ ms. The coherence time during which the data is taken in the experiment is of order $1$ ms. We can therefore set $J \sim 1$ and consider about one unit of time. 


Not surprisingly, the experimental timescales are currently too short to observe or refute the asymptotic scalings predicted by our theory for $d = 3$ and $\alpha = 3$. Hence, we analyze the short time behavior of the stochastic model. 

First, we give a rough estimate for the size of the coherent spin cluster that develops after one unit of time. In the language of the stochastic model, this is the number of occupied spins. Initially, there is only one spin in the occupied state. This is the initial operator $Z_i^a$ in the local OTOC. According to the rules of the stochastic model, the probability to spread the occupation decays as $\frac{1}{r^6}$. At short times, the long-range part is negligible and we may truncate to nearest neighbor interactions. When the support of $Z_i^a(t)$ spreads to a spin, the operator associated with it will quickly reach equilibrium, leading to a roughly equal probability to be $X$, $Y$, $Z$, or the identity. Therefore at equilibrium, each site has $3/4$ probability to be occupied (3 Pauli matrices out of 4 single-site Hermitian operators). For a molecule with $16$ spins, there will be on average $12$ spins occupied in equilibrium. 

How many molecules can reach equilibrium after one unit of time? There are 256 spin interactions between the two molecules. It then takes $16 / 256 = 1/16$ unit of time for one molecule to spread to another molecule. But that only takes into account the process where the identity operator becomes non-identity. Since there are 3 non-identity Pauli matrices and one identity matrix, there is $\frac{1}{3}$ of the rate that converts non-identity operators into an identity operator. This is also why there are $12$ instead of $16$ spins occupied on average in equilibrium. With this correction, it should take $\frac{1}{16} \times \frac{4}{3} = \frac{1}{12}$ unit of time to equilibrate a new molecule. Hence, in one unit of time, the system can equilibrate $12$ molecules. 
In addition, the coefficient $\frac{3}{16}$ in Eq.~\eqref{eq:H_D_ab} decreases the number to $12 \frac{3}{16} \sim 2.25$ molecules. This is the linear dimension of the occupied cluster. In 3 dimensions, there will be roughly 
\begin{equation}
2.25^3 \times 12 \sim 136 \sim 10^2
\end{equation}
occupied spins. The global OTOC is the area under the local OTOC, which in the stochastic model is the average number of spins occupied. Compared with the multiple quantum coherence measurement results\cite{sanchez_perturbation_2020,sanchez_clustering_2014}, the $10^2$ estimate is consistent with the data.

We emphasize that the above estimation depends sensitively on the parameter choices. For example, our stochastic model predicts that the cluster size grows faster than any power of time in the limit of large time (with the approximation of instantaneous dipolar interactions). We also truncated the interaction to nearest neighbors and ignored the lattice structure. To obtain a more accurate description of the dynamics, we carried out a Monte Carlo simulation of the stochastic process on the fcc lattice using the experimental parameters. 

The results are shown in Fig.~\ref{fig:fcc_data}, where we normalize the time to have a unit of 0.4 ms in our estimation. In Fig.~\ref{fig:alvarez_data}, the coherence time is about 3 units of time, so our estimation would give $136 \times 3^3 \sim 4 \times 10^3$ as the final cluster size, which is consistent with the scale of the data. The best fit of the stochastic model to the experiment\cite{alvarez_localization-delocalization_2015} ($H_{DQ}$ evolution) corresponds to taking $J \sim 1.76$ and shift the time by about $-0.87$ unit. The fit is quite close to the experimental data points. One possible interpretation is that we recalibrate the time after local thermalization beyond which the stochastic approximation is valid. This assumption is subject to test with future experimental data, especially if one can probe several units of time. In Fig.~\ref{fig:sanchez_data}, we fit the stochastic model with the experiment\cite{sanchez_perturbation_2020} ($H_{YY}$ evolution) with $J \sim 2.7$ and shifting the time by about $-1.48$ unit. We can see that this relatively early time growth can still be roughly captured by our model, but the fit is not as good as with the larger cluster sizes.

\subsection{The Weakly Polarized State Approximation}
\label{subsec:weakly_polarized}

Before moving on, let us comment on one approximation used in the above analysis. In the NMR setup to measure the OTOC, one of the time evolved operators $Z(t)$ comes from the high temperature expansion of the time evolved density matrix $\rho(t)$. In a Zeeman field in the $z$ direction, the initial density matrix can be expanded as
\begin{equation}
\label{eq:weakly_polarized}
  \rho(0) \sim e^{-\frac{\frac{\gamma}{2} B Z  }{k_B T}} \sim  \I -  \frac{\gamma B Z}{2k_B T}, 
\end{equation}
where $\I$ is omitted later in the calculation, resulting in the schematic $\rho(t) \sim Z(-t)$. 

Since $Z$ is a many-body operator, this expansion is formally only valid if the system size is sufficiently small. The gyro-magnetic ratio in $1$ T 
magnetic field is about $2.4 mK$ for the protons in adamantane. At room temperature, the (dimensionless) coefficient in front of $Z$ is of order $10^{-5}$. Hence the expansion is valid when the operator norm of $Z$ is smaller than $10^5$. This sets an upper limit on the cluster size (the global OTOC).

When the scale of the global OTOC is beyond $10^5$, the weakly polarized state assumption in Eq.~\eqref{eq:weakly_polarized} fails, and one needs to consider the high temperature expansion for each spin separately,
\begin{equation}
\rho(0) \sim \prod_i \left( \I -  \frac{\gamma B Z_i}{2k_B T}  \right),
\end{equation} 
which contains higher order monomials of $Z_i$. This is closer to the situation encountered in the polar molecule setup in Sec.~\ref{sec:polar_m}, where the initial state is typically a polarized pure state.


\section{Polar Molecules}
\label{sec:polar_m}

We now turn to another physical realization of dipolar interactions via polar molecules and consider the possibility of experiments similar to those in adamantane and other NMR systems. 

One way to create such gas of polar molecules is by laser cooling neutral atoms like Rubidium (Rb) and Potassium (K) to a few hundred nK and then removing their binding energy\cite{ni_high_2008,yan_observation_2013,hazzard_many-body_2014}. The resulting molecules will be in the rotationally and vibrational ground state. This ground state is then taken to be the $|\!\!\downarrow\rangle$ state of a pseudo-spin degree of freedom. The $|\!\! \uparrow \rangle$ can be taken as one of the rotational excited states\cite{gadway_strongly_2016}. The effective Hamiltonian of the pseudo-spin includes an electric dipolar interaction,
\begin{equation}
 H = \sum_{ij} D_{ij} ( J_{\perp} (X_i X_j + Y_i Y_j ) + J_z Z_i Z_j  ) ,
\end{equation}
where $D_{ij}$ has been defined in Eq.~\eqref{eq:D_ij}. For ${}^{40}{\rm K}{}^{87}{\rm Rb}$ molecule, $J_{\perp}$ is nonzero even without an external electric field, while $J_z$ can be tuned by an applied electric field. Provided $J_z$ and $J_\perp$ can be tuned appropriately, global spin rotations can again be used to effectively invert the Hamiltonian dynamics. (A case where this inversion is not possible via just global rotations is $J_z=J_\perp$ since the interactions are SU($2$) symmetric.) 

The experimental controls available in the polar molecule case are similar to the NMR setting. The $|\! \downarrow \rangle $ population, or in other words $\tr( \rho Z )$ can be directly measured. Global pseudo-spin rotations can be performed by microwave pulses. There have also been Ramsey spectroscopy experiments with oscillatory spin echo signals showing direct manifestations of the dipole interaction\cite{gadway_strongly_2016}. 

These similarities prompt us to propose that global OTOCs can also be probed via polar molecules using a very similar set of pulses as in the nuclear spin experiments. In particular, given an initial state $\rho$, the procedure is to measure the phase rotated quantity $\tr( e^{i\phi X} e^{-i H t} \rho e^{ i H t} e^{ -i \phi X} e^{-iHt} Z e^{iHt} )$, and then compute via post-processing its second order derivative with respect to $\phi$. The result will be proportional to $- \tr( [X ,\rho(t)][ X , Z(-t)] )$, with $\rho(t) = e^{-i Ht}\rho e^{i H t}$ and $Z(-t) = e^{-i H t} Z e^{i Ht}$, see App.~\ref{app:nmr_review}. Note that here we are assuming that the dynamics do not conserve total $Z$, otherwise $Z(t)$ is just $Z(0)$. As in the NMR context, this can be circumvented by rotating the frame of the interaction.

In the NMR analysis, the high temperature expansion of the mixed state enables us to rewrite $\rho(t)$ as a constant times $Z(-t)$, thus identifying the measured quantity as the global OTOC. By contrast, in the polar molecule case, it is experimentally easiest to begin with a pure state. Suppose $\rho$ is the all $\up$ eigenstate of $Z$. Then initial density matrix can be written as
\begin{equation}
\rho(0) = \frac{1}{2^N}\prod_{i} (1 + Z_i  ).
\end{equation}
The expression is a sum of homogeneous polynomials of $Z_i$, in which the first term is proportional to $\I$, the second term is proportional to $Z = \sum_i Z_i$, \etc. Truncating to the second term gives us the global OTOC as in the case of nuclear spins. Higher order polynomials of $Z_i$ create extra off-diagonal terms such as 
\begin{equation}
- \tr( [ Z_a(t), Z_b] [Z_a(t), Z_c Z_d] ) / 2^N.
\end{equation}
According to our prior argument, assuming $Z_a(t)$ is a random-like operator as in Sec.~\ref{sec:otoc_lc}, each individual off-diagonal term will be exponentially suppressed as $2^{-N_{op}/2}$. There are $N_{op}^2$ terms that need to be taken into account (since both $Z_c$ and $Z_d$ must lie within the effective light cone for the term to have a chance of being significantly non-zero). Each term is exponentially suppressed in $N_{op}$, so after a short time, the exponential suppression easily overwhelms the polynomial number of choices and the off-diagonal terms give a negligible contribution. Given this argument, we expect that even the pure state will give the global OTOC, up to an overall constant, at long times. 

We now estimate the requirements needed to probe the long-time regime in experiments with KRb molecules. In this case, $J_{\perp}$ is about $2\pi \times 104  \sim 650$ Hz. The coherence time shown in the Ramsey spectroscopy experiment is of order 10 ms. Hence the coherence time is about 10 units of time, an order of magnitude larger than the nuclear spin experiment. However, unlike in the nuclear spin case, experimental realizations to date involve a dilute lattice of spins, with many lattice sites empty. Moreover, the number of molecules in the optical lattices is about $10^5$, which constrains the largest spin cluster that can be formed in the evolution. 

Previous experiments achieved a filling factor of less than 10\%, which is in sharp contrast to the $16$ spins on each site in the nuclear spin experiment. As expected, the low occupancy significantly hinders the spreading, although the long-range nature of the dipolar interaction moderates this slowdown to some extent. To give a crude estimate, imagine a sphere surrounding one molecule. The volume of the sphere is $\frac{4\pi}{3} r_0^3 \approx 4 r_0^3$. Taking the occupancy to be 5\%, a volume of $4 r_0^3 = 20$ has only one site occupied by a molecule. On average, the nearest neighbor interaction is reduced by a factor of $\frac{1}{r_0^6} = \frac{1}{25}$ (in the classical stochastic model, the rate is $\frac{1}{r^6}$ rather than $\frac{1}{r^3}$ due to dephasing). Hence 10 units of time can only populate a cluster of size $10/25 \times \frac{3}{4} = 0.3$, which is barely one spin. 
Keeping with this estimate, the linear size of the cluster is $10 \times (4p)^2 \times \frac{3}{4} = 120 p^2$. The volume is $120^3 p^6$. Thus the thresholds of occupancy to reach cluster sizes of $10$, $10^2$, $10^3$ are $13.4\%$, $19.7.\% $ and $28.9\%$ respectively. 

\begin{figure}[h]
\centering
\includegraphics[width=0.8\columnwidth]{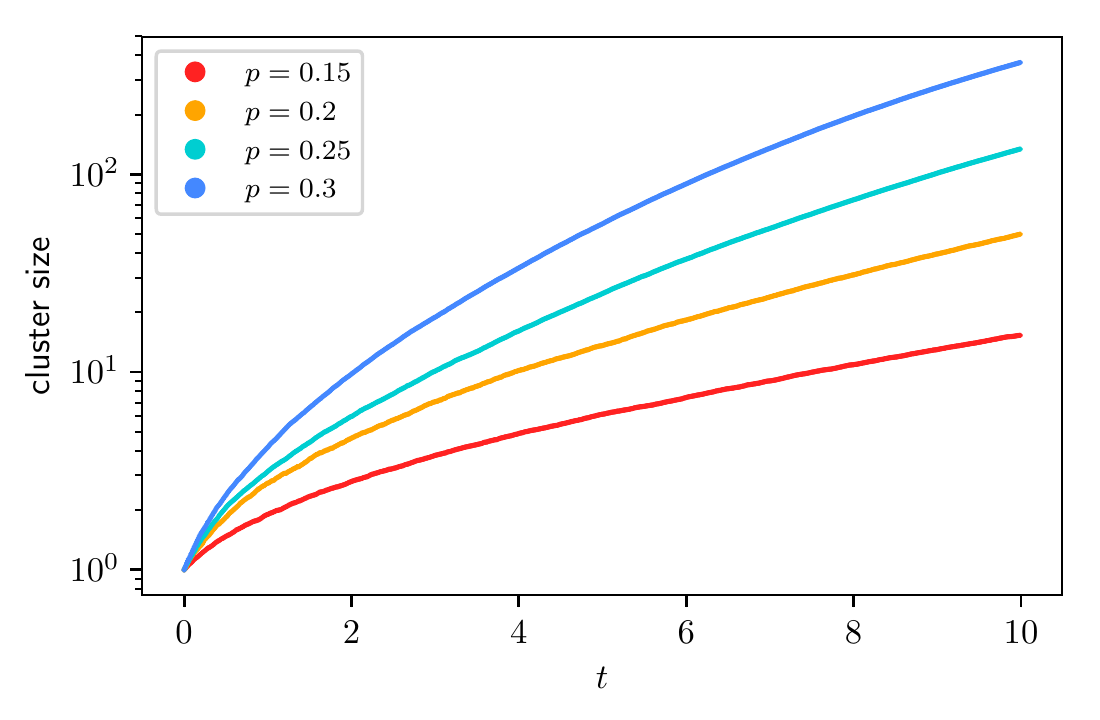}
\caption{The numerical data for the growth of the global OTOC in our stochastic model using the KRb polar molecule parameters. $p$ is the occupation fraction.}
\label{fig:KRb}
\end{figure}

We numerically simulate the stochastic process on a simple cubic lattice for $p \in [15\%, 30\%]$. The global OTOC does match the order of magnitude of our estimation, see Fig.~\ref{fig:KRb}

In summary, the coherence time (in units where $\hbar = 1$) of the dipolar molecule systems is roughly one order of magnitude larger than in the NMR system, but the relatively low density of occupation hinders rapid operator growth. However, the cluster size has a $6$th power dependence with respect to the occupancy $p$ and $3$rd power dependence with respect to the coherence time. Hence, reaching a cluster of size $10^3$ requires only a moderate increase in occupancy or coherence time. Assuming the presently available factor of $10$ enhancement in the coherence time, we estimate the threshold to see significant operator growth to be $p\approx 30\%$. Beyond this density, one should be able to observe some of the growth patterns of the global OTOC in the polar molecule system.


\section{Discussion and Conclusion}

We showed that OTOCs of the global spins give the shape of the light-cone probed by local OTOCs, with the assumption that certain ``off-diagonal'' terms could be neglected. We argued that this is the case in quantum chaotic systems, and verified this property in a concrete random circuit model (App.~\ref{app:cross_term}) and with small scale exact diagonalization numerics (App.~\ref{app:off_diag}). As a corollary, the global OTOC is proportional to the area under the local OTOC curve. 

If the interactions are local, then the local OTOC will expand at the butterfly velocity and the area enclosed for a $d$-dimensional system is $(v_B t )^d$. From the asymptotic light cone and tail scalings in Tab.~\ref{tab:model1p}, this result can be extended to systems with long-range interactions, when $\alpha \ge d + \frac{1}{2} $. In that regime, the light cone is still linear, and the local OTOC's wavefront broadening is slower than linear. In contrast, when the long-range interaction exponent $\alpha < d + \frac{1}{2}$, the local OTOC's asymptotic light cone can be super-linear, and hence the growth of the global OTOC can be faster than $t^d$. When  $\alpha \in (d, d+ \frac{1}{2})$, assuming the tail distributions in Tab.~\ref{tab:model1p} are correct in all dimensions, the local OTOC scales like
\begin{equation}
\left( \frac{r}{t^{\frac{1}{2\alpha - 2d}}} \right)^{ - 2\alpha } 
= \frac{t^{\frac{\alpha }{\alpha - d}}}{r^{2\alpha} }
\end{equation}
Integration over the separation $r$ in $d$ dimension brings in a constant; the time dependence is $t^{\frac{\alpha}{\alpha - d}}$. 

With these theoretical preparations, we examined experiments with solid adamantane, which consists of adamantane molecules each with $16$ nuclear spins, arranged in a face-centered-cubic lattice. We showed that previous stochastic $Kn$ space approaches predict an exponential growth of the (global and local) OTOCs, but these approaches neglect the spatial structure of the interaction. Due to the fast molecular tumbling, we get a simple point-dipole model, and further simplify this model to a Brownian model which retains only the power-law character of the interactions and the number of spins per lattice site. 

This model corresponds to $\alpha = d =3$, in which case the scaling function is $\left( \frac{r}{t^{\frac{1}{4d} \log_2 t}} \right)^{ - 2d} $. This gives a time dependence of the global OTOC of $t^{\frac{1}{2}\log_2 t}$, which is faster than any power of time. This is a remarkable prediction which would be extremely interesting to observe in an experiment if the coherence time allows. In particular, this result indicates that the asymptotic rate of operator spreading with dipoles proceeds infinitely fast when the speed of light is neglected. 

At present, experiments have only probed the relatively short-time regime of many-body dynamics. We managed to estimate and match the order of magnitude of the global OTOC given the coherence time. Furthermore, using a two-parameter numerical simulation of our stochastic model, we could get remarkable agreement with the experimental global OTOC curve for adamantane up to cluster sizes of order $10^4$. Our result gives an interpretation of the mysterious $t^{4.36}$ power law fit in the experimental data. And we predict the above faster-than-polynomial growth of global OTOCs at longer times.

There are also a few important complications in the comparison with experiments. First, we ignored any effects of dissipation, coupling to the lattice, and so on, which are present and important. Our theory here assumes ideal evolution, and while experiments are pushing to longer many-body coherence times, it would be very interesting to supplement our theory with dissipative effects. Second, our theory is predicated on a hypothesis of emergent universality in chaotic systems, i.e. that asymptotic time dependence of the global OTOC is characterized by the dimensionality of the system and the power-law exponent of the interactions independent of other system-specific features. Even conservation laws are not expected to strongly modify the leading growth behavior of the local OTOC, which is what controls the global OTOC. But at the relatively short times accessible in current experiments, all the details of the system can matter. Our universal theory makes the cleanest predictions at somewhat longer time scales, so it would be interesting to study in more detail particular Hamiltonians, e.g. standard truncated dipolar vs double quantum models. There are also observations of ``localization'' effects \cite{alvarez_localization-delocalization_2015,alvarez_localization_2011,alvarez_nmr_2010,alvarez_quantum_2013} when the forward and backward evolutions are perturbed to no longer match by adding a small term to the forward Hamiltonian that conserves total $Z$. It would be interesting to see how the eventual saturation of the global OTOC (or the observed tendency) can be described in our operator spreading theory.

Generalizing beyond nuclear spin systems, we noted that the capabilities required to engineer a many-body dipole Hamiltonian and its forward/backward evolution are present in other contexts. In particular, ultra-cold polar molecules confined in an optical lattice have electric dipole interactions and similar global control can also be achieved by microwave pulse sequences. Reported polar molecule experiments have a longer (dimensionless) coherence time but also exhibit a relatively low occupancy of the lattice which hinders operator spreading. We argued based on simple estimates and our stochastic model that if the occupancy of each site can be modestly increased, say to about 30\% or more, then extrapolations of existing experimental configurations should be able to probe the global OTOC dynamics predicted by our theory.

Building on these developments, there are a number of additional directions for further work. First, it might be possible to better match with experiments on different compounds, e.g. Ref.~\onlinecite{sanchez_clustering_2014}, by incorporating appropriate conservation laws into the stochastic model. Alternatively, experiments on adamantane might be modified to explicitly realize a random circuit model if one can study the system at lower temperatures where the tumbling time is comparable to the nearest neighbor dipole interaction strength. Second, it would be interesting to explore the role of dimensionality, for example, in quasi-one-dimensional systems, and look for crossovers to three-dimensional behavior. Third, given the relatively large value of the number of dipoles per site in adamantane, it is interesting to explore various so-called large $N$ models, which feature many degrees of freedom per site, and which are often analytically tractable. Fourth, NV centers also provide a tempting platform to explore this physics, and it would be interesting to develop a concrete proposal in that context.


\acknowledgements
TZ was supported by a postdoctoral fellowship from the Gordon and Betty Moore Foundation, under the EPiQS initiative, Grant GBMF4304, at the Kavli Institute for Theoretical Physics. 
TZ is currently supported as a postdoctoral researcher from NTT Research Award AGMT DTD 9.24.20 and the Massachusetts Institute of Technology. BGS acknowledges support from the Simons Foundation via the It From Qubit Collaboration. This research is supported in part by the National Science Foundation under Grant No. NSF PHY-1748958. This work was supported by a grant to the KITP from the Simons Foundation (\#216179). 

\appendix

\section{The diagonal approximation of the global OTOC} 
\label{app:cross_term}

In Sec.~\ref{sec:otoc_lc} of the main text, we argued that the global OTOC can be well approximated by the diagonal terms---the local OTOCs:
\begin{equation}
\label{eq:app_global_local}
\begin{aligned}
-\tr( [Z(t), Z]^2 ) &= - \sum_{abcd} \tr( [Z_a(t), Z_b][Z_c(t), Z_d] ) \\
&\approx -\sum_{ab} \tr( [Z_a(t), Z_b][Z_a(t), Z_b] ) .
\end{aligned}
\end{equation}
Hence the global OTOC measures the area under the local OTOC curve. 

\begin{figure}[h]
\centering
\subfigure[]{
  \label{fig:ruc}	
  \includegraphics[width=0.46\columnwidth]{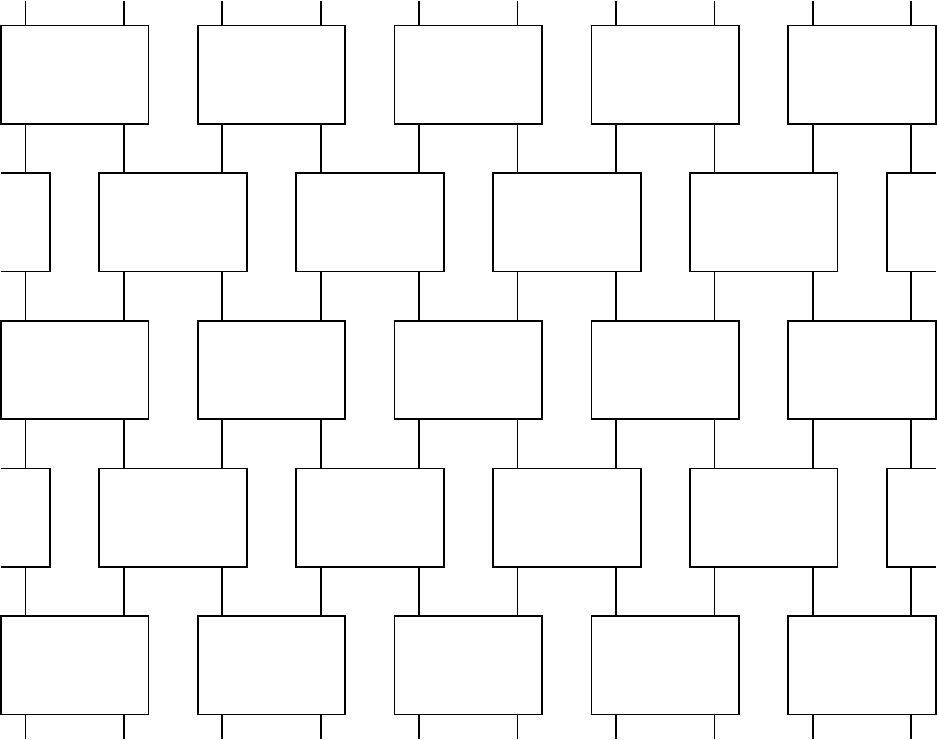}
}
\subfigure[]{
  \label{fig:fb}	
  \includegraphics[width=0.46\columnwidth]{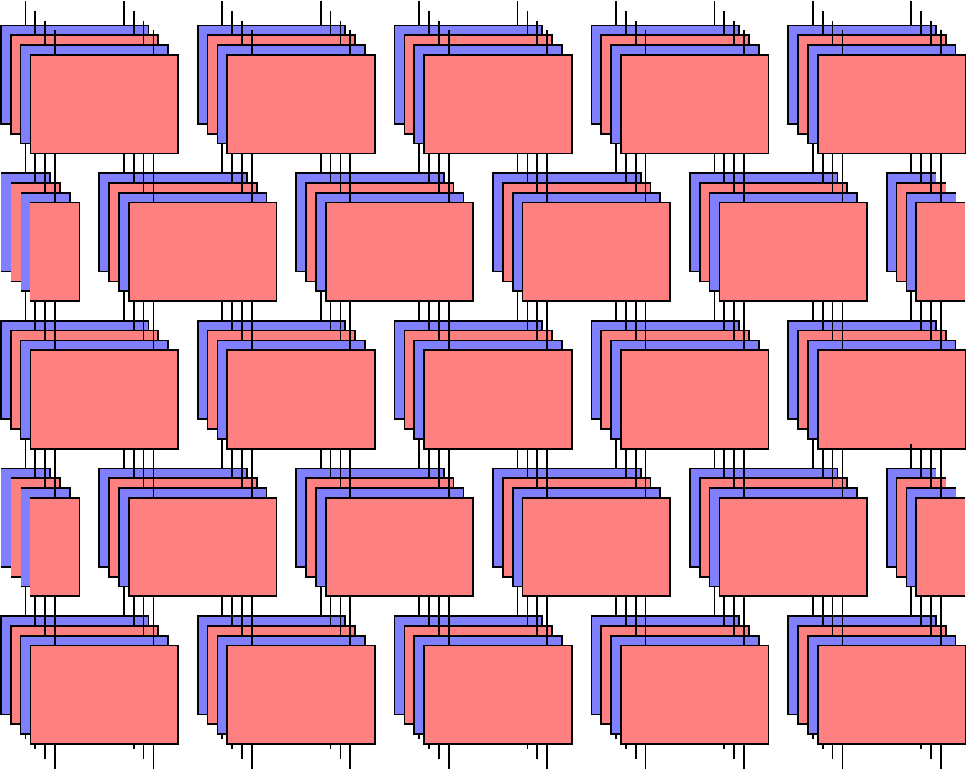}
}
\caption{Unitary evolution matrix in the form of a quantum circuit. (a) The structure of the circuit. The four-leg tensor is a unitary matrix (gate) on two sites. The structure models local interactions. (b) The forward and backward evolutions involved in the computation of OTOC. Red/blue gates represent forward/backward unitary evolution.}
\label{fig:ruc_fb}
\end{figure}

In this appendix, we present a more rigorous calculation to show why the off-diagonal terms can be neglected. We assume that the system is evolved by a unitary circuit with a structure shown in Fig.~\ref{fig:ruc}. Then the entanglement membrane picture \cite{zhou_entanglement_2020,zhou_emergent_2019,von_keyserlingk_operator_2018,jonay_coarse-grained_2018,nahum_quantum_2017,nahum_operator_2018} developed in Ref.~\onlinecite{zhou_entanglement_2020} becomes a useful tool to estimate the off-diagonal terms for a generic chaotic evolution without random averaging. This justifies our claim for short-range interacting systems. We then generalize the estimate to the long-range case.

\subsection{Systems with local interactions}

In this subsection, we assume the time evolution operator is modeled by a unitary circuit with the structure shown in Fig.~\ref{fig:ruc}. The only requirement for the gate choice is that the whole circuit is non-integrable and in a crude sense chaotic. This applies to systems with or without lattice/time translation symmetries. 

For each term in the expansion of the global OTOC in Eq.~\eqref{eq:app_global_local}, the product of the commutator can be written as four terms 
\begin{equation}
\begin{aligned}
&\tr( [Z_a(t), Z_b][Z_c(t), Z_d] ) = \\
&- \tr( Z_a(t) Z_b Z_c(t) Z_d )  + \tr( Z_a(t)  Z_c(t) Z_d  Z_b) \\
&+ \tr( Z_a(t)  Z_b Z_d Z_c(t) )  - \tr( Z_a(t)  Z_d Z_c(t)   Z_b) 
\end{aligned}
\end{equation}
Each term contains two forward and two backward evolutions as shown in Fig.~\ref{fig:ruc_fb}. The operator insertion and traces bring in boundary conditions that contract with the 4-layer structure in Fig.~\ref{fig:ruc_fb}. On site without the operator insertion, we have $\fineq[-0.4ex][0.25][0.8]{\idst[0][0][][][r]} $ at the bottom and $\fineq[-0.4ex][0.25][0.8]{\swapst[0][0][][][]} $ on the top. At site $a/c$, we have
\begin{equation}
 \left\lbrace
\begin{aligned}
  & \fineq[-.6ex][0.5][0.5]{\idst[0][0][$Z_a$][][r]} \otimes \fineq[-.6ex][0.5][0.5]{\idst[0][0][][$Z_c$][r]}  & \quad  a \ne c\\
  & \fineq[-.6ex][0.5][0.5]{\idst[0][0][$Z_a$][$Z_a$][r]} & \quad a = c \\
\end{aligned} \right. 
\end{equation}
They connect with the tensor in Fig.~\ref{fig:ruc_fb} at the bottom. At site $b/d$, we have
\begin{equation}
\left\lbrace
\begin{aligned}
  & \left(  \fineq[-.8ex][0.5][0.5]{\swapst[0][0][$Z_b$][]} - \fineq[-.8ex][0.5][0.5]{\swapst[0][0][][$Z_b$]}  \right) \otimes 
  \left(  \fineq[-.8ex][0.5][0.5]{\swapst[0][0][$Z_d$][]} - \fineq[-.8ex][0.5][0.5]{\swapst[0][0][][$Z_d$]}  \right) & b \ne d  \\
  & 2(\fineq[-.8ex][0.5][0.5]{\swapst[0][0][][]}- \fineq[-.8ex][0.5][0.5]{\swapst[0][0][$Z_b$][$Z_b$]}  ) & \quad b = d  \\
\end{aligned} \right. 
\end{equation}
They connect with the tensor in Fig.~\ref{fig:ruc_fb} on the top.

Our approach converts each term into a statistical mechanical problem of interacting spins. Following the notation in Ref.~\onlinecite{zhou_entanglement_2020}, and assuming the local Hilbert space dimension is $q$ (for spin-$\frac{1}{2}$ $q = 2$), we define 
\begin{equation}
\begin{aligned}
 | + \rangle  &= |\fineq[-.6ex][0.25][0.5]{\idst[0][0][][][r]} \rangle  \quad  | - \rangle  = | \fineq[-.6ex][0.25][0.5]{\swapst[0][0][][][r]} \rangle \\
 | +^* \rangle &= \frac{1}{q^2 - 1}( |+ \rangle - \frac{1}{q} |- \rangle ) \\
 | -^* \rangle &= \frac{1}{q^2 - 1}( |- \rangle - \frac{1}{q} |+ \rangle ) \\
\end{aligned}
\end{equation}
where $|+^* \rangle $ and $|-^* \rangle $ are the corresponding dual states of $|+ \rangle $ and $|- \rangle $. For states on two sites, we have the dual states to be
\begin{equation}
\begin{aligned}
  |( ++)^* \rangle  &= \frac{1}{q^4 - 1} ( |++ \rangle  - \frac{1}{q^2} | -- \rangle  )\\
  |( ++)^* \rangle  &= \frac{1}{q^4 - 1} ( |-- \rangle  - \frac{1}{q^2} | ++ \rangle  )
\end{aligned}
\end{equation}
We have the corresponding kets similarly constructed from $\langle \fineq[-0.4ex][0.25][0.8]{\idst[0][0][][][]}|$ and $\langle \fineq[-0.4ex][0.25][0.8]{\swapst[0][0][][][]}|$. 

\begin{figure}[h]
\centering
\includegraphics[width=0.8\columnwidth]{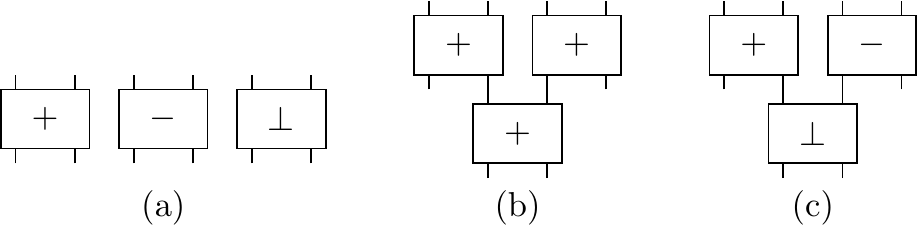}
\caption{(a) Three choices of spins according to the decomposition in Eq.~\eqref{eq:u4_perp}. Rules of the spins: (b) The same spins on the top two gates forces the same spin below them (the weight is $1$) (c) $\perp$ spin can only occur at the domain wall -- below different spins or another $\perp$ spins. }
\label{fig:u_gate_rule}
\end{figure}

With these facilities, the random average of two forward gates and two backward gates under the Haar ensemble can be written as
\begin{equation}
\label{eq:u4_aver}
\overline{u\otimes u^* \otimes u \otimes u^* } = |( ++)^* \rangle \langle ++| + |( --)^* \rangle \langle --|
\end{equation}
In Ref.~\onlinecite{zhou_entanglement_2020}, we generalize this expression to systems without randomness, so that there is an additional term in this expression
\begin{equation}
\label{eq:u4_perp}
u\otimes u^* \otimes u \otimes u^* = |( ++)^* \rangle \langle ++| + |( --)^* \rangle \langle --| + "\perp" 
\end{equation}
Here the $\perp$ state is the difference of the LHS and the first two terms of the RHS. It is a tensor that depends on the gate $u$. Each gate in the circuit then has three choices, $+$, $-$ and $\perp$ according to the terms in Eq.~\eqref{eq:u4_perp}, see Fig.~\ref{fig:u_gate_rule} (a). The OTOC then becomes a partition function of those spins on each gate, and the boundary conditions are given above (we adopt a more convenient convention and rewrite them in Eq.~\eqref{eq:ac_bd_cond} and Eq.~\eqref{eq:bd_bd_cond} below).

The expression on the LHS of Eq.~\eqref{eq:u4_aver} is time reversal invariant. However, the RHS has a preferred direction of time, due to the choice of the non-orthogonal states $++$ and $--$. In this convention, it is more convenient to turn the whole diagram upside down, with  $a$ and $c$ on the top, $b$ and $d$ at the bottom. To avoid confusion, we rewrite the boundary condition at $a/c$ as 

\begin{equation}
\label{eq:ac_bd_cond}
 \left\lbrace
\begin{aligned}
  & \fineq[-.6ex][0.5][0.5]{\idst[0][0][$O_a$][][]} \otimes \fineq[-.6ex][0.5][0.5]{\idst[0][0][][$O_c$][]}  & \quad  a \ne c\\
  & \fineq[-.6ex][0.5][0.5]{\idst[0][0][$O_a$][$O_a$][]} & \quad a = c \\
\end{aligned} \right. 
\end{equation}
and at $b/d$ as
\begin{equation}
\label{eq:bd_bd_cond}
\left\lbrace
\begin{aligned}
  & \left(  \fineq[-.8ex][0.5][0.5]{\swapst[0][0][$O_b$][][r]} - \fineq[-.8ex][0.5][0.5]{\swapst[0][0][][$O_b$][r]}  \right) \otimes 
  \left(  \fineq[-.8ex][0.5][0.5]{\swapst[0][0][$O_d$][][r]} - \fineq[-.8ex][0.5][0.5]{\swapst[0][0][][$O_d$][r]}  \right) & b \ne d  \\
  & 2(\fineq[-.8ex][0.5][0.5]{\swapst[0][0][$O^2_b$][][r]}- \fineq[-.8ex][0.5][0.5]{\swapst[0][0][$O_b$][$O_b$][r]}  ) & \quad b = d  \\
\end{aligned} \right. 
\end{equation}
with the more general traceless operators $O_{a, b,c,d}$. 

There are rules for the spin assignment for each gate. When the spins of neighboring gates are the same, then it forces the gate below to have the same spin. Such a structure has weight $1$. When the spins of neighboring gates are different, the spin below can be either $+$, $-$ or $\perp$. The first two choices form a perfect domain wall, which has weight $\frac{q}{q^2 +1}$ (for spin-$\frac{1}{2}$, it is $\frac{2}{5}$). The $\perp$ spin can only occur beneath a domain wall or other $\perp$ spins, see examples in Fig.~\ref{fig:u_gate_rule} (b) and (c).

We then deal with different choices of $a$,$b$,$c$ and $d$. 

1. local OTOC: $a = c$, $b = d$

\begin{figure}[h]
\centering
\includegraphics[width=0.8\columnwidth]{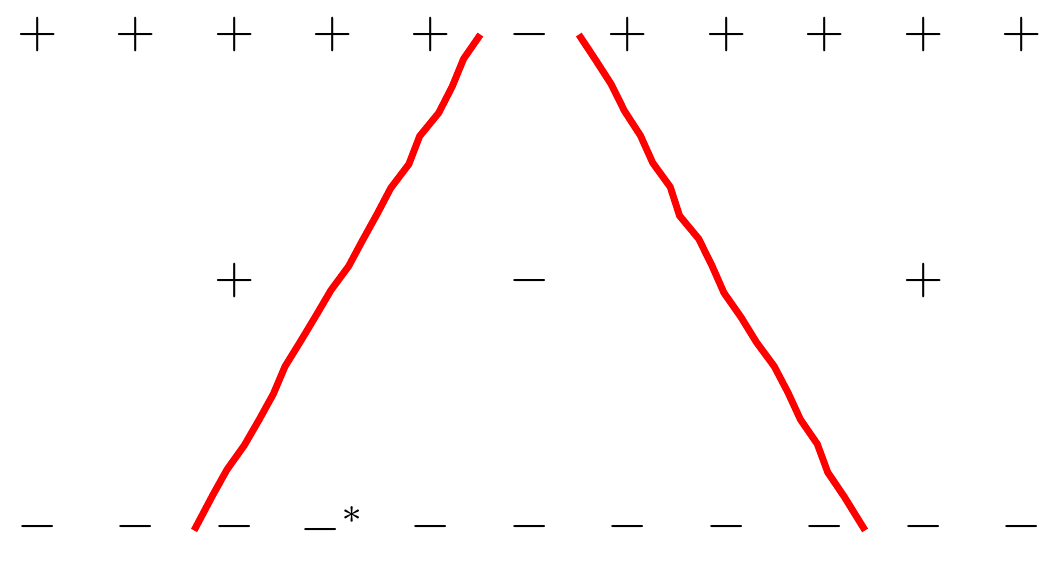}
\caption{Domain wall configuration for $a = c$, $b = d$. The operator insertion point $a$ at the bottom forces a $-$ spin. There are two domain walls emitted from the top.}
\label{fig:case_1}
\end{figure}

The local OTOC has been analyzed in Ref.~\onlinecite{zhou_entanglement_2020}. Here we review the calculation and set a benchmark for the other cases. 

We consider a more general local OTOC:
\begin{equation}
- \tr( [O_a, O_b]^2 ) 
\end{equation}
The boundary conditions are
\begin{equation}
\begin{aligned}
 \fineq[-.6ex][0.5][0.5]{\idst[0][0][$O_a$][$O_a$]} , \quad 2(\fineq[-.8ex][0.5][0.5]{\swapst[0][0][$O_b^2$][][r]}- \fineq[-.8ex][0.5][0.5]{\swapst[0][0][$O_b$][$O_b$][r]}  ).  \\
\end{aligned}
\end{equation}
To simplify the result, we can random average the single site traceless operator $O_a\rightarrow V_a O_a V_a^{\dagger} $. This amounts to contract $|+^* \rangle  \langle  + |  + |-^* \rangle  \langle  -|$ with the boundary loops at site $a$ and site $b$. The state at site $b$ becomes 
\begin{equation}
2q \tr( O_b^2 )| -^* \rangle  
\end{equation}
i.e. the spin at $b$ will force a $-$ spin above it. Then the top site $a$ must have be a $-$ spin, otherwise an all $+$ boundary condition can be pushed to the bottom with unitarity property and the whole quantity vanishes. So we have
\begin{equation}
  \frac{\tr( O_a^2) }{q^2 - 1} \langle  - | 
\end{equation}
In summary, the top boundary has two domain walls emitted at the two sides of $a$. The bottom boundary condition favors $-$ spins, so the two domain walls tend to have a larger $-$ domain. However domain wall with larger slope (here defined to be horizontal distance divided by vertical distance, which has the dimension of velocity) also costs energy. The equilibrium is reached when both of the domain walls are stretched as slope $v_B$, see Fig.~\ref{fig:case_1}. The domain wall fluctuations with a region of size $\sqrt{t}$. When site $b$ is outside the slope $v_B$ of site $a$, it will then force the domain wall to have slope more than $v_B$, resulting in the exponential decay of the OTOC.

2. $a = c$ and $b \ne d $

The case for $a \ne c$ and $b = d$ is the same, since
\begin{equation}
\tr( [Z_a(t), Z_b] [Z_c(t), Z_d] ) = \tr( [Z_a, Z_b(-t)] [Z_c, Z_d(-t)] )
\end{equation}

\begin{figure}[h]
\centering
\includegraphics[width=0.8\columnwidth]{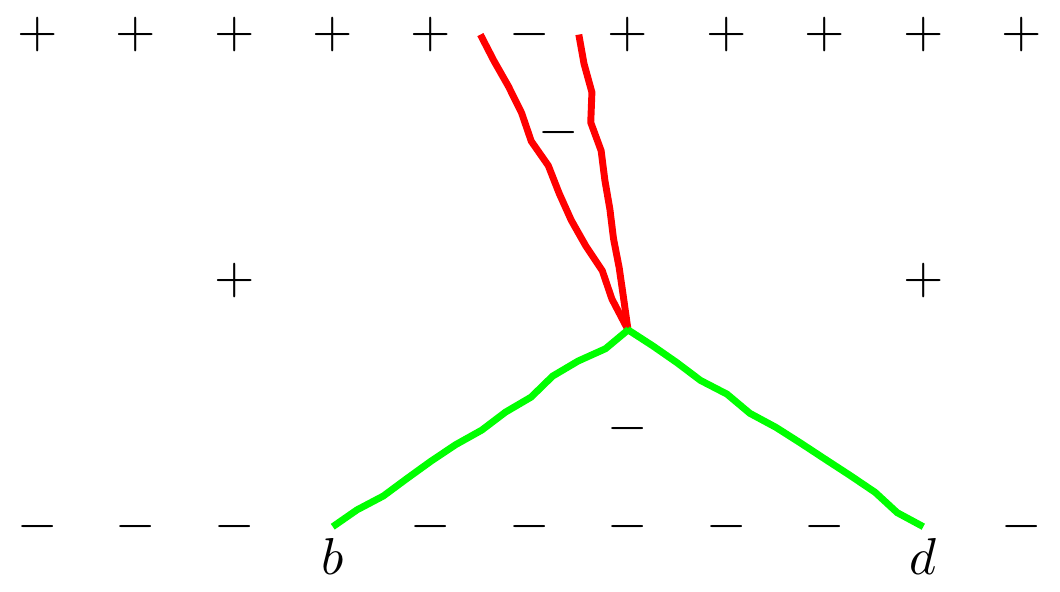}
\caption{Domain wall configuration for $a \ne c$, $b = d$. There are $\perp$ clusters connecting $a, b, c, d$.  }
\label{fig:case_2}
\end{figure}

The boundary conditions are
\begin{equation}
\begin{aligned}
\fineq[-.6ex][0.5][0.5]{\idst[0][0][$O_a$][$O_a$]}
\quad 
\left(  \fineq[-.8ex][0.5][0.5]{\swapst[0][0][$O_b$][][r]} - \fineq[-.8ex][0.5][0.5]{\swapst[0][0][][$O_b$][r]}  \right) \otimes \left(  \fineq[-.8ex][0.5][0.5]{\swapst[0][0][$O_d$][][r]} - \fineq[-.8ex][0.5][0.5]{\swapst[0][0][][$O_d$][r]}  \right)
\end{aligned}
\end{equation}
The boundary condition at either site $b$ or $d$ can only accept a $\perp$ spin above it (the contraction with either $+$ or $-$ is zero). Hence if we average over the operator at site $a$, it has to be a $-$ spin so that domain wall and $\perp$ spin can be produced. We arrive at the boundary spin configurations in Fig.~\ref{fig:case_2}. 

There is another special feature for the boundary conditions at $b$ and $d$ -- it is odd under then exchange of the first and second copies of $u\otimes u^*$. This amounts to interchanges the two circles in the boundary condition, which generates a minus sign. So for example, if a configuration surround site $b$ with only $+, -$ spins, isolating it from site $b$, such as the following
\begin{equation}
\fineq[-0.8ex][0.6][1]{
\ugate[0][0][][$$];
  \ugate[-1][1.5][][$+$];
  \ugate[1][1.5][][$$];
  \ugate[0][3][][$+$];
  \ugate[2][3][][$-$];
  \node[below] () at (0,0) {$b$};
  \node[below] () at (1,0) {$-$};
  \node[below] () at (-1,1.5) {$-$};
  \node[below] () at (2,1.5) {$-$};
  \node[below] () at (3,3) {$-$};
  \draw[dashed,red, line width = 1pt] (-0.5,0.0)--++(0,1.5)--++(1,0)--++(0,1.5)--++(2,0)--++(0,-1.5);
}
\end{equation}
then the region enclosed by the red dash line has $\pm$ as its boundary conditions. Its weight has a symmetry by interchanging the first and second copies of $u\otimes u^*$. On the global scale, this symmetry can be viewed the cyclic property of the trace
\begin{equation}
\label{eq:O_a_O_b_eg}
\tr( [O_a(t), O_b] O_a(t) ) = \tr( O_a(t) [O_a(t), O_b] )
\end{equation}
The unitaries and $\pm$ boundary conditions are invariant. However, the boundary condition contributes a minus sign. This indicates that the weight of such diagram is zero, just like Eq.~\eqref{eq:O_a_O_b_eg}. Therefore the $\perp$ clusters of sites $b$ and $d$ have to meet in the bulk. 

If the distance $x_{bd}$ between $b$ and $d$ is greater than $2t$, then there is no possibility for a $\perp$ cluster to connect them. We can then restrict $x_{bd}$ to $2t$. A typical configuration is shown in Fig.~\ref{fig:case_2}. The two green curves represent the $\perp$ cluster that connects sites $b$ and $d$. The two red lines represent the domain walls that seed the $\perp$ cluster. Compared to a local OTOC, the diagram suffers from two main suppressions. One is that the $-$ domain connecting the bottom boundary has size $x_{bd}$, while a local OTOC has at least $2 v_B t$. This brings in a factor of $q^{- (2v_B t - x_{bd})}$. Another suppression comes from the $\perp$ cluster. Its relative weight with respect to an ordinary domain wall is $q^{-t_{\perp}}$ where $t_{\perp}$ is the persistent time of a cluster. In this case, the suppression factor is $q^{- x_{bd}}$. The red and green two-segment domain wall is also not optimal, but we neglect this factor. Overall, the diagram in Fig.~\ref{fig:case_2} can be a factor of $q^{- (2v_B t - x_{bd}} q^{- x_{bd}} = q^{-2 v_B t}$ smaller than a local OTOC. Even if there can be $(2 t)^2$ terms, the contribution is still negligible than the $v_B t$ local OTOCs. 

3. $a \ne c$, $b \ne d$
\begin{figure}[h]
\centering
\includegraphics[width=0.8\columnwidth]{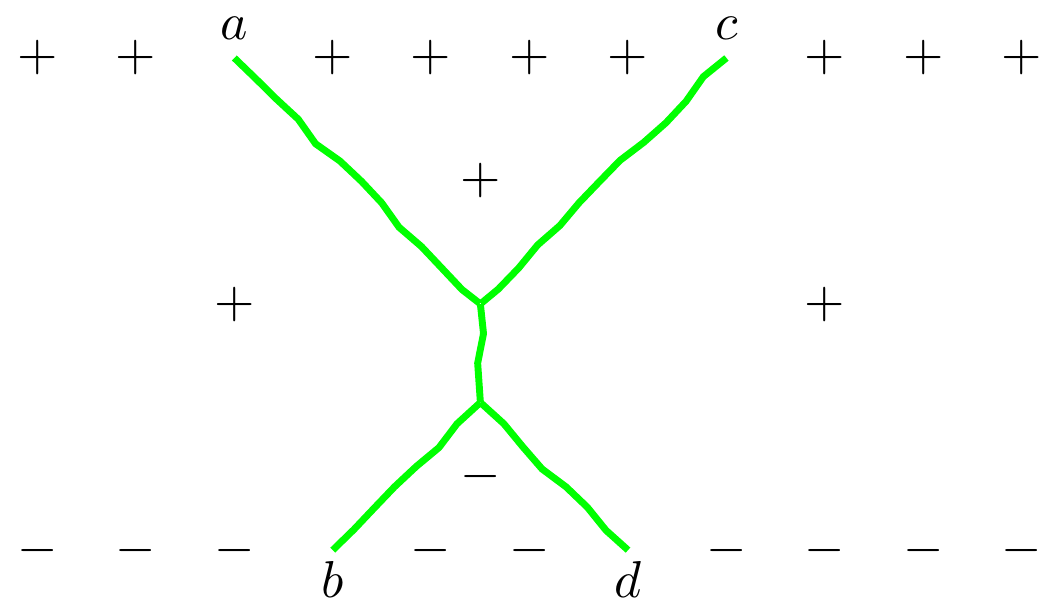}
\caption{Membrane configuration for $a \ne c$, $b \ne d$. There is a large $\perp$ cluster connecting $a$, $b$, $c$, $d$, resulting in an overall $q^{-t}$ decay. }
\label{fig:case_3}
\end{figure}

We have boundary conditions
\begin{equation}
\fineq[-.6ex][0.5][0.5]{\idst[0][0][$O_a$][][]} \otimes \fineq[-.6ex][0.5][0.5]{\idst[0][0][][$O_c$][]} 
\end{equation}
and
\begin{equation}
\left(  \fineq[-.8ex][0.5][0.5]{\swapst[0][0][$O_b$][][r]} - \fineq[-.8ex][0.5][0.5]{\swapst[0][0][][$O_b$][r]}  \right) \otimes \left(  \fineq[-.8ex][0.5][0.5]{\swapst[0][0][$O_d$][][r]} - \fineq[-.8ex][0.5][0.5]{\swapst[0][0][][$O_d$][r]}  \right)
\end{equation}

The analysis for $b$ and $d$ is the same. They will be connected by a $\perp$ cluster. Now sites $a$ and $c$ have to be connected by a $\perp$ cluster spin. Since $\perp$ spin can be ended by a domain of the $+$ or $-$ spins, the $\perp$ clusters of site $a$ and $c$ have to join or meet the $\perp$ clusters of $b$ and $d$. Such a large connected cluster gives a suppression of $q^{-t}$. Fixing the position of $a$, in order for all the $\perp$ clusters to meet, there are at most $(2t)^3$ terms. Hence the sum of all diagrams in Fig.~\ref{fig:case_3} is at least $t^3 q^{-t}$ smaller than the local OTOC, which is negligible in large $t$. 

We conclude that when evolution is given by a local chaotic circuit, the asymptotic scaling of the global OTOC can be very well approximated by the sum of the local OTOCs. For local interactions, it is just the size of the light cone $2v_B t$. 

\subsection{Generalization to Long-range Interactions}

We only consider the off-diagonal terms here. 

In the analysis of the local interaction, we see that when $b \ne d$, the $\perp$ cluster brings in a factor of $q^{ -( v_B t - x_{bd})}$. For the long-range interaction, they can be generalized to $q^{ -(2x_{\rm LC} - x_{bd})}$ in case 2. Summing over different choices of $d$ within a regime of $2x_{\rm LC}$ sites gives an order $1$ factor. There is still a summation of $b$ within a regime of $2x_{\rm LC}$ sites. The off-diagonal terms are real numbers that can be positive or negative. Assuming these are random numbers, then the typical amplitude of the sum is $\sqrt{ x_{LC} }$, still smaller than $x_{LC}$, which is a lower bound for the sum of the local OTOCs. So case 2 is negligible. 

    For case 3, there is still a $\perp$ cluster connecting the four sites, although the spins can spread long locally. But it is safe to say there is at least one $\perp$ spin at each time slice. Then the suppressing factor of $q^{-t}$ still works. There are at most $x_{LC}^3$ sites when fixing site $a$, so the sum is at the sale of $x_{LC}^3 q^{-t}$. Since light cone spreads at most with a stretched exponential (for $\alpha > 0.5$), this sum for case 3 will be negligible for large $t$. 

Therefore we still expect the sum of the local OTOCs to dominate.



\section{NMR experimental review}
\label{app:nmr_review}

Nuclear magnetic resonance (NMR) is a standard technology that uses nuclear spin as the degree of freedom to study interacting quantum magnetism in and out of equilibrium. In this appendix, we review at a high level the experimental procedures to measure the global OTOC in materials like adamantane and some standard theoretical interpretations of the data. Throughout the appendix, we use $I_{iz} = \frac{1}{2} Z_i$ to represent the nuclear spin-$\frac{1}{2}$ operator at site $i$, and $I_z = \frac{1}{2} \sum_{i} Z_i$ for the total spin operator.

\subsection{The single quantum coherence}
In a typical solid state NMR experiment, a material is exposed to a strong uniform magnetic field. The energy scale of this Zeeman interaction is much larger than any other scale in the problem except the temperature, hence in equilibrium the nuclear spins are polarized in the $z$ direction---the direction of the magnetic field. For proton nuclear spin
\begin{equation}
H_{\rm Zeeman} = - \gamma I_z B_0 = - \gamma \frac{1}{2} Z B_0 
\end{equation}
where $\gamma$ is the gyromagnetic ratio. If $B_0$ is $1T$, $\gamma B$ corresponds to $2\pi \times 42.6 \text{ M Hz}$ or $2mK$ in energy. Hence at room temperature, the initial density matrix at thermal equilibrium can be expanded in the high temperature limit
\begin{equation}
\rho = \frac{e^{ + \frac{\gamma B_0}{2k_B T} Z }}{\tr( e^{ + \frac{\gamma B_0}{2k_B T} Z } )} \propto \I + \frac{\gamma B_0}{2k_B T} Z
\end{equation}
Since in the correlators below, the $\I$ part of the density matrix gives zero contribution, oftentimes the density matrix is written as $\rho = Z$. 

Modulated radio-frequency waves can exert a magnetic field in the $x$ direction on top of the Larmor precession. When the radio-frequency wave is removed, the magnetization will decay to its equilibrium value through spin-spin or spin-lattice relaxation processes. The $x, y$ magnetization can generate induction in the coil, and reading out the free induction signal can tell us $\tr( \rho X )$ and $\tr( \rho Y)$. The measurement of $\tr( \rho Z )$ can be converted to the $X$, $Y$ magnetization by first imposing a spin rotation pulse---the $\frac{\pi}{2}$ pulse---and measuring the free induction signal afterward. 

There are also internal interactions, on the scale of kHz. The most prominent one for protons is the dipole interaction. But since the Zeeman field corresponds to an energy scale of $10^3$ kHz, the dipole interaction is well-approximated by secular form,
\begin{equation}
  H_{\rm int}  = \sum_{i\ne j }D_{ij} ( 3 I_{zi} I_{zj} - \vec{I}_i \cdot \vec{I}_j ),
\end{equation}
where
\begin{equation}
\label{eq:D_ij}
D_{ij} = \frac{\gamma^2 \hbar}{r_{ij}^3} \frac{3 \cos^2 \theta_{ij} - 1}{2} . 
\end{equation}
Other interactions, such as the chemical shift, scalar coupling and quadrupole coupling either vanish for protons or are much smaller. 

With this setup, one can measure the magnetization of the time evolved state, for example 
\begin{equation}
\tr( e^{i Ht} \rho e^{ -i Ht}  X) 
\end{equation}
Since the $X$ operator changes the total $Z$ eigenvalue by $\pm 1$, the measurement only probes the matrix elements of $\rho(t)$ slightly away from the diagonal. Hence, it is called the single quantum coherence.

\subsection{Multiple quantum coherence}

The multiple quantum coherence corresponds to the expectation values of operators that change the total $Z$ eigenvalue by more than $1$. We can systematically decompose the density matrix as
\begin{equation}
\rho = \sum_{n }\rho_n ,
\end{equation}
where the $n$-quantum coherence component satisfies
\begin{equation}
\label{eq:def_rho_n}
e^{ i \phi I_z } \rho_n e^{ - i\phi I_z } = \rho_n e^{ i n\phi }.
\end{equation}
Formally, MQC can be defined as
\begin{equation}
g_n = \frac{1}{\tr( I_z^2 )} \tr( \rho_n \rho_{-n} ) 
\end{equation}

Experimentally, one can add a twist after a time evolution to measure the Fourier transform of the multiple quantum coherence.
\begin{equation}
\label{eq:I_phi_t}
I(\phi, t ) = \frac{1}{\tr( I_z^2 )} \tr( e^{i\phi I_z} \rho(t) e^{ i \phi I_z} e^{iHt} I_z e^{-iHt} ) 
\end{equation}
In fact, expanding $\rho(t) $ and using the property in Eq.~\eqref{eq:def_rho_n}, we have
\begin{equation}
I( \phi, t ) = \frac{1}{\tr( I_z^2 )} \tr( \sum_n \rho_n e^{i n\phi} \sum_m \rho_m ) = \sum_n g_n e^{i n\phi} .
\end{equation}
On the other hand
\begin{equation}
\begin{aligned}
\sum_n n^2 g_n &= -\partial_\phi^2 I(\phi, t )\Big|_{\phi = 0}  \\
&= \frac{1}{\tr( I_z^2 )} \tr( [I_z , [I_z ,\rho(t)]] I_z(t) ) \\
&= - \frac{1}{\tr( I_z^2 )} \tr( [I_z ,I_z(t)][ I_z , I_z(t)] ) .
\end{aligned}
\end{equation}
Therefore if we sample $I(\phi, t)$ at discrete values of $\phi$, we can do an inverse Fourier transform to figure out the multiple quantum coherence $g_n$, whose second moment is the global OTOC. 

\subsection{Engineering of the backward time evolution}

With the presence of the external radio-frequency wave, the total Hamiltonian in the rotating frame is 
\begin{equation}
H = H_{\rm int} + H_{\rm rf} (t) 
\end{equation}
in which the latter can be time dependent. 

The analysis is usually carried out in the toggling frame. Define $U_{\rm rf}(t) = \mathcal{T} e^{ -i \int_0^t H_{\rm rf} (t')dt' } $, the toggling frame Hamiltonian is defined as
\begin{equation}
H_{\rm tf}(t) = U_{\rm rf}^{\dagger}(t) H_{\rm int} U(t)_{\rm rf} 
\end{equation}
so that
\begin{equation}
\mathcal{T} e^{ - \int_0^t H(t') dt' } = U_{\rm rf} (t) \mathcal{T} e^{ - \int_0^t H_{\rm tf}(t') dt' } 
\end{equation}
If the pulse is periodic, then $U_{\rm rf} = 1$ at those periods. So if we make measurements at those time points, the evolution is determined by the toggling frame Hamiltonian. The time independent effective Hamiltonian can be worked out by a Magus expansion. At the lowest order, the effective Hamiltonian is the average of the toggling frame Hamiltonian
\begin{equation}
H_{\rm eff} = \frac{1}{T} \int_0^t H_{\rm tf}(t') dt' 
\end{equation}
This is the basis to engineer interacting Hamiltonians in the NMR system. 

In the 80s, pulse sequences with four $\frac{\pi}{2}$ pulses were used to transform the original dipolar Hamiltonian to the double quantum Hamiltonian
\begin{equation}
H_{\rm DQ} = \sum_{ij} D_{ij} ( X_i X_j - Y_i Y_j ).
\end{equation}
Since the double quantum Hamiltonian is an operator of second order quantum coherence, a rotation by $\pi$ can create a minus sign. Thus an additional $\pi$ pulse on top of the original pulse sequence can create $-H_{\rm DQ}$, enabling backward time evolution. 

The experiment in the main text that we cite used a different approach. It is an eight pulse sequence with parameter $\delta$ in the time interval of each pulse. It can create the dipolar Hamiltonian in $Y$ direction with strength proportional to $\delta$, 
\begin{equation}
H_{\rm YY} = \delta \sum_{ij} D_{ij}( Y_i Y_j - Z_iZ_j - X_iX_j ) 
\end{equation}
Thus, by changing the sign of $\delta$, which amounts to changing the time interval between the pulses, one can obtain $- H_{\rm YY}$ and the backward time evolution. 

It is therefore technically possible to measure $I(\phi, t )$ in Eq.~\eqref{eq:I_phi_t} in an experiment.


\section{The Kn space stochastic process}
\label{app:Kn}

The NMR community has developed simplified models for multiple quantum coherence. It is a stochastic process in the $Kn$ space. Essentially, each multiple quantum coherence component of the density matrix is further decomposed as
\begin{equation}
\rho_n = \sum_{K} \rho_{Kn}.
\end{equation}
Using the Pauli string basis for operators, the number $K$ is the number of Pauli operators in the string. In the main text, we introduced this number as the effective size of the spin system. It is generically time dependent. The time evolution will transfer the operator from a smaller $K$ to large $K$. One can then view this as a stochastic process in the $Kn$ space, where the transition probability is determined by the number of interaction terms connecting the states. The assumption here is that all states with the same $K$ and $n$ are equally likely and the transition can occur when the Hamiltonian allows. A finer multiple quantum coherence, or the probability of staying at state $K,n$ is given by $g_{Kn}$. Clearly, $\sum_K g_{Kn} = g_n$. 

Ref.~\onlinecite{munowitz_multiplequantum_1987} used the double quantum Hamiltonian as an example. Defining 
\begin{equation}
Q_{Kn} = \sum_{c_+= n}^{K } { K \choose c_{+} }  { K- c_{+} \choose c_+ - n },
\end{equation}
the transition probability can be written as
\begin{equation}
\begin{aligned}
W_{K+1,n\pm 2, Kn} &= \frac{K ( N - K)}{N - 1} \frac{Q_{K-1,n} + Q_{K-1, n\pm 1} }{Q_{Kn}} \\
W_{K-1,n\pm 2, Kn} &= \frac{K ( K - 1)}{N - 1} \frac{Q_{K-2,n\pm 2} + Q_{K-2, n\pm 1} }{Q_{Kn}} 
\end{aligned}
\end{equation}
where $N$ is the total number of spins. 

\begin{figure}[h]
\centering
\subfigure[]{
  \label{fig:N6}	
  \includegraphics[width=0.8\columnwidth]{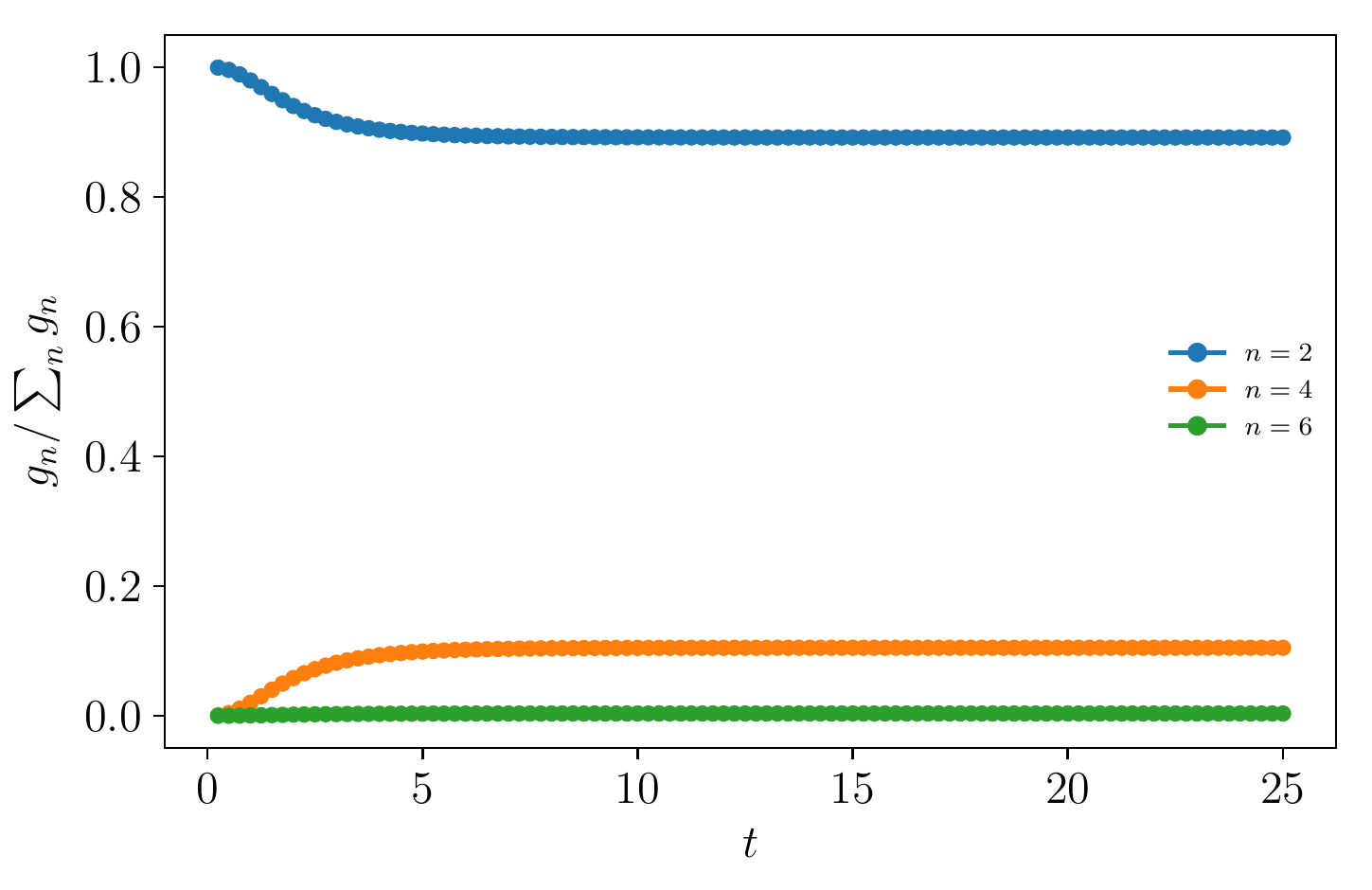}
}
\subfigure[]{
  \label{fig:N21}	
  \includegraphics[width=0.8\columnwidth]{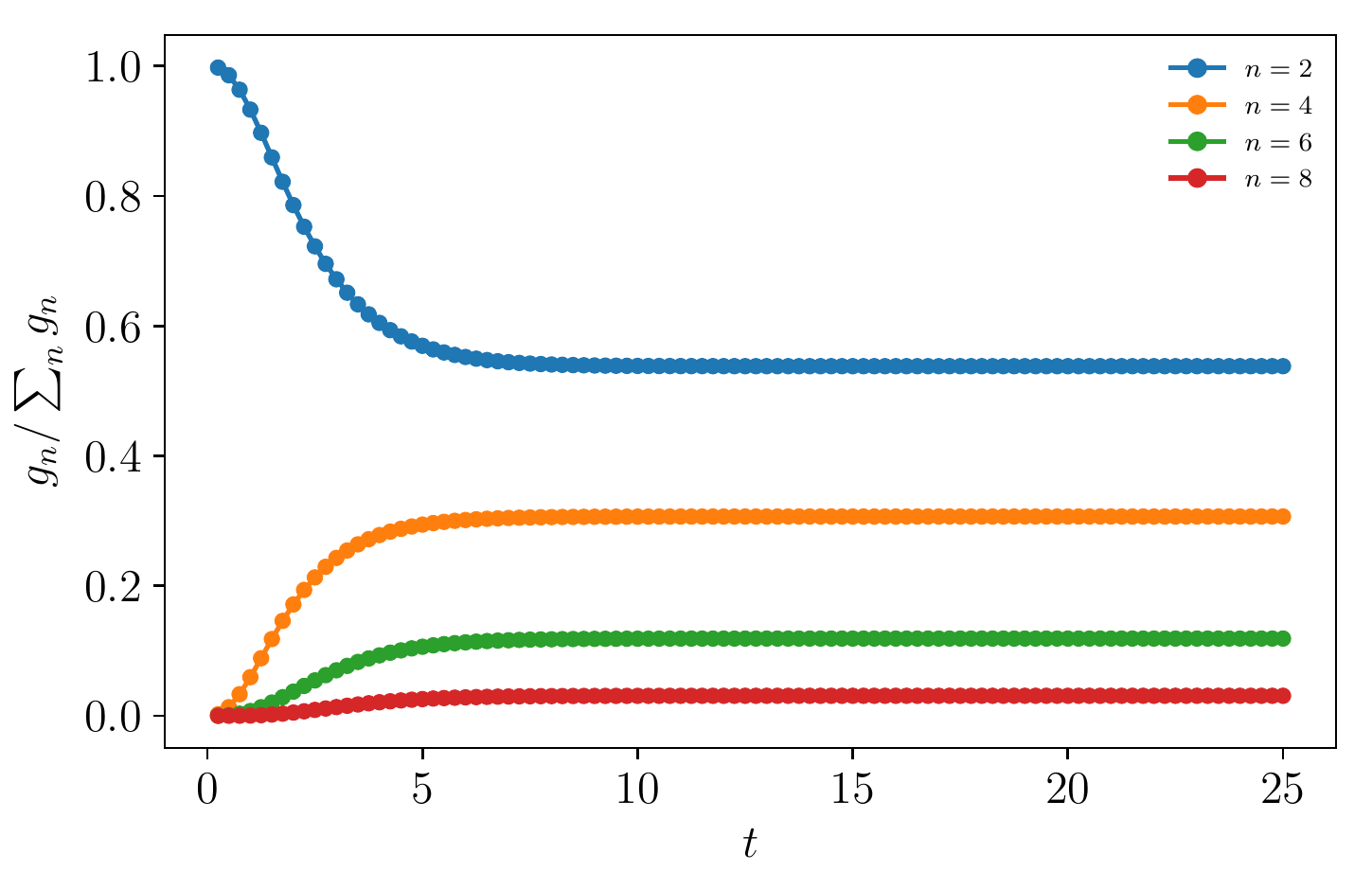}
}
\caption{Numerical results of the normalized multiple quantum coherence $g_n / \sum_n g_n$ for (a) $N = 6$ and (b) $N = 21$ spins.}
\label{fig:check_Kn}
\end{figure}

We simulate this process and reproduce the multiple quantum coherence for $N = 6$ and $21$ sites, see Fig.~\ref{fig:N6} and Fig.~\ref{fig:N21}.

 When we increase the number of sites to a few hundred, we observe that the OTOC grows exponentially in time (Fig.~\ref{fig:Kn}). Hence the $Kn$ space model, which ignores the spatial structure of the interactions, gives exponential growth of the OTOC.


\section{Computations of off-diagonal OTOCs}
\label{app:off_diag}

In this appendix, we provide further evidence that off-diagonal OTOCs are negligible in a variety of models. For example, for holographic CFTs one can extend the results of Ref.~\onlinecite{shenker_black_2014} by mapping off-diagonal OTOCs at non-zero energy density to certain two-sided correlations in a black hole spacetime where the operators are inserted at different spatial locations. Using, for example, a geodesic approximation to the correlator, one can then verify that off-diagonal OTOCs decay exponentially with the separation between operators. We can also study this question in a variety of lattice models using exact diagonalization and Krylov techniques. 

To illustrate the basic physics, we consider a spin model, studied at finite size using exact evolution of the many-body quantum state. The model is a long-range version of the well studied kicked Ising model. It is a Floquet model with a single period of time evolution generated by $U = U_I U_K$ with
\begin{equation}
    U_K = \exp\left( i b \sum_r \sigma_r^x\right)
\end{equation}
and
\begin{equation}
    U_I = \exp\left( i J \sum_{r,d} \frac{1}{d^\alpha} \sigma_r^z \sigma^z_{r+d} + i \sum_r h_r \sigma^z_r\right).
\end{equation}
The couplings $h_r$ are random and drawn from a Gaussian distribution with mean zero and standard deviation $h$.

We choose this model because in the local case it is a model of strong quantum chaos~\cite{Bertini_2018}. In particular, when $\alpha=\infty$ (local interactions) and $J = b = \pi/4$, the model is at the dual unitary point and exhibits a number of exact features characteristic of quantum chaos.

\begin{figure}[h]
\centering
\includegraphics[width=\columnwidth]{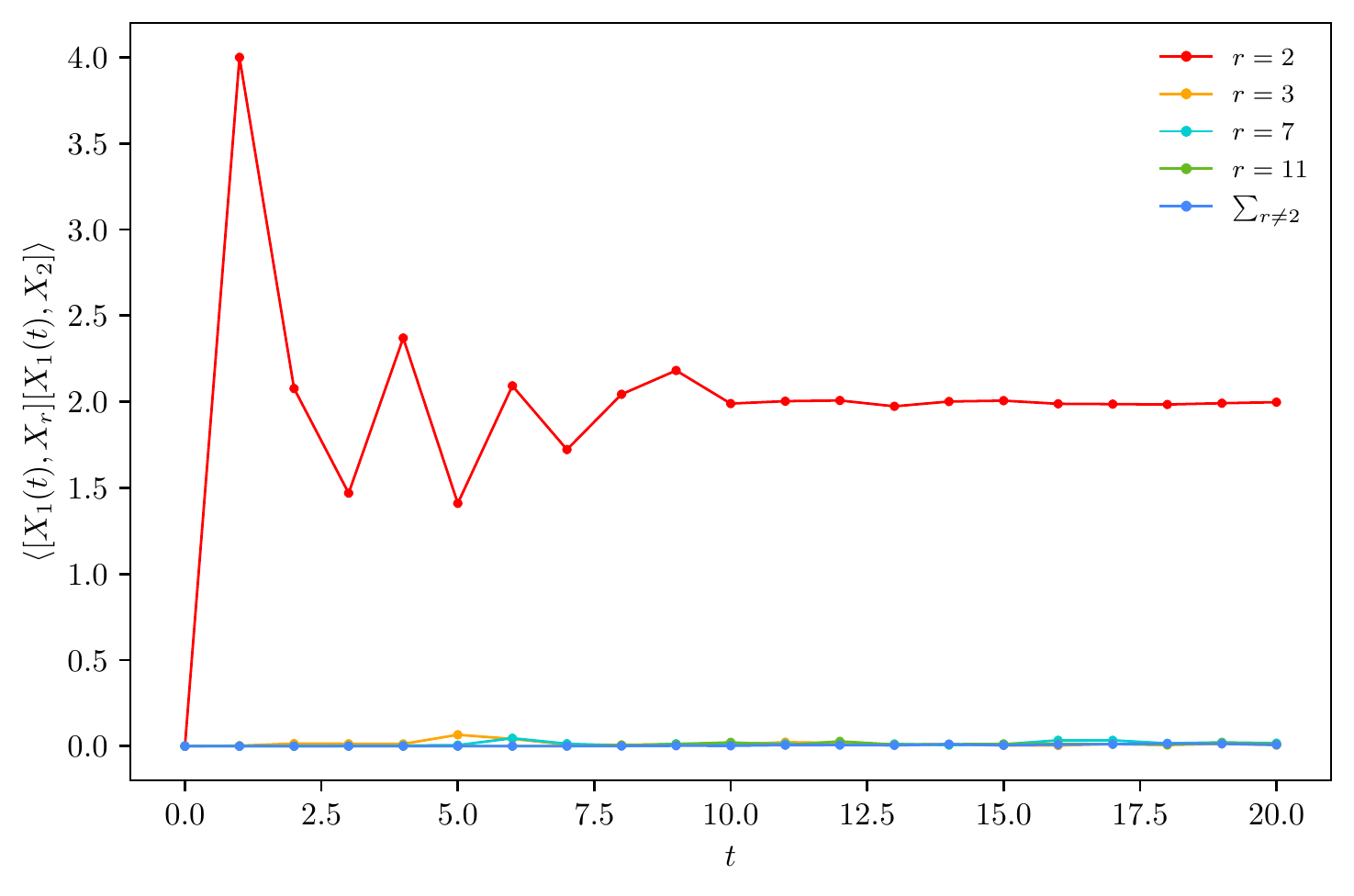}
\caption{Numerical computation of the diaognal and off-diaognal OTOCs of local operators in a 1d system of linear size $L = 14$. $X_i(t)$ is the local Pauli $X$ operator at site $i$ and time $t$ }
\label{fig:L14_num}
\end{figure}

Here we consider a long-range version of the model, still with $J=b=\pi/4$ and now with $\alpha<\infty$. As a simple diagnostic, we compute
\begin{equation}
    |\langle  [X_1(t), X_r(t) ][X_1(t), X_2(t) ] \rangle |,
\end{equation}
where the quantum average is taken over a random state in Hilbert space. This would reduce to a trace in the maximally mixed state if we also averaged over the choice of random state, but these data are for a single realization of the random state. The diagonal term corresponds to $r = 2$, which gives order $1$ value; the off-diagonal terms and their sum is two orders of magnitude smaller, see Fig.~\ref{fig:L14_num}. This indicates the OTOCs of global operators can be approximated by diagonal OTOCs of local operators, which is interpreted as the area under the local OTOC curve.

\bibliographystyle{apsrev4-1}
\bibliography{nmr_otoc_paper}

\begin{thebibliography}{71}%
\makeatletter
\providecommand \@ifxundefined [1]{%
 \@ifx{#1\undefined}
}%
\providecommand \@ifnum [1]{%
 \ifnum #1\expandafter \@firstoftwo
 \else \expandafter \@secondoftwo
 \fi
}%
\providecommand \@ifx [1]{%
 \ifx #1\expandafter \@firstoftwo
 \else \expandafter \@secondoftwo
 \fi
}%
\providecommand \natexlab [1]{#1}%
\providecommand \enquote  [1]{``#1''}%
\providecommand \bibnamefont  [1]{#1}%
\providecommand \bibfnamefont [1]{#1}%
\providecommand \citenamefont [1]{#1}%
\providecommand \href@noop [0]{\@secondoftwo}%
\providecommand \href [0]{\begingroup \@sanitize@url \@href}%
\providecommand \@href[1]{\@@startlink{#1}\@@href}%
\providecommand \@@href[1]{\endgroup#1\@@endlink}%
\providecommand \@sanitize@url [0]{\catcode `\\12\catcode `\$12\catcode
  `\&12\catcode `\#12\catcode `\^12\catcode `\_12\catcode `\%12\relax}%
\providecommand \@@startlink[1]{}%
\providecommand \@@endlink[0]{}%
\providecommand \url  [0]{\begingroup\@sanitize@url \@url }%
\providecommand \@url [1]{\endgroup\@href {#1}{\urlprefix }}%
\providecommand \urlprefix  [0]{URL }%
\providecommand \Eprint [0]{\href }%
\providecommand \doibase [0]{http://dx.doi.org/}%
\providecommand \selectlanguage [0]{\@gobble}%
\providecommand \bibinfo  [0]{\@secondoftwo}%
\providecommand \bibfield  [0]{\@secondoftwo}%
\providecommand \translation [1]{[#1]}%
\providecommand \BibitemOpen [0]{}%
\providecommand \bibitemStop [0]{}%
\providecommand \bibitemNoStop [0]{.\EOS\space}%
\providecommand \EOS [0]{\spacefactor3000\relax}%
\providecommand \BibitemShut  [1]{\csname bibitem#1\endcsname}%
\let\auto@bib@innerbib\@empty
\bibitem [{\citenamefont {Larkin}\ and\ \citenamefont
  {Ovchinnikov}(1969)}]{larkin_quasiclassical_1969}%
  \BibitemOpen
  \bibfield  {author} {\bibinfo {author} {\bibfnamefont {A.~I.}\ \bibnamefont
  {Larkin}}\ and\ \bibinfo {author} {\bibfnamefont {Y.~N.}\ \bibnamefont
  {Ovchinnikov}},\ }\bibfield  {title} {\bibinfo {title} {Quasiclassical
  {{Method}} in the {{Theory}} of {{Superconductivity}}},\ }\href@noop {}
  {\bibfield  {journal} {\bibinfo  {journal} {Soviet J. Exp. Theor. Phys.}\
  }\textbf {\bibinfo {volume} {28}},\ \bibinfo {pages} {1200} (\bibinfo {year}
  {1969})}\BibitemShut {NoStop}%
\bibitem [{\citenamefont {Shenker}\ and\ \citenamefont
  {Stanford}(2014)}]{shenker_black_2014}%
  \BibitemOpen
  \bibfield  {author} {\bibinfo {author} {\bibfnamefont {S.~H.}\ \bibnamefont
  {Shenker}}\ and\ \bibinfo {author} {\bibfnamefont {D.}~\bibnamefont
  {Stanford}},\ }\bibfield  {title} {{\selectlanguage {en}\bibinfo {title}
  {Black holes and the butterfly effect},\ }}\href {\doibase
  10.1007/jhep03(2014)067} {\bibfield  {journal} {\bibinfo  {journal} {J. High
  Energ. Phys.}\ }\textbf {\bibinfo {volume} {2014}},\ \bibinfo {pages} {67}
  (\bibinfo {year} {2014})}\BibitemShut {NoStop}%
\bibitem [{\citenamefont {Kitaev}(2002)}]{kitaev2015}%
  \BibitemOpen
  \bibfield  {author} {\bibinfo {author} {\bibfnamefont {A.}~\bibnamefont
  {Kitaev}},\ }\href {\doibase 10.1090/psapm/058/1922902} {\bibinfo {title}
  {Topological quantum codes and anyons},\ } (\bibinfo {year} {2002}),\
  \bibinfo {note} {talks at KITP, April 7, 2015 and May 27, 2015}\BibitemShut
  {NoStop}%
\bibitem [{\citenamefont {Maldacena}\ \emph {et~al.}(2016)\citenamefont
  {Maldacena}, \citenamefont {Shenker},\ and\ \citenamefont
  {Stanford}}]{maldacena_bound_2016}%
  \BibitemOpen
  \bibfield  {author} {\bibinfo {author} {\bibfnamefont {J.}~\bibnamefont
  {Maldacena}}, \bibinfo {author} {\bibfnamefont {S.~H.}\ \bibnamefont
  {Shenker}}, \ and\ \bibinfo {author} {\bibfnamefont {D.}~\bibnamefont
  {Stanford}},\ }\bibfield  {title} {{\selectlanguage {en}\bibinfo {title} {A
  bound on chaos},\ }}\href {\doibase 10.1007/jhep08(2016)106} {\bibfield
  {journal} {\bibinfo  {journal} {J. High Energ. Phys.}\ }\textbf {\bibinfo
  {volume} {2016}},\ \bibinfo {pages} {106} (\bibinfo {year}
  {2016})}\BibitemShut {NoStop}%
\bibitem [{\citenamefont {Shenker}\ and\ \citenamefont
  {Stanford}(2015)}]{shenker_stringy_2014}%
  \BibitemOpen
  \bibfield  {author} {\bibinfo {author} {\bibfnamefont {S.~H.}\ \bibnamefont
  {Shenker}}\ and\ \bibinfo {author} {\bibfnamefont {D.}~\bibnamefont
  {Stanford}},\ }\bibfield  {title} {\bibinfo {title} {Stringy effects in
  scrambling},\ }\href {\doibase 10.1007/jhep05(2015)132} {\bibfield  {journal}
  {\bibinfo  {journal} {J. High Energ. Phys.}\ }\textbf {\bibinfo {volume}
  {2015}} (\bibinfo {year} {2015}),\ 10.1007/jhep05(2015)132}\BibitemShut
  {NoStop}%
\bibitem [{\citenamefont {Nahum}\ \emph {et~al.}(2018)\citenamefont {Nahum},
  \citenamefont {Vijay},\ and\ \citenamefont {Haah}}]{nahum_operator_2018}%
  \BibitemOpen
  \bibfield  {author} {\bibinfo {author} {\bibfnamefont {A.}~\bibnamefont
  {Nahum}}, \bibinfo {author} {\bibfnamefont {S.}~\bibnamefont {Vijay}}, \ and\
  \bibinfo {author} {\bibfnamefont {J.}~\bibnamefont {Haah}},\ }\bibfield
  {title} {\bibinfo {title} {Operator spreading in random unitary circuits},\
  }\href {\doibase 10.1103/physrevx.8.021014} {\bibfield  {journal} {\bibinfo
  {journal} {Phys. Rev. X}\ }\textbf {\bibinfo {volume} {8}},\ \bibinfo {pages}
  {021014} (\bibinfo {year} {2018})}\BibitemShut {NoStop}%
\bibitem [{\citenamefont {von Keyserlingk}\ \emph {et~al.}(2018)\citenamefont
  {von Keyserlingk}, \citenamefont {Rakovszky}, \citenamefont {Pollmann},\ and\
  \citenamefont {Sondhi}}]{von_keyserlingk_operator_2018}%
  \BibitemOpen
  \bibfield  {author} {\bibinfo {author} {\bibfnamefont {C.~W.}\ \bibnamefont
  {von Keyserlingk}}, \bibinfo {author} {\bibfnamefont {T.}~\bibnamefont
  {Rakovszky}}, \bibinfo {author} {\bibfnamefont {F.}~\bibnamefont {Pollmann}},
  \ and\ \bibinfo {author} {\bibfnamefont {S.~L.}\ \bibnamefont {Sondhi}},\
  }\bibfield  {title} {\bibinfo {title} {Operator hydrodynamics, {OTOCs,} and
  entanglement growth in systems without conservation laws},\ }\href {\doibase
  10.1103/physrevx.8.021013} {\bibfield  {journal} {\bibinfo  {journal} {Phys.
  Rev. X}\ }\textbf {\bibinfo {volume} {8}},\ \bibinfo {pages} {021013}
  (\bibinfo {year} {2018})}\BibitemShut {NoStop}%
\bibitem [{\citenamefont {Aleiner}\ \emph {et~al.}(2016)\citenamefont
  {Aleiner}, \citenamefont {Faoro},\ and\ \citenamefont
  {Ioffe}}]{aleiner_microscopic_2016}%
  \BibitemOpen
  \bibfield  {author} {\bibinfo {author} {\bibfnamefont {I.~L.}\ \bibnamefont
  {Aleiner}}, \bibinfo {author} {\bibfnamefont {L.}~\bibnamefont {Faoro}}, \
  and\ \bibinfo {author} {\bibfnamefont {L.~B.}\ \bibnamefont {Ioffe}},\
  }\bibfield  {title} {\bibinfo {title} {Microscopic model of quantum butterfly
  effect: {Out-of-time-order} correlators and traveling combustion waves},\
  }\href {\doibase 10.1016/j.aop.2016.09.006} {\bibfield  {journal} {\bibinfo
  {journal} {Ann. Phys-new. York.}\ }\textbf {\bibinfo {volume} {375}},\
  \bibinfo {pages} {378} (\bibinfo {year} {2016})}\BibitemShut {NoStop}%
\bibitem [{\citenamefont {Xu}\ and\ \citenamefont
  {Swingle}(2019)}]{xu_locality_2018}%
  \BibitemOpen
  \bibfield  {author} {\bibinfo {author} {\bibfnamefont {S.}~\bibnamefont
  {Xu}}\ and\ \bibinfo {author} {\bibfnamefont {B.}~\bibnamefont {Swingle}},\
  }\bibfield  {title} {\bibinfo {title} {Locality, {{Quantum Fluctuations}},
  and {{Scrambling}}},\ }\href {\doibase 10.1103/PhysRevX.9.031048} {\bibfield
  {journal} {\bibinfo  {journal} {Phys. Rev. X}\ }\textbf {\bibinfo {volume}
  {9}},\ \bibinfo {pages} {031048} (\bibinfo {year} {2019})}\BibitemShut
  {NoStop}%
\bibitem [{\citenamefont {Davison}\ \emph {et~al.}(2017)\citenamefont
  {Davison}, \citenamefont {Fu}, \citenamefont {Georges}, \citenamefont {Gu},
  \citenamefont {Jensen},\ and\ \citenamefont
  {Sachdev}}]{davison_thermoelectric_2017}%
  \BibitemOpen
  \bibfield  {author} {\bibinfo {author} {\bibfnamefont {R.~A.}\ \bibnamefont
  {Davison}}, \bibinfo {author} {\bibfnamefont {W.}~\bibnamefont {Fu}},
  \bibinfo {author} {\bibfnamefont {A.}~\bibnamefont {Georges}}, \bibinfo
  {author} {\bibfnamefont {Y.}~\bibnamefont {Gu}}, \bibinfo {author}
  {\bibfnamefont {K.}~\bibnamefont {Jensen}}, \ and\ \bibinfo {author}
  {\bibfnamefont {S.}~\bibnamefont {Sachdev}},\ }\bibfield  {title} {\bibinfo
  {title} {Thermoelectric transport in disordered metals without
  quasiparticles: {The} sachdev-ye-kitaev models and holography},\ }\href
  {\doibase 10.1103/physrevb.95.155131} {\bibfield  {journal} {\bibinfo
  {journal} {Phys. Rev. B}\ }\textbf {\bibinfo {volume} {95}},\ \bibinfo
  {pages} {155131} (\bibinfo {year} {2017})}\BibitemShut {NoStop}%
\bibitem [{\citenamefont {Gu}\ \emph {et~al.}(2017)\citenamefont {Gu},
  \citenamefont {Lucas},\ and\ \citenamefont {Qi}}]{gu_energy_2017}%
  \BibitemOpen
  \bibfield  {author} {\bibinfo {author} {\bibfnamefont {Y.}~\bibnamefont
  {Gu}}, \bibinfo {author} {\bibfnamefont {A.}~\bibnamefont {Lucas}}, \ and\
  \bibinfo {author} {\bibfnamefont {X.-L.}\ \bibnamefont {Qi}},\ }\bibfield
  {title} {{\selectlanguage {en}\bibinfo {title} {Energy diffusion and the
  butterfly effect in inhomogeneous sachdev-ye-kitaev chains},\ }}\href
  {\doibase 10.21468/scipostphys.2.3.018} {\bibfield  {journal} {\bibinfo
  {journal} {SciPost Phys.}\ }\textbf {\bibinfo {volume} {2}},\ \bibinfo
  {pages} {018} (\bibinfo {year} {2017})}\BibitemShut {NoStop}%
\bibitem [{\citenamefont {Liao}\ and\ \citenamefont
  {Galitski}(2018)}]{liao_nonlinear_2018}%
  \BibitemOpen
  \bibfield  {author} {\bibinfo {author} {\bibfnamefont {Y.}~\bibnamefont
  {Liao}}\ and\ \bibinfo {author} {\bibfnamefont {V.}~\bibnamefont
  {Galitski}},\ }\bibfield  {title} {\bibinfo {title} {Nonlinear sigma model
  approach to many-body quantum chaos: {Regularized} and unregularized
  out-of-time-ordered correlators},\ }\href {\doibase
  10.1103/physrevb.98.205124} {\bibfield  {journal} {\bibinfo  {journal} {Phys.
  Rev. B}\ }\textbf {\bibinfo {volume} {98}},\ \bibinfo {pages} {205124}
  (\bibinfo {year} {2018})}\BibitemShut {NoStop}%
\bibitem [{\citenamefont {Zhou}\ and\ \citenamefont
  {Chen}(2019)}]{zhou_operator_2019}%
  \BibitemOpen
  \bibfield  {author} {\bibinfo {author} {\bibfnamefont {T.}~\bibnamefont
  {Zhou}}\ and\ \bibinfo {author} {\bibfnamefont {X.}~\bibnamefont {Chen}},\
  }\bibfield  {title} {\bibinfo {title} {Operator dynamics in a {Brownian}
  quantum circuit},\ }\href {\doibase 10.1103/physreve.99.052212} {\bibfield
  {journal} {\bibinfo  {journal} {Phys. Rev. E}\ }\textbf {\bibinfo {volume}
  {99}},\ \bibinfo {pages} {052212} (\bibinfo {year} {2019})}\BibitemShut
  {NoStop}%
\bibitem [{\citenamefont {Islam}\ \emph {et~al.}(2013)\citenamefont {Islam},
  \citenamefont {Senko}, \citenamefont {Campbell}, \citenamefont {Korenblit},
  \citenamefont {Smith}, \citenamefont {Lee}, \citenamefont {Edwards},
  \citenamefont {Wang}, \citenamefont {Freericks},\ and\ \citenamefont
  {Monroe}}]{Monroe13}%
  \BibitemOpen
  \bibfield  {author} {\bibinfo {author} {\bibfnamefont {R.}~\bibnamefont
  {Islam}}, \bibinfo {author} {\bibfnamefont {C.}~\bibnamefont {Senko}},
  \bibinfo {author} {\bibfnamefont {W.~C.}\ \bibnamefont {Campbell}}, \bibinfo
  {author} {\bibfnamefont {S.}~\bibnamefont {Korenblit}}, \bibinfo {author}
  {\bibfnamefont {J.}~\bibnamefont {Smith}}, \bibinfo {author} {\bibfnamefont
  {A.}~\bibnamefont {Lee}}, \bibinfo {author} {\bibfnamefont {E.~E.}\
  \bibnamefont {Edwards}}, \bibinfo {author} {\bibfnamefont {C.-C.~J.}\
  \bibnamefont {Wang}}, \bibinfo {author} {\bibfnamefont {J.~K.}\ \bibnamefont
  {Freericks}}, \ and\ \bibinfo {author} {\bibfnamefont {C.}~\bibnamefont
  {Monroe}},\ }\bibfield  {title} {\bibinfo {title} {Emergence and frustration
  of magnetism with variable-range interactions in a quantum simulator},\
  }\href {\doibase 10.1126/science.1232296} {\bibfield  {journal} {\bibinfo
  {journal} {Science}\ }\textbf {\bibinfo {volume} {340}},\ \bibinfo {pages}
  {583} (\bibinfo {year} {2013})}\BibitemShut {NoStop}%
\bibitem [{\citenamefont {Yan}\ \emph {et~al.}(2013{\natexlab{a}})\citenamefont
  {Yan}, \citenamefont {Moses}, \citenamefont {Gadway}, \citenamefont {Covey},
  \citenamefont {Hazzard}, \citenamefont {Rey}, \citenamefont {Jin},\ and\
  \citenamefont {Ye}}]{yan_observation_2013}%
  \BibitemOpen
  \bibfield  {author} {\bibinfo {author} {\bibfnamefont {B.}~\bibnamefont
  {Yan}}, \bibinfo {author} {\bibfnamefont {S.~A.}\ \bibnamefont {Moses}},
  \bibinfo {author} {\bibfnamefont {B.}~\bibnamefont {Gadway}}, \bibinfo
  {author} {\bibfnamefont {J.~P.}\ \bibnamefont {Covey}}, \bibinfo {author}
  {\bibfnamefont {K.~R.~A.}\ \bibnamefont {Hazzard}}, \bibinfo {author}
  {\bibfnamefont {A.~M.}\ \bibnamefont {Rey}}, \bibinfo {author} {\bibfnamefont
  {D.~S.}\ \bibnamefont {Jin}}, \ and\ \bibinfo {author} {\bibfnamefont
  {J.}~\bibnamefont {Ye}},\ }\bibfield  {title} {{\selectlanguage {en}\bibinfo
  {title} {Observation of dipolar spin-exchange interactions with
  lattice-confined polar molecules},\ }}\href {\doibase 10.1038/nature12483}
  {\bibfield  {journal} {\bibinfo  {journal} {Nature}\ }\textbf {\bibinfo
  {volume} {501}},\ \bibinfo {pages} {521} (\bibinfo {year}
  {2013}{\natexlab{a}})}\BibitemShut {NoStop}%
\bibitem [{\citenamefont {Britton}\ \emph {et~al.}(2012)\citenamefont
  {Britton}, \citenamefont {Sawyer}, \citenamefont {Keith}, \citenamefont
  {Wang}, \citenamefont {Freericks}, \citenamefont {Uys}, \citenamefont
  {Biercuk},\ and\ \citenamefont {Bollinger}}]{Britton12}%
  \BibitemOpen
  \bibfield  {author} {\bibinfo {author} {\bibfnamefont {J.~W.}\ \bibnamefont
  {Britton}}, \bibinfo {author} {\bibfnamefont {B.~C.}\ \bibnamefont {Sawyer}},
  \bibinfo {author} {\bibfnamefont {A.~C.}\ \bibnamefont {Keith}}, \bibinfo
  {author} {\bibfnamefont {C.-C.~J.}\ \bibnamefont {Wang}}, \bibinfo {author}
  {\bibfnamefont {J.~K.}\ \bibnamefont {Freericks}}, \bibinfo {author}
  {\bibfnamefont {H.}~\bibnamefont {Uys}}, \bibinfo {author} {\bibfnamefont
  {M.~J.}\ \bibnamefont {Biercuk}}, \ and\ \bibinfo {author} {\bibfnamefont
  {J.~J.}\ \bibnamefont {Bollinger}},\ }\bibfield  {title} {\bibinfo {title}
  {Engineered two-dimensional ising interactions in a trapped-ion quantum
  simulator with hundreds of spins},\ }\href {\doibase 10.1038/nature10981}
  {\bibfield  {journal} {\bibinfo  {journal} {Nature}\ }\textbf {\bibinfo
  {volume} {484}},\ \bibinfo {pages} {489} (\bibinfo {year}
  {2012})}\BibitemShut {NoStop}%
\bibitem [{\citenamefont {Blatt}\ and\ \citenamefont {Roos}(2012)}]{Blatt12}%
  \BibitemOpen
  \bibfield  {author} {\bibinfo {author} {\bibfnamefont {R.}~\bibnamefont
  {Blatt}}\ and\ \bibinfo {author} {\bibfnamefont {C.~F.}\ \bibnamefont
  {Roos}},\ }\bibfield  {title} {\bibinfo {title} {Quantum simulations with
  trapped ions},\ }\href {\doibase 10.1038/nphys2252} {\bibfield  {journal}
  {\bibinfo  {journal} {Nature Phys}\ }\textbf {\bibinfo {volume} {8}},\
  \bibinfo {pages} {277} (\bibinfo {year} {2012})}\BibitemShut {NoStop}%
\bibitem [{\citenamefont {Choi}\ \emph {et~al.}(2017)\citenamefont {Choi},
  \citenamefont {Choi}, \citenamefont {Landig}, \citenamefont {Kucsko},
  \citenamefont {Zhou}, \citenamefont {Isoya}, \citenamefont {Jelezko},
  \citenamefont {Onoda}, \citenamefont {Sumiya}, \citenamefont {Khemani},
  \citenamefont {von Keyserlingk}, \citenamefont {Yao}, \citenamefont
  {Demler},\ and\ \citenamefont {Lukin}}]{Lukin17}%
  \BibitemOpen
  \bibfield  {author} {\bibinfo {author} {\bibfnamefont {S.}~\bibnamefont
  {Choi}}, \bibinfo {author} {\bibfnamefont {J.}~\bibnamefont {Choi}}, \bibinfo
  {author} {\bibfnamefont {R.}~\bibnamefont {Landig}}, \bibinfo {author}
  {\bibfnamefont {G.}~\bibnamefont {Kucsko}}, \bibinfo {author} {\bibfnamefont
  {H.}~\bibnamefont {Zhou}}, \bibinfo {author} {\bibfnamefont {J.}~\bibnamefont
  {Isoya}}, \bibinfo {author} {\bibfnamefont {F.}~\bibnamefont {Jelezko}},
  \bibinfo {author} {\bibfnamefont {S.}~\bibnamefont {Onoda}}, \bibinfo
  {author} {\bibfnamefont {H.}~\bibnamefont {Sumiya}}, \bibinfo {author}
  {\bibfnamefont {V.}~\bibnamefont {Khemani}}, \bibinfo {author} {\bibfnamefont
  {C.}~\bibnamefont {von Keyserlingk}}, \bibinfo {author} {\bibfnamefont
  {N.~Y.}\ \bibnamefont {Yao}}, \bibinfo {author} {\bibfnamefont
  {E.}~\bibnamefont {Demler}}, \ and\ \bibinfo {author} {\bibfnamefont {M.~D.}\
  \bibnamefont {Lukin}},\ }\bibfield  {title} {\bibinfo {title} {Observation of
  discrete time-crystalline order in a disordered dipolar many-body system},\
  }\href {\doibase 10.1038/nature21426} {\bibfield  {journal} {\bibinfo
  {journal} {Nature}\ }\textbf {\bibinfo {volume} {543}},\ \bibinfo {pages}
  {221} (\bibinfo {year} {2017})}\BibitemShut {NoStop}%
\bibitem [{\citenamefont {Gärttner}\ \emph {et~al.}(2017)\citenamefont
  {Gärttner}, \citenamefont {Bohnet}, \citenamefont {Safavi-Naini},
  \citenamefont {Wall}, \citenamefont {Bollinger},\ and\ \citenamefont
  {Rey}}]{Bollinger17}%
  \BibitemOpen
  \bibfield  {author} {\bibinfo {author} {\bibfnamefont {M.}~\bibnamefont
  {Gärttner}}, \bibinfo {author} {\bibfnamefont {J.~G.}\ \bibnamefont
  {Bohnet}}, \bibinfo {author} {\bibfnamefont {A.}~\bibnamefont
  {Safavi-Naini}}, \bibinfo {author} {\bibfnamefont {M.~L.}\ \bibnamefont
  {Wall}}, \bibinfo {author} {\bibfnamefont {J.~J.}\ \bibnamefont {Bollinger}},
  \ and\ \bibinfo {author} {\bibfnamefont {A.~M.}\ \bibnamefont {Rey}},\
  }\bibfield  {title} {\bibinfo {title} {Measuring out-of-time-order
  correlations and multiple quantum spectra in a trapped-ion quantum magnet},\
  }\href {\doibase 10.1038/nphys4119} {\bibfield  {journal} {\bibinfo
  {journal} {Nature Phys}\ }\textbf {\bibinfo {volume} {13}},\ \bibinfo {pages}
  {781} (\bibinfo {year} {2017})}\BibitemShut {NoStop}%
\bibitem [{\citenamefont {Zhang}\ \emph {et~al.}(2009)\citenamefont {Zhang},
  \citenamefont {Cappellaro}, \citenamefont {Antler}, \citenamefont {Pepper},
  \citenamefont {Cory}, \citenamefont {Dobrovitski}, \citenamefont
  {Ramanathan},\ and\ \citenamefont {Viola}}]{PhysRevA.80.052323}%
  \BibitemOpen
  \bibfield  {author} {\bibinfo {author} {\bibfnamefont {W.}~\bibnamefont
  {Zhang}}, \bibinfo {author} {\bibfnamefont {P.}~\bibnamefont {Cappellaro}},
  \bibinfo {author} {\bibfnamefont {N.}~\bibnamefont {Antler}}, \bibinfo
  {author} {\bibfnamefont {B.}~\bibnamefont {Pepper}}, \bibinfo {author}
  {\bibfnamefont {D.~G.}\ \bibnamefont {Cory}}, \bibinfo {author}
  {\bibfnamefont {V.~V.}\ \bibnamefont {Dobrovitski}}, \bibinfo {author}
  {\bibfnamefont {C.}~\bibnamefont {Ramanathan}}, \ and\ \bibinfo {author}
  {\bibfnamefont {L.}~\bibnamefont {Viola}},\ }\bibfield  {title} {\bibinfo
  {title} {{NMR} multiple quantum coherences in quasi-one-dimensional spin
  systems: {Comparison} with ideal spin-chain dynamics},\ }\href {\doibase
  10.1103/physreva.80.052323} {\bibfield  {journal} {\bibinfo  {journal} {Phys.
  Rev. A}\ }\textbf {\bibinfo {volume} {80}},\ \bibinfo {pages} {052323}
  (\bibinfo {year} {2009})}\BibitemShut {NoStop}%
\bibitem [{\citenamefont {Wei}\ \emph {et~al.}(2018)\citenamefont {Wei},
  \citenamefont {Ramanathan},\ and\ \citenamefont
  {Cappellaro}}]{wei_exploring_2018}%
  \BibitemOpen
  \bibfield  {author} {\bibinfo {author} {\bibfnamefont {K.~X.}\ \bibnamefont
  {Wei}}, \bibinfo {author} {\bibfnamefont {C.}~\bibnamefont {Ramanathan}}, \
  and\ \bibinfo {author} {\bibfnamefont {P.}~\bibnamefont {Cappellaro}},\
  }\bibfield  {title} {\bibinfo {title} {Exploring localization in nuclear spin
  chains},\ }\href {\doibase 10.1103/physrevlett.120.070501} {\bibfield
  {journal} {\bibinfo  {journal} {Phys. Rev. Lett.}\ }\textbf {\bibinfo
  {volume} {120}},\ \bibinfo {pages} {070501} (\bibinfo {year}
  {2018})}\BibitemShut {NoStop}%
\bibitem [{\citenamefont {Wei}\ \emph {et~al.}(2019)\citenamefont {Wei},
  \citenamefont {Peng}, \citenamefont {Shtanko}, \citenamefont {Marvian},
  \citenamefont {Lloyd}, \citenamefont {Ramanathan},\ and\ \citenamefont
  {Cappellaro}}]{wei_emergent_2019}%
  \BibitemOpen
  \bibfield  {author} {\bibinfo {author} {\bibfnamefont {K.~X.}\ \bibnamefont
  {Wei}}, \bibinfo {author} {\bibfnamefont {P.}~\bibnamefont {Peng}}, \bibinfo
  {author} {\bibfnamefont {O.}~\bibnamefont {Shtanko}}, \bibinfo {author}
  {\bibfnamefont {I.}~\bibnamefont {Marvian}}, \bibinfo {author} {\bibfnamefont
  {S.}~\bibnamefont {Lloyd}}, \bibinfo {author} {\bibfnamefont
  {C.}~\bibnamefont {Ramanathan}}, \ and\ \bibinfo {author} {\bibfnamefont
  {P.}~\bibnamefont {Cappellaro}},\ }\bibfield  {title} {\bibinfo {title}
  {Emergent prethermalization signatures in out-of-time ordered correlations},\
  }\href {\doibase 10.1103/physrevlett.123.090605} {\bibfield  {journal}
  {\bibinfo  {journal} {Phys. Rev. Lett.}\ }\textbf {\bibinfo {volume} {123}},\
  \bibinfo {pages} {090605} (\bibinfo {year} {2019})}\BibitemShut {NoStop}%
\bibitem [{\citenamefont {Sánchez}\ \emph
  {et~al.}(2014{\natexlab{a}})\citenamefont {Sánchez}, \citenamefont {Acosta},
  \citenamefont {Levstein}, \citenamefont {Pastawski},\ and\ \citenamefont
  {Chattah}}]{sanchez_clustering_2014}%
  \BibitemOpen
  \bibfield  {author} {\bibinfo {author} {\bibfnamefont {C.~M.}\ \bibnamefont
  {Sánchez}}, \bibinfo {author} {\bibfnamefont {R.~H.}\ \bibnamefont
  {Acosta}}, \bibinfo {author} {\bibfnamefont {P.~R.}\ \bibnamefont
  {Levstein}}, \bibinfo {author} {\bibfnamefont {H.~M.}\ \bibnamefont
  {Pastawski}}, \ and\ \bibinfo {author} {\bibfnamefont {A.~K.}\ \bibnamefont
  {Chattah}},\ }\bibfield  {title} {\bibinfo {title} {Clustering and
  decoherence of correlated spins under double quantum dynamics},\ }\href
  {\doibase 10.1103/physreva.90.042122} {\bibfield  {journal} {\bibinfo
  {journal} {Phys. Rev. A}\ }\textbf {\bibinfo {volume} {90}},\ \bibinfo
  {pages} {042122} (\bibinfo {year} {2014}{\natexlab{a}})}\BibitemShut
  {NoStop}%
\bibitem [{\citenamefont {Álvarez}\ \emph {et~al.}(2015)\citenamefont
  {Álvarez}, \citenamefont {Suter},\ and\ \citenamefont
  {Kaiser}}]{alvarez_localization-delocalization_2015}%
  \BibitemOpen
  \bibfield  {author} {\bibinfo {author} {\bibfnamefont {G.~A.}\ \bibnamefont
  {Álvarez}}, \bibinfo {author} {\bibfnamefont {D.}~\bibnamefont {Suter}}, \
  and\ \bibinfo {author} {\bibfnamefont {R.}~\bibnamefont {Kaiser}},\
  }\bibfield  {title} {{\selectlanguage {en}\bibinfo {title}
  {Localization-delocalization transition in the dynamics of dipolar-coupled
  nuclear spins},\ }}\href {\doibase 10.1126/science.1261160} {\bibfield
  {journal} {\bibinfo  {journal} {Science}\ }\textbf {\bibinfo {volume}
  {349}},\ \bibinfo {pages} {846} (\bibinfo {year} {2015})}\BibitemShut
  {NoStop}%
\bibitem [{\citenamefont {Sánchez}\ \emph {et~al.}(2020)\citenamefont
  {Sánchez}, \citenamefont {Chattah}, \citenamefont {Wei}, \citenamefont
  {Buljubasich}, \citenamefont {Cappellaro},\ and\ \citenamefont
  {Pastawski}}]{sanchez_perturbation_2020}%
  \BibitemOpen
  \bibfield  {author} {\bibinfo {author} {\bibfnamefont {C.~M.}\ \bibnamefont
  {Sánchez}}, \bibinfo {author} {\bibfnamefont {A.~K.}\ \bibnamefont
  {Chattah}}, \bibinfo {author} {\bibfnamefont {K.~X.}\ \bibnamefont {Wei}},
  \bibinfo {author} {\bibfnamefont {L.}~\bibnamefont {Buljubasich}}, \bibinfo
  {author} {\bibfnamefont {P.}~\bibnamefont {Cappellaro}}, \ and\ \bibinfo
  {author} {\bibfnamefont {H.~M.}\ \bibnamefont {Pastawski}},\ }\bibfield
  {title} {\bibinfo {title} {Perturbation independent decay of the loschmidt
  echo in a many-body system},\ }\href {\doibase
  10.1103/physrevlett.124.030601} {\bibfield  {journal} {\bibinfo  {journal}
  {Phys. Rev. Lett.}\ }\textbf {\bibinfo {volume} {124}},\ \bibinfo {pages}
  {030601} (\bibinfo {year} {2020})}\BibitemShut {NoStop}%
\bibitem [{\citenamefont {Zhou}\ \emph {et~al.}(2020)\citenamefont {Zhou},
  \citenamefont {Xu}, \citenamefont {Chen}, \citenamefont {Guo},\ and\
  \citenamefont {Swingle}}]{zhou_operator_2020}%
  \BibitemOpen
  \bibfield  {author} {\bibinfo {author} {\bibfnamefont {T.}~\bibnamefont
  {Zhou}}, \bibinfo {author} {\bibfnamefont {S.}~\bibnamefont {Xu}}, \bibinfo
  {author} {\bibfnamefont {X.}~\bibnamefont {Chen}}, \bibinfo {author}
  {\bibfnamefont {A.}~\bibnamefont {Guo}}, \ and\ \bibinfo {author}
  {\bibfnamefont {B.}~\bibnamefont {Swingle}},\ }\bibfield  {title} {\bibinfo
  {title} {Operator l\'evy flight: {Light} cones in chaotic long-range
  interacting systems},\ }\href {\doibase 10.1103/physrevlett.124.180601}
  {\bibfield  {journal} {\bibinfo  {journal} {Phys. Rev. Lett.}\ }\textbf
  {\bibinfo {volume} {124}},\ \bibinfo {pages} {180601} (\bibinfo {year}
  {2020})}\BibitemShut {NoStop}%
\bibitem [{\citenamefont {Chen}\ and\ \citenamefont
  {Zhou}(2019)}]{chen_quantum_2019}%
  \BibitemOpen
  \bibfield  {author} {\bibinfo {author} {\bibfnamefont {X.}~\bibnamefont
  {Chen}}\ and\ \bibinfo {author} {\bibfnamefont {T.}~\bibnamefont {Zhou}},\
  }\bibfield  {title} {\bibinfo {title} {Quantum chaos dynamics in long-range
  power law interaction systems},\ }\href {\doibase
  10.1103/physrevb.100.064305} {\bibfield  {journal} {\bibinfo  {journal}
  {Phys. Rev. B}\ }\textbf {\bibinfo {volume} {100}},\ \bibinfo {pages}
  {064305} (\bibinfo {year} {2019})}\BibitemShut {NoStop}%
\bibitem [{\citenamefont {Mezei}(2017)}]{mezei_entanglement_2016}%
  \BibitemOpen
  \bibfield  {author} {\bibinfo {author} {\bibfnamefont {M.}~\bibnamefont
  {Mezei}},\ }\bibfield  {title} {\bibinfo {title} {On entanglement spreading
  from holography},\ }\href {\doibase 10.1007/jhep05(2017)064} {\bibfield
  {journal} {\bibinfo  {journal} {J. High Energ. Phys.}\ }\textbf {\bibinfo
  {volume} {2017}} (\bibinfo {year} {2017}),\
  10.1007/jhep05(2017)064}\BibitemShut {NoStop}%
\bibitem [{\citenamefont {Qi}\ and\ \citenamefont
  {Streicher}(2019)}]{qi_quantum_2018}%
  \BibitemOpen
  \bibfield  {author} {\bibinfo {author} {\bibfnamefont {X.-L.}\ \bibnamefont
  {Qi}}\ and\ \bibinfo {author} {\bibfnamefont {A.}~\bibnamefont {Streicher}},\
  }\bibfield  {title} {\bibinfo {title} {Quantum epidemiology: {Operator}
  growth, thermal effects, and {SYK}},\ }\href {\doibase
  10.1007/jhep08(2019)012} {\bibfield  {journal} {\bibinfo  {journal} {J. High
  Energ. Phys.}\ }\textbf {\bibinfo {volume} {2019}} (\bibinfo {year} {2019}),\
  10.1007/jhep08(2019)012}\BibitemShut {NoStop}%
\bibitem [{\citenamefont {Roberts}\ \emph {et~al.}(2018)\citenamefont
  {Roberts}, \citenamefont {Stanford},\ and\ \citenamefont
  {Streicher}}]{roberts_operator_2018}%
  \BibitemOpen
  \bibfield  {author} {\bibinfo {author} {\bibfnamefont {D.~A.}\ \bibnamefont
  {Roberts}}, \bibinfo {author} {\bibfnamefont {D.}~\bibnamefont {Stanford}}, \
  and\ \bibinfo {author} {\bibfnamefont {A.}~\bibnamefont {Streicher}},\
  }\bibfield  {title} {{\selectlanguage {en}\bibinfo {title} {Operator growth
  in the {SYK} model},\ }}\href {\doibase 10.1007/jhep06(2018)122} {\bibfield
  {journal} {\bibinfo  {journal} {J. High Energ. Phys.}\ }\textbf {\bibinfo
  {volume} {2018}},\ \bibinfo {pages} {122} (\bibinfo {year}
  {2018})}\BibitemShut {NoStop}%
\bibitem [{\citenamefont {Swingle}\ \emph {et~al.}(2016)\citenamefont
  {Swingle}, \citenamefont {Bentsen}, \citenamefont {Schleier-Smith},\ and\
  \citenamefont {Hayden}}]{swingle_measuring_2016}%
  \BibitemOpen
  \bibfield  {author} {\bibinfo {author} {\bibfnamefont {B.}~\bibnamefont
  {Swingle}}, \bibinfo {author} {\bibfnamefont {G.}~\bibnamefont {Bentsen}},
  \bibinfo {author} {\bibfnamefont {M.}~\bibnamefont {Schleier-Smith}}, \ and\
  \bibinfo {author} {\bibfnamefont {P.}~\bibnamefont {Hayden}},\ }\bibfield
  {title} {\bibinfo {title} {Measuring the scrambling of quantum information},\
  }\href {\doibase 10.1103/physreva.94.040302} {\bibfield  {journal} {\bibinfo
  {journal} {Phys. Rev. A}\ }\textbf {\bibinfo {volume} {94}} (\bibinfo {year}
  {2016}),\ 10.1103/physreva.94.040302}\BibitemShut {NoStop}%
\bibitem [{\citenamefont {Yao}\ \emph {et~al.}(2016)\citenamefont {Yao},
  \citenamefont {Grusdt}, \citenamefont {Swingle}, \citenamefont {Lukin},
  \citenamefont {{Stamper-Kurn}}, \citenamefont {Moore},\ and\ \citenamefont
  {Demler}}]{yao_interferometric_2016-1}%
  \BibitemOpen
  \bibfield  {author} {\bibinfo {author} {\bibfnamefont {N.~Y.}\ \bibnamefont
  {Yao}}, \bibinfo {author} {\bibfnamefont {F.}~\bibnamefont {Grusdt}},
  \bibinfo {author} {\bibfnamefont {B.}~\bibnamefont {Swingle}}, \bibinfo
  {author} {\bibfnamefont {M.~D.}\ \bibnamefont {Lukin}}, \bibinfo {author}
  {\bibfnamefont {D.~M.}\ \bibnamefont {{Stamper-Kurn}}}, \bibinfo {author}
  {\bibfnamefont {J.~E.}\ \bibnamefont {Moore}}, \ and\ \bibinfo {author}
  {\bibfnamefont {E.~A.}\ \bibnamefont {Demler}},\ }\bibfield  {title}
  {\bibinfo {title} {Interferometric {{Approach}} to {{Probing Fast
  Scrambling}}},\ }\href@noop {} {\bibfield  {journal} {\bibinfo  {journal}
  {arXiv:1607.01801 [cond-mat, physics:hep-th, physics:quant-ph]}\ } (\bibinfo
  {year} {2016})}\BibitemShut {NoStop}%
\bibitem [{\citenamefont {Vermersch}\ \emph {et~al.}(2019)\citenamefont
  {Vermersch}, \citenamefont {Elben}, \citenamefont {Sieberer}, \citenamefont
  {Yao},\ and\ \citenamefont {Zoller}}]{vermersch_probing_2019}%
  \BibitemOpen
  \bibfield  {author} {\bibinfo {author} {\bibfnamefont {B.}~\bibnamefont
  {Vermersch}}, \bibinfo {author} {\bibfnamefont {A.}~\bibnamefont {Elben}},
  \bibinfo {author} {\bibfnamefont {L.~M.}\ \bibnamefont {Sieberer}}, \bibinfo
  {author} {\bibfnamefont {N.~Y.}\ \bibnamefont {Yao}}, \ and\ \bibinfo
  {author} {\bibfnamefont {P.}~\bibnamefont {Zoller}},\ }\bibfield  {title}
  {\bibinfo {title} {Probing scrambling using statistical correlations between
  randomized measurements},\ }\href {\doibase 10.1103/physrevx.9.021061}
  {\bibfield  {journal} {\bibinfo  {journal} {Phys. Rev. X}\ }\textbf {\bibinfo
  {volume} {9}},\ \bibinfo {pages} {021061} (\bibinfo {year}
  {2019})}\BibitemShut {NoStop}%
\bibitem [{\citenamefont {Baum}\ and\ \citenamefont
  {Pines}(1986)}]{baum_nmr_1986}%
  \BibitemOpen
  \bibfield  {author} {\bibinfo {author} {\bibfnamefont {J.}~\bibnamefont
  {Baum}}\ and\ \bibinfo {author} {\bibfnamefont {A.}~\bibnamefont {Pines}},\
  }\bibfield  {title} {\bibinfo {title} {{NMR} studies of clustering in
  solids},\ }\href {\doibase 10.1021/ja00284a001} {\bibfield  {journal}
  {\bibinfo  {journal} {J. Am. Chem. Soc.}\ }\textbf {\bibinfo {volume}
  {108}},\ \bibinfo {pages} {7447} (\bibinfo {year} {1986})}\BibitemShut
  {NoStop}%
\bibitem [{\citenamefont {Yen}\ and\ \citenamefont
  {Pines}(1983)}]{yen_multiplequantum_1983}%
  \BibitemOpen
  \bibfield  {author} {\bibinfo {author} {\bibfnamefont {Y.}~\bibnamefont
  {Yen}}\ and\ \bibinfo {author} {\bibfnamefont {A.}~\bibnamefont {Pines}},\
  }\bibfield  {title} {\bibinfo {title} {Multiple-quantum {NMR} in solids},\
  }\href {\doibase 10.1063/1.445185} {\bibfield  {journal} {\bibinfo  {journal}
  {J. Chem. Phys.}\ }\textbf {\bibinfo {volume} {78}},\ \bibinfo {pages} {3579}
  (\bibinfo {year} {1983})}\BibitemShut {NoStop}%
\bibitem [{\citenamefont {Munowitz}\ and\ \citenamefont
  {Pines}(1986)}]{munowitz_multiple-quantum_1986}%
  \BibitemOpen
  \bibfield  {author} {\bibinfo {author} {\bibfnamefont {M.}~\bibnamefont
  {Munowitz}}\ and\ \bibinfo {author} {\bibfnamefont {A.}~\bibnamefont
  {Pines}},\ }\bibfield  {title} {{\selectlanguage {en}\bibinfo {title}
  {Multiple-quantum nuclear magnetic resonance spectroscopy},\ }}\href
  {\doibase 10.1126/science.233.4763.525} {\bibfield  {journal} {\bibinfo
  {journal} {Science}\ }\textbf {\bibinfo {volume} {233}},\ \bibinfo {pages}
  {525} (\bibinfo {year} {1986})}\BibitemShut {NoStop}%
\bibitem [{\citenamefont {Ernst}\ \emph {et~al.}(1998)\citenamefont {Ernst},
  \citenamefont {Meier}, \citenamefont {Tomaselli},\ and\ \citenamefont
  {Pines}}]{ernst_time-reversal_1998}%
  \BibitemOpen
  \bibfield  {author} {\bibinfo {author} {\bibfnamefont {M.}~\bibnamefont
  {Ernst}}, \bibinfo {author} {\bibfnamefont {B.~H.}\ \bibnamefont {Meier}},
  \bibinfo {author} {\bibfnamefont {M.}~\bibnamefont {Tomaselli}}, \ and\
  \bibinfo {author} {\bibfnamefont {A.}~\bibnamefont {Pines}},\ }\bibfield
  {title} {\bibinfo {title} {Time-reversal of cross-polarization in nuclear
  magnetic resonance},\ }\href {\doibase 10.1063/1.476435} {\bibfield
  {journal} {\bibinfo  {journal} {J. Chem. Phys.}\ }\textbf {\bibinfo {volume}
  {108}},\ \bibinfo {pages} {9611} (\bibinfo {year} {1998})}\BibitemShut
  {NoStop}%
\bibitem [{\citenamefont {Rhim}\ \emph {et~al.}(1971)\citenamefont {Rhim},
  \citenamefont {Pines},\ and\ \citenamefont
  {Waugh}}]{rhim_time-reversal_1971}%
  \BibitemOpen
  \bibfield  {author} {\bibinfo {author} {\bibfnamefont {W.-K.}\ \bibnamefont
  {Rhim}}, \bibinfo {author} {\bibfnamefont {A.}~\bibnamefont {Pines}}, \ and\
  \bibinfo {author} {\bibfnamefont {J.~S.}\ \bibnamefont {Waugh}},\ }\bibfield
  {title} {\bibinfo {title} {Time-reversal experiments in dipolar-coupled spin
  systems},\ }\href {\doibase 10.1103/physrevb.3.684} {\bibfield  {journal}
  {\bibinfo  {journal} {Phys. Rev. B}\ }\textbf {\bibinfo {volume} {3}},\
  \bibinfo {pages} {684} (\bibinfo {year} {1971})}\BibitemShut {NoStop}%
\bibitem [{\citenamefont {Suter}\ \emph {et~al.}(1987)\citenamefont {Suter},
  \citenamefont {Liu}, \citenamefont {Baum},\ and\ \citenamefont
  {Pines}}]{suter_multiple_1987}%
  \BibitemOpen
  \bibfield  {author} {\bibinfo {author} {\bibfnamefont {D.}~\bibnamefont
  {Suter}}, \bibinfo {author} {\bibfnamefont {S.}~\bibnamefont {Liu}}, \bibinfo
  {author} {\bibfnamefont {J.}~\bibnamefont {Baum}}, \ and\ \bibinfo {author}
  {\bibfnamefont {A.}~\bibnamefont {Pines}},\ }\bibfield  {title} {\bibinfo
  {title} {Multiple quantum {NMR} excitation with a one-quantum hamiltonian},\
  }\href {\doibase 10.1016/0301-0104(87)80023-x} {\bibfield  {journal}
  {\bibinfo  {journal} {Chem. Phys.}\ }\textbf {\bibinfo {volume} {114}},\
  \bibinfo {pages} {103} (\bibinfo {year} {1987})}\BibitemShut {NoStop}%
\bibitem [{\citenamefont {Schnell}\ and\ \citenamefont
  {Spiess}(2001)}]{schnell_high-resolution_2001}%
  \BibitemOpen
  \bibfield  {author} {\bibinfo {author} {\bibfnamefont {I.}~\bibnamefont
  {Schnell}}\ and\ \bibinfo {author} {\bibfnamefont {H.~W.}\ \bibnamefont
  {Spiess}},\ }\bibfield  {title} {{\selectlanguage {en}\bibinfo {title}
  {High-resolution {1H} {NMR} spectroscopy in the solid state: {Very} fast
  sample rotation and multiple-quantum coherences},\ }}\href {\doibase
  10.1006/jmre.2001.2336} {\bibfield  {journal} {\bibinfo  {journal} {J. Magn.
  Reson.}\ }\textbf {\bibinfo {volume} {151}},\ \bibinfo {pages} {153}
  (\bibinfo {year} {2001})}\BibitemShut {NoStop}%
\bibitem [{\citenamefont {Keselman}\ \emph {et~al.}(2021)\citenamefont
  {Keselman}, \citenamefont {Nie},\ and\ \citenamefont
  {Berg}}]{keselman_scrambling_2021}%
  \BibitemOpen
  \bibfield  {author} {\bibinfo {author} {\bibfnamefont {A.}~\bibnamefont
  {Keselman}}, \bibinfo {author} {\bibfnamefont {L.}~\bibnamefont {Nie}}, \
  and\ \bibinfo {author} {\bibfnamefont {E.}~\bibnamefont {Berg}},\ }\bibfield
  {title} {\bibinfo {title} {Scrambling and lyapunov exponent in spatially
  extended systems},\ }\href {\doibase 10.1103/physrevb.103.l121111} {\bibfield
   {journal} {\bibinfo  {journal} {Phys. Rev. B}\ }\textbf {\bibinfo {volume}
  {103}},\ \bibinfo {pages} {L121111} (\bibinfo {year} {2021})}\BibitemShut
  {NoStop}%
\bibitem [{\citenamefont {Kukuljan}\ \emph {et~al.}(2017)\citenamefont
  {Kukuljan}, \citenamefont {Grozdanov},\ and\ \citenamefont
  {Prosen}}]{kukuljan_weak_2017}%
  \BibitemOpen
  \bibfield  {author} {\bibinfo {author} {\bibfnamefont {I.}~\bibnamefont
  {Kukuljan}}, \bibinfo {author} {\bibfnamefont {S.}~\bibnamefont {Grozdanov}},
  \ and\ \bibinfo {author} {\bibfnamefont {T.}~\bibnamefont {Prosen}},\
  }\bibfield  {title} {\bibinfo {title} {Weak quantum chaos},\ }\href {\doibase
  10.1103/physrevb.96.060301} {\bibfield  {journal} {\bibinfo  {journal} {Phys.
  Rev. B}\ }\textbf {\bibinfo {volume} {96}},\ \bibinfo {pages} {060301}
  (\bibinfo {year} {2017})}\BibitemShut {NoStop}%
\bibitem [{\citenamefont {Munowitz}\ \emph {et~al.}(1987)\citenamefont
  {Munowitz}, \citenamefont {Pines},\ and\ \citenamefont
  {Mehring}}]{munowitz_multiplequantum_1987}%
  \BibitemOpen
  \bibfield  {author} {\bibinfo {author} {\bibfnamefont {M.}~\bibnamefont
  {Munowitz}}, \bibinfo {author} {\bibfnamefont {A.}~\bibnamefont {Pines}}, \
  and\ \bibinfo {author} {\bibfnamefont {M.}~\bibnamefont {Mehring}},\
  }\bibfield  {title} {\bibinfo {title} {Multiple-quantum dynamics in {NMR:}
  {A} directed walk through liouville space},\ }\href {\doibase
  10.1063/1.452028} {\bibfield  {journal} {\bibinfo  {journal} {J. Chem.
  Phys.}\ }\textbf {\bibinfo {volume} {86}},\ \bibinfo {pages} {3172} (\bibinfo
  {year} {1987})}\BibitemShut {NoStop}%
\bibitem [{\citenamefont {Yan}\ \emph {et~al.}(2013{\natexlab{b}})\citenamefont
  {Yan}, \citenamefont {Moses}, \citenamefont {Gadway}, \citenamefont {Covey},
  \citenamefont {Hazzard}, \citenamefont {Rey}, \citenamefont {Jin},\ and\
  \citenamefont {Ye}}]{Ye13}%
  \BibitemOpen
  \bibfield  {author} {\bibinfo {author} {\bibfnamefont {B.}~\bibnamefont
  {Yan}}, \bibinfo {author} {\bibfnamefont {S.~A.}\ \bibnamefont {Moses}},
  \bibinfo {author} {\bibfnamefont {B.}~\bibnamefont {Gadway}}, \bibinfo
  {author} {\bibfnamefont {J.~P.}\ \bibnamefont {Covey}}, \bibinfo {author}
  {\bibfnamefont {K.~R.~A.}\ \bibnamefont {Hazzard}}, \bibinfo {author}
  {\bibfnamefont {A.~M.}\ \bibnamefont {Rey}}, \bibinfo {author} {\bibfnamefont
  {D.~S.}\ \bibnamefont {Jin}}, \ and\ \bibinfo {author} {\bibfnamefont
  {J.}~\bibnamefont {Ye}},\ }\bibfield  {title} {\bibinfo {title} {Observation
  of dipolar spin-exchange interactions with lattice-confined polar
  molecules},\ }\href {\doibase 10.1038/nature12483} {\bibfield  {journal}
  {\bibinfo  {journal} {Nature}\ }\textbf {\bibinfo {volume} {501}},\ \bibinfo
  {pages} {521} (\bibinfo {year} {2013}{\natexlab{b}})}\BibitemShut {NoStop}%
\bibitem [{\citenamefont {Hazzard}\ \emph {et~al.}(2014)\citenamefont
  {Hazzard}, \citenamefont {Gadway}, \citenamefont {Foss-Feig}, \citenamefont
  {Yan}, \citenamefont {Moses}, \citenamefont {Covey}, \citenamefont {Yao},
  \citenamefont {Lukin}, \citenamefont {Ye}, \citenamefont {Jin},\ and\
  \citenamefont {Rey}}]{hazzard_many-body_2014}%
  \BibitemOpen
  \bibfield  {author} {\bibinfo {author} {\bibfnamefont {K.~R.~A.}\
  \bibnamefont {Hazzard}}, \bibinfo {author} {\bibfnamefont {B.}~\bibnamefont
  {Gadway}}, \bibinfo {author} {\bibfnamefont {M.}~\bibnamefont {Foss-Feig}},
  \bibinfo {author} {\bibfnamefont {B.}~\bibnamefont {Yan}}, \bibinfo {author}
  {\bibfnamefont {S.~A.}\ \bibnamefont {Moses}}, \bibinfo {author}
  {\bibfnamefont {J.~P.}\ \bibnamefont {Covey}}, \bibinfo {author}
  {\bibfnamefont {N.~Y.}\ \bibnamefont {Yao}}, \bibinfo {author} {\bibfnamefont
  {M.~D.}\ \bibnamefont {Lukin}}, \bibinfo {author} {\bibfnamefont
  {J.}~\bibnamefont {Ye}}, \bibinfo {author} {\bibfnamefont {D.~S.}\
  \bibnamefont {Jin}}, \ and\ \bibinfo {author} {\bibfnamefont {A.~M.}\
  \bibnamefont {Rey}},\ }\bibfield  {title} {\bibinfo {title} {Many-body
  dynamics of dipolar molecules in an optical lattice},\ }\href {\doibase
  10.1103/physrevlett.113.195302} {\bibfield  {journal} {\bibinfo  {journal}
  {Phys. Rev. Lett.}\ }\textbf {\bibinfo {volume} {113}},\ \bibinfo {pages}
  {195302} (\bibinfo {year} {2014})}\BibitemShut {NoStop}%
\bibitem [{Note1()}]{Note1}%
  \BibitemOpen
  \bibinfo {note} {$a$ can label the spatial coordinate as well as internal
  degrees of freedom}\BibitemShut {NoStop}%
\bibitem [{\citenamefont {Baum}\ \emph {et~al.}(1985)\citenamefont {Baum},
  \citenamefont {Munowitz}, \citenamefont {Garroway},\ and\ \citenamefont
  {Pines}}]{baum_multiplequantum_1985}%
  \BibitemOpen
  \bibfield  {author} {\bibinfo {author} {\bibfnamefont {J.}~\bibnamefont
  {Baum}}, \bibinfo {author} {\bibfnamefont {M.}~\bibnamefont {Munowitz}},
  \bibinfo {author} {\bibfnamefont {A.~N.}\ \bibnamefont {Garroway}}, \ and\
  \bibinfo {author} {\bibfnamefont {A.}~\bibnamefont {Pines}},\ }\bibfield
  {title} {\bibinfo {title} {Multiple-quantum dynamics in solid state {NMR}},\
  }\href {\doibase 10.1063/1.449344} {\bibfield  {journal} {\bibinfo  {journal}
  {J. Chem. Phys.}\ }\textbf {\bibinfo {volume} {83}},\ \bibinfo {pages} {2015}
  (\bibinfo {year} {1985})}\BibitemShut {NoStop}%
\bibitem [{\citenamefont {Cho}\ and\ \citenamefont
  {Yesinowski}(1996)}]{cho_h_1996}%
  \BibitemOpen
  \bibfield  {author} {\bibinfo {author} {\bibfnamefont {G.}~\bibnamefont
  {Cho}}\ and\ \bibinfo {author} {\bibfnamefont {J.~P.}\ \bibnamefont
  {Yesinowski}},\ }\bibfield  {title} {\bibinfo {title} {H and {19F}
  multiple-quantum {NMR} dynamics in quasi-one-dimensional spin clusters in
  apatites},\ }\href {\doibase 10.1021/jp9614815} {\bibfield  {journal}
  {\bibinfo  {journal} {J. Phys. Chem.}\ }\textbf {\bibinfo {volume} {100}},\
  \bibinfo {pages} {15716} (\bibinfo {year} {1996})}\BibitemShut {NoStop}%
\bibitem [{\citenamefont {Munowitz}\ and\ \citenamefont
  {Pines}(2007)}]{prigogine_principles_2007}%
  \BibitemOpen
  \bibfield  {author} {\bibinfo {author} {\bibfnamefont {M.}~\bibnamefont
  {Munowitz}}\ and\ \bibinfo {author} {\bibfnamefont {A.}~\bibnamefont
  {Pines}},\ }in\ \href {\doibase 10.1002/9780470142929.ch1} {{\selectlanguage
  {en}\emph {\bibinfo {booktitle} {Advances in Chemical Physics}}}},\ \bibinfo
  {editor} {edited by\ \bibinfo {editor} {\bibfnamefont {I.}~\bibnamefont
  {Prigogine}}\ and\ \bibinfo {editor} {\bibfnamefont {S.~A.}\ \bibnamefont
  {Rice}}}\ (\bibinfo  {publisher} {John Wiley \& Sons, Inc.},\ \bibinfo
  {address} {{Hoboken, NJ, USA}},\ \bibinfo {year} {2007})\ pp.\ \bibinfo
  {pages} {1--152}\BibitemShut {NoStop}%
\bibitem [{\citenamefont {Sánchez}\ \emph {et~al.}(2007)\citenamefont
  {Sánchez}, \citenamefont {Pastawski},\ and\ \citenamefont
  {Levstein}}]{sanchez_time_2007}%
  \BibitemOpen
  \bibfield  {author} {\bibinfo {author} {\bibfnamefont {C.~M.}\ \bibnamefont
  {Sánchez}}, \bibinfo {author} {\bibfnamefont {H.~M.}\ \bibnamefont
  {Pastawski}}, \ and\ \bibinfo {author} {\bibfnamefont {P.~R.}\ \bibnamefont
  {Levstein}},\ }\bibfield  {title} {{\selectlanguage {en}\bibinfo {title}
  {Time evolution of multiple quantum coherences in {NMR}},\ }}\href {\doibase
  10.1016/j.physb.2007.04.092} {\bibfield  {journal} {\bibinfo  {journal}
  {Physica B}\ }\textbf {\bibinfo {volume} {398}},\ \bibinfo {pages} {472}
  (\bibinfo {year} {2007})}\BibitemShut {NoStop}%
\bibitem [{\citenamefont {Sánchez}\ \emph {et~al.}(2017)\citenamefont
  {Sánchez}, \citenamefont {Buljubasich}, \citenamefont {Pastawski},\ and\
  \citenamefont {Chattah}}]{sanchez_evolution_2017}%
  \BibitemOpen
  \bibfield  {author} {\bibinfo {author} {\bibfnamefont {C.~M.}\ \bibnamefont
  {Sánchez}}, \bibinfo {author} {\bibfnamefont {L.}~\bibnamefont
  {Buljubasich}}, \bibinfo {author} {\bibfnamefont {H.~M.}\ \bibnamefont
  {Pastawski}}, \ and\ \bibinfo {author} {\bibfnamefont {A.~K.}\ \bibnamefont
  {Chattah}},\ }\bibfield  {title} {{\selectlanguage {en}\bibinfo {title}
  {Evolution of multiple quantum coherences with scaled dipolar
  {Hamiltonian}},\ }}\href {\doibase 10.1016/j.jmr.2017.05.009} {\bibfield
  {journal} {\bibinfo  {journal} {J. Magn. Reson.}\ }\textbf {\bibinfo {volume}
  {281}},\ \bibinfo {pages} {75} (\bibinfo {year} {2017})}\BibitemShut
  {NoStop}%
\bibitem [{\citenamefont {Jalabert}\ and\ \citenamefont
  {Pastawski}(2001)}]{PhysRevLett.86.2490}%
  \BibitemOpen
  \bibfield  {author} {\bibinfo {author} {\bibfnamefont {R.~A.}\ \bibnamefont
  {Jalabert}}\ and\ \bibinfo {author} {\bibfnamefont {H.~M.}\ \bibnamefont
  {Pastawski}},\ }\bibfield  {title} {\bibinfo {title} {Environment-independent
  decoherence rate in classically chaotic systems},\ }\href {\doibase
  10.1103/physrevlett.86.2490} {\bibfield  {journal} {\bibinfo  {journal}
  {Phys. Rev. Lett.}\ }\textbf {\bibinfo {volume} {86}},\ \bibinfo {pages}
  {2490} (\bibinfo {year} {2001})}\BibitemShut {NoStop}%
\bibitem [{\citenamefont {Khitrin}(1997)}]{khitrin_growth_1997}%
  \BibitemOpen
  \bibfield  {author} {\bibinfo {author} {\bibfnamefont {A.}~\bibnamefont
  {Khitrin}},\ }\bibfield  {title} {{\selectlanguage {en}\bibinfo {title}
  {Growth of {NMR} multiple-quantum coherences in quasi-one-dimensional
  systems},\ }}\href {\doibase 10.1016/s0009-2614(97)00661-1} {\bibfield
  {journal} {\bibinfo  {journal} {Chem. Phys. Lett.}\ }\textbf {\bibinfo
  {volume} {274}},\ \bibinfo {pages} {217} (\bibinfo {year}
  {1997})}\BibitemShut {NoStop}%
\bibitem [{Note2()}]{Note2}%
  \BibitemOpen
  \bibinfo {note} {Other variants that take the distribution to be a
  superposition of Gaussian functions with different cluster sizes also predict
  exponential growth~\cite
  {sanchez_quantum_2016,sanchez_clustering_2014-1}.}\BibitemShut {Stop}%
\bibitem [{\citenamefont {Levy}\ and\ \citenamefont
  {Gleason}(1992)}]{levy_multiple_1992}%
  \BibitemOpen
  \bibfield  {author} {\bibinfo {author} {\bibfnamefont {D.~H.}\ \bibnamefont
  {Levy}}\ and\ \bibinfo {author} {\bibfnamefont {K.~K.}\ \bibnamefont
  {Gleason}},\ }\bibfield  {title} {\bibinfo {title} {Multiple quantum nuclear
  magnetic resonance as a probe for the dimensionality of hydrogen in
  polycrystalline powders and diamond films},\ }\href {\doibase
  10.1021/j100199a056} {\bibfield  {journal} {\bibinfo  {journal} {J. Phys.
  Chem.}\ }\textbf {\bibinfo {volume} {96}},\ \bibinfo {pages} {8125} (\bibinfo
  {year} {1992})}\BibitemShut {NoStop}%
\bibitem [{\citenamefont {Dom{\'i}nguez}\ and\ \citenamefont
  {{\'A}lvarez}(2021)}]{dominguez_dynamics_2021}%
  \BibitemOpen
  \bibfield  {author} {\bibinfo {author} {\bibfnamefont {F.~D.}\ \bibnamefont
  {Dom{\'i}nguez}}\ and\ \bibinfo {author} {\bibfnamefont {G.~A.}\ \bibnamefont
  {{\'A}lvarez}},\ }\bibfield  {title} {\bibinfo {title} {Dynamics of quantum
  information scrambling under decoherence effects},\ }\href@noop {} {\bibfield
   {journal} {\bibinfo  {journal} {arXiv:2107.03870 [cond-mat,
  physics:quant-ph]}\ } (\bibinfo {year} {2021})}\BibitemShut {NoStop}%
\bibitem [{\citenamefont {Álvarez}\ \emph {et~al.}(2013)\citenamefont
  {Álvarez}, \citenamefont {Kaiser},\ and\ \citenamefont
  {Suter}}]{alvarez_quantum_2013}%
  \BibitemOpen
  \bibfield  {author} {\bibinfo {author} {\bibfnamefont {G.~A.}\ \bibnamefont
  {Álvarez}}, \bibinfo {author} {\bibfnamefont {R.}~\bibnamefont {Kaiser}}, \
  and\ \bibinfo {author} {\bibfnamefont {D.}~\bibnamefont {Suter}},\ }\bibfield
   {title} {\bibinfo {title} {Quantum simulations of localization effects with
  dipolar interactions},\ }\href {\doibase 10.1002/andp.201300096} {\bibfield
  {journal} {\bibinfo  {journal} {Ann. Phys.}\ }\textbf {\bibinfo {volume}
  {525}},\ \bibinfo {pages} {833} (\bibinfo {year} {2013})}\BibitemShut
  {NoStop}%
\bibitem [{\citenamefont {Keleş}\ \emph {et~al.}(2019)\citenamefont {Keleş},
  \citenamefont {Zhao},\ and\ \citenamefont {Liu}}]{keles_scrambling_2019}%
  \BibitemOpen
  \bibfield  {author} {\bibinfo {author} {\bibfnamefont {A.}~\bibnamefont
  {Keleş}}, \bibinfo {author} {\bibfnamefont {E.}~\bibnamefont {Zhao}}, \ and\
  \bibinfo {author} {\bibfnamefont {W.~V.}\ \bibnamefont {Liu}},\ }\bibfield
  {title} {\bibinfo {title} {Scrambling dynamics and many-body chaos in a
  random dipolar spin model},\ }\href {\doibase 10.1103/physreva.99.053620}
  {\bibfield  {journal} {\bibinfo  {journal} {Phys. Rev. A}\ }\textbf {\bibinfo
  {volume} {99}},\ \bibinfo {pages} {053620} (\bibinfo {year}
  {2019})}\BibitemShut {NoStop}%
\bibitem [{\citenamefont {Hallatschek}\ and\ \citenamefont
  {Fisher}(2014)}]{hallatschek_acceleration_2014}%
  \BibitemOpen
  \bibfield  {author} {\bibinfo {author} {\bibfnamefont {O.}~\bibnamefont
  {Hallatschek}}\ and\ \bibinfo {author} {\bibfnamefont {D.~S.}\ \bibnamefont
  {Fisher}},\ }\bibfield  {title} {{\selectlanguage {en}\bibinfo {title}
  {Acceleration of evolutionary spread by long-range dispersal},\ }}\href
  {\doibase 10.1073/pnas.1404663111} {\bibfield  {journal} {\bibinfo  {journal}
  {Proc Natl Acad Sci USA}\ }\textbf {\bibinfo {volume} {111}},\ \bibinfo
  {pages} {E4911} (\bibinfo {year} {2014})}\BibitemShut {NoStop}%
\bibitem [{\citenamefont {Chatterjee}\ and\ \citenamefont
  {S.~Dey}(2015)}]{chatterjee_multiple_2013}%
  \BibitemOpen
  \bibfield  {author} {\bibinfo {author} {\bibfnamefont {S.}~\bibnamefont
  {Chatterjee}}\ and\ \bibinfo {author} {\bibfnamefont {P.}~\bibnamefont
  {S.~Dey}},\ }\bibfield  {title} {\bibinfo {title} {Multiple phase transitions
  in long-range first-passage percolation on square lattices},\ }\href
  {\doibase 10.1002/cpa.21571} {\bibfield  {journal} {\bibinfo  {journal}
  {Commun. Pur. Appl. Math.}\ }\textbf {\bibinfo {volume} {69}},\ \bibinfo
  {pages} {203} (\bibinfo {year} {2015})}\BibitemShut {NoStop}%
\bibitem [{\citenamefont {Ni}\ \emph {et~al.}(2008)\citenamefont {Ni},
  \citenamefont {Ospelkaus}, \citenamefont {de~Miranda}, \citenamefont {Pe'er},
  \citenamefont {Neyenhuis}, \citenamefont {Zirbel}, \citenamefont
  {Kotochigova}, \citenamefont {Julienne}, \citenamefont {Jin},\ and\
  \citenamefont {Ye}}]{ni_high_2008}%
  \BibitemOpen
  \bibfield  {author} {\bibinfo {author} {\bibfnamefont {K.-K.}\ \bibnamefont
  {Ni}}, \bibinfo {author} {\bibfnamefont {S.}~\bibnamefont {Ospelkaus}},
  \bibinfo {author} {\bibfnamefont {M.~H.~G.}\ \bibnamefont {de~Miranda}},
  \bibinfo {author} {\bibfnamefont {A.}~\bibnamefont {Pe'er}}, \bibinfo
  {author} {\bibfnamefont {B.}~\bibnamefont {Neyenhuis}}, \bibinfo {author}
  {\bibfnamefont {J.~J.}\ \bibnamefont {Zirbel}}, \bibinfo {author}
  {\bibfnamefont {S.}~\bibnamefont {Kotochigova}}, \bibinfo {author}
  {\bibfnamefont {P.~S.}\ \bibnamefont {Julienne}}, \bibinfo {author}
  {\bibfnamefont {D.~S.}\ \bibnamefont {Jin}}, \ and\ \bibinfo {author}
  {\bibfnamefont {J.}~\bibnamefont {Ye}},\ }\bibfield  {title}
  {{\selectlanguage {en}\bibinfo {title} {A high phase-space-density gas of
  polar molecules},\ }}\href {\doibase 10.1126/science.1163861} {\bibfield
  {journal} {\bibinfo  {journal} {Science}\ }\textbf {\bibinfo {volume}
  {322}},\ \bibinfo {pages} {231} (\bibinfo {year} {2008})}\BibitemShut
  {NoStop}%
\bibitem [{\citenamefont {Gadway}\ and\ \citenamefont
  {Yan}(2016)}]{gadway_strongly_2016}%
  \BibitemOpen
  \bibfield  {author} {\bibinfo {author} {\bibfnamefont {B.}~\bibnamefont
  {Gadway}}\ and\ \bibinfo {author} {\bibfnamefont {B.}~\bibnamefont {Yan}},\
  }\bibfield  {title} {\bibinfo {title} {Strongly interacting ultracold polar
  molecules},\ }\href {\doibase 10.1088/0953-4075/49/15/152002} {\bibfield
  {journal} {\bibinfo  {journal} {J. Phys. B: At. Mol. Opt. Phys.}\ }\textbf
  {\bibinfo {volume} {49}},\ \bibinfo {pages} {152002} (\bibinfo {year}
  {2016})}\BibitemShut {NoStop}%
\bibitem [{\citenamefont {Álvarez}\ and\ \citenamefont
  {Suter}(2011)}]{alvarez_localization_2011}%
  \BibitemOpen
  \bibfield  {author} {\bibinfo {author} {\bibfnamefont {G.~A.}\ \bibnamefont
  {Álvarez}}\ and\ \bibinfo {author} {\bibfnamefont {D.}~\bibnamefont
  {Suter}},\ }\bibfield  {title} {\bibinfo {title} {Localization effects
  induced by decoherence in superpositions of many-spin quantum states},\
  }\href {\doibase 10.1103/physreva.84.012320} {\bibfield  {journal} {\bibinfo
  {journal} {Phys. Rev. A}\ }\textbf {\bibinfo {volume} {84}},\ \bibinfo
  {pages} {012320} (\bibinfo {year} {2011})}\BibitemShut {NoStop}%
\bibitem [{\citenamefont {Álvarez}\ and\ \citenamefont
  {Suter}(2010)}]{alvarez_nmr_2010}%
  \BibitemOpen
  \bibfield  {author} {\bibinfo {author} {\bibfnamefont {G.~A.}\ \bibnamefont
  {Álvarez}}\ and\ \bibinfo {author} {\bibfnamefont {D.}~\bibnamefont
  {Suter}},\ }\bibfield  {title} {\bibinfo {title} {{NMR} quantum simulation of
  localization effects induced by decoherence},\ }\href {\doibase
  10.1103/physrevlett.104.230403} {\bibfield  {journal} {\bibinfo  {journal}
  {Phys. Rev. Lett.}\ }\textbf {\bibinfo {volume} {104}},\ \bibinfo {pages}
  {230403} (\bibinfo {year} {2010})}\BibitemShut {NoStop}%
\bibitem [{\citenamefont {Zhou}\ and\ \citenamefont
  {Nahum}(2020)}]{zhou_entanglement_2020}%
  \BibitemOpen
  \bibfield  {author} {\bibinfo {author} {\bibfnamefont {T.}~\bibnamefont
  {Zhou}}\ and\ \bibinfo {author} {\bibfnamefont {A.}~\bibnamefont {Nahum}},\
  }\bibfield  {title} {\bibinfo {title} {Entanglement membrane in chaotic
  many-body systems},\ }\href {\doibase 10.1103/physrevx.10.031066} {\bibfield
  {journal} {\bibinfo  {journal} {Phys. Rev. X}\ }\textbf {\bibinfo {volume}
  {10}},\ \bibinfo {pages} {031066} (\bibinfo {year} {2020})}\BibitemShut
  {NoStop}%
\bibitem [{\citenamefont {Zhou}\ and\ \citenamefont
  {Nahum}(2019)}]{zhou_emergent_2019}%
  \BibitemOpen
  \bibfield  {author} {\bibinfo {author} {\bibfnamefont {T.}~\bibnamefont
  {Zhou}}\ and\ \bibinfo {author} {\bibfnamefont {A.}~\bibnamefont {Nahum}},\
  }\bibfield  {title} {\bibinfo {title} {Emergent statistical mechanics of
  entanglement in random unitary circuits},\ }\href {\doibase
  10.1103/physrevb.99.174205} {\bibfield  {journal} {\bibinfo  {journal} {Phys.
  Rev. B}\ }\textbf {\bibinfo {volume} {99}},\ \bibinfo {pages} {174205}
  (\bibinfo {year} {2019})}\BibitemShut {NoStop}%
\bibitem [{\citenamefont {Jonay}\ \emph {et~al.}(2018)\citenamefont {Jonay},
  \citenamefont {Huse},\ and\ \citenamefont
  {Nahum}}]{jonay_coarse-grained_2018}%
  \BibitemOpen
  \bibfield  {author} {\bibinfo {author} {\bibfnamefont {C.}~\bibnamefont
  {Jonay}}, \bibinfo {author} {\bibfnamefont {D.~A.}\ \bibnamefont {Huse}}, \
  and\ \bibinfo {author} {\bibfnamefont {A.}~\bibnamefont {Nahum}},\ }\bibfield
   {title} {\bibinfo {title} {Coarse-grained dynamics of operator and state
  entanglement},\ }\href@noop {} {\bibfield  {journal} {\bibinfo  {journal}
  {arXiv:1803.00089 [cond-mat, physics:hep-th, physics:nlin,
  physics:quant-ph]}\ } (\bibinfo {year} {2018})}\BibitemShut {NoStop}%
\bibitem [{\citenamefont {Nahum}\ \emph {et~al.}(2017)\citenamefont {Nahum},
  \citenamefont {Ruhman}, \citenamefont {Vijay},\ and\ \citenamefont
  {Haah}}]{nahum_quantum_2017}%
  \BibitemOpen
  \bibfield  {author} {\bibinfo {author} {\bibfnamefont {A.}~\bibnamefont
  {Nahum}}, \bibinfo {author} {\bibfnamefont {J.}~\bibnamefont {Ruhman}},
  \bibinfo {author} {\bibfnamefont {S.}~\bibnamefont {Vijay}}, \ and\ \bibinfo
  {author} {\bibfnamefont {J.}~\bibnamefont {Haah}},\ }\bibfield  {title}
  {\bibinfo {title} {Quantum entanglement growth under random unitary
  dynamics},\ }\href {\doibase 10.1103/physrevx.7.031016} {\bibfield  {journal}
  {\bibinfo  {journal} {Phys. Rev. X}\ }\textbf {\bibinfo {volume} {7}},\
  \bibinfo {pages} {031016} (\bibinfo {year} {2017})}\BibitemShut {NoStop}%
\bibitem [{\citenamefont {Bertini}\ \emph {et~al.}(2018)\citenamefont
  {Bertini}, \citenamefont {Kos},\ and\ \citenamefont {Prosen}}]{Bertini_2018}%
  \BibitemOpen
  \bibfield  {author} {\bibinfo {author} {\bibfnamefont {B.}~\bibnamefont
  {Bertini}}, \bibinfo {author} {\bibfnamefont {P.}~\bibnamefont {Kos}}, \ and\
  \bibinfo {author} {\bibfnamefont {T.}~\bibnamefont {Prosen}},\ }\bibfield
  {title} {\bibinfo {title} {Exact spectral form factor in a minimal model of
  many-body quantum chaos},\ }\href {\doibase 10.1103/physrevlett.121.264101}
  {\bibfield  {journal} {\bibinfo  {journal} {Phys. Rev. Lett.}\ }\textbf
  {\bibinfo {volume} {121}} (\bibinfo {year} {2018}),\
  10.1103/physrevlett.121.264101}\BibitemShut {NoStop}%
\bibitem [{\citenamefont {Sánchez}\ \emph {et~al.}(2016)\citenamefont
  {Sánchez}, \citenamefont {Levstein}, \citenamefont {Buljubasich},
  \citenamefont {Pastawski},\ and\ \citenamefont
  {Chattah}}]{sanchez_quantum_2016}%
  \BibitemOpen
  \bibfield  {author} {\bibinfo {author} {\bibfnamefont {C.~M.}\ \bibnamefont
  {Sánchez}}, \bibinfo {author} {\bibfnamefont {P.~R.}\ \bibnamefont
  {Levstein}}, \bibinfo {author} {\bibfnamefont {L.}~\bibnamefont
  {Buljubasich}}, \bibinfo {author} {\bibfnamefont {H.~M.}\ \bibnamefont
  {Pastawski}}, \ and\ \bibinfo {author} {\bibfnamefont {A.~K.}\ \bibnamefont
  {Chattah}},\ }\bibfield  {title} {\bibinfo {title} {Quantum dynamics of
  excitations and decoherence in many-spin systems detected with loschmidt
  echoes: {Its} relation to their spreading through the {Hilbert} space},\
  }\href {\doibase 10.1098/rsta.2015.0155} {\bibfield  {journal} {\bibinfo
  {journal} {Phil. Trans. R. Soc. A.}\ }\textbf {\bibinfo {volume} {374}},\
  \bibinfo {pages} {20150155} (\bibinfo {year} {2016})}\BibitemShut {NoStop}%
\bibitem [{\citenamefont {Sánchez}\ \emph
  {et~al.}(2014{\natexlab{b}})\citenamefont {Sánchez}, \citenamefont {Acosta},
  \citenamefont {Levstein}, \citenamefont {Pastawski},\ and\ \citenamefont
  {Chattah}}]{sanchez_clustering_2014-1}%
  \BibitemOpen
  \bibfield  {author} {\bibinfo {author} {\bibfnamefont {C.~M.}\ \bibnamefont
  {Sánchez}}, \bibinfo {author} {\bibfnamefont {R.~H.}\ \bibnamefont
  {Acosta}}, \bibinfo {author} {\bibfnamefont {P.~R.}\ \bibnamefont
  {Levstein}}, \bibinfo {author} {\bibfnamefont {H.~M.}\ \bibnamefont
  {Pastawski}}, \ and\ \bibinfo {author} {\bibfnamefont {A.~K.}\ \bibnamefont
  {Chattah}},\ }\bibfield  {title} {\bibinfo {title} {Clustering and
  decoherence of correlated spins under double quantum dynamics},\ }\href
  {\doibase 10.1103/physreva.90.042122} {\bibfield  {journal} {\bibinfo
  {journal} {Phys. Rev. A}\ }\textbf {\bibinfo {volume} {90}},\ \bibinfo
  {pages} {042122} (\bibinfo {year} {2014}{\natexlab{b}})}\BibitemShut
  {NoStop}%
\end{thebibliography}%

\end{document}